\documentclass[aps,prl,twocolumn,superscriptaddress,reprint,showpacs,longbibliography,amsmath,amssymb,nofootinbib]{revtex4-2}
\usepackage[utf8]{inputenc}
\usepackage[T1]{fontenc}
\usepackage{microtype}
\usepackage{graphicx}
\usepackage{mathbbol}
\usepackage{amssymb}
\usepackage{comment}
\usepackage[usenames,dvipsnames,svgnames,table]{xcolor}
\usepackage[colorlinks,linkcolor=blue,citecolor=blue,urlcolor=blue]{hyperref}
\usepackage{siunitx}
\usepackage{mathtools}
\usepackage[caption=false]{subfig}
\usepackage{comment}
\usepackage{gensymb}

\def\equationautorefname~#1\null{Eq.~(#1)\null}

\renewcommand{\Re}{\mathrm{Re}}
\renewcommand{\Im}{\mathrm{Im}}
\DeclareMathOperator{\cov}{cov}

\newcommand{\braket}[1]{\ensuremath{\langle{#1}\rangle}}
\newcommand{\bra}[1]{\langle #1 |}
\newcommand{\ket}[1]{| #1 \rangle}

\newcommand{\om}{\omega_\mathrm{m}}
\newcommand{\cm}{c_\mathrm{m}}
\newcommand{\phim}{\phi_\mathrm{m}}
\newcommand{\hbdg}{\mathcal{H}}
\newcommand{\hsd}{\mathcal{H}_{\text{SD}}}
\newcommand{\hsct}{\mathcal{H}_{\text{SCT}}}
\newcommand{\gpt}{G\mathcal{PT}}
\newcommand{\pglt}{\mathcal{P}_{gl}\mathcal{T}}

\newcommand{\hs}{\mathcal{H}_{\text{SD}}}
\newcommand{\kin}{\kappa_{\mathrm{in}}}

\def\equationautorefname~#1\null{Eq.~(#1)\null}

\font\myfont=cmr12 at 13pt
\title{\myfont Non-Hermitian chiral phononics through optomechanically-induced squeezing}

	\begin{abstract}
	Imposing chirality on a physical system engenders unconventional energy flow and responses, such as the Aharonov-Bohm effect and the topological quantum Hall phase for electrons in a symmetry-breaking magnetic field. Recently, great interest has arisen in combining that principle with broken Hermiticity to explore novel topological phases and applications. Here, we report unique phononic states formed when combining the controlled breaking of time-reversal symmetry with non-Hermitian dynamics, both induced through time-modulated radiation pressure forces in small nano-optomechanical networks. We observe chiral energy flow among mechanical resonators in a synthetic dimension and Aharonov-Bohm tuning of their hybridised modes. Introducing particle-non-conserving squeezing interactions, we discover a non-Hermitian Aharonov-Bohm effect in ring-shaped networks in which mechanical quasiparticles experience parametric gain. The resulting nontrivial complex mode spectra indicate flux-tuning of squeezing, exceptional points, instabilities and unidirectional phononic amplification. This rich new phenomenology points the way to the exploration of new non-Hermitian topological bosonic phases and applications in sensing and transport that exploit spatiotemporal symmetry breaking.
	\end{abstract}

\begin{document}
	\title{Non-Hermitian chiral phononics through optomechanically-induced squeezing}
	%
	\author{Javier del Pino}
	\thanks{J. d. P. and J. J. S. contributed  equally to this work.}
	\affiliation{Center for Nanophotonics, AMOLF, Science Park 104, 1098 XG Amsterdam, The Netherlands}
	\affiliation{Institute for Theoretical Physics, ETH Zürich, 8093 Zürich, Switzerland}
	\author{Jesse J. Slim}
	\thanks{J. d. P. and J. J. S. contributed equally to this work.}
	\affiliation{Center for Nanophotonics, AMOLF, Science Park 104, 1098 XG Amsterdam, The Netherlands}
	\author{Ewold Verhagen}
	\email{verhagen@amolf.nl}
	\affiliation{Center for Nanophotonics, AMOLF, Science Park 104, 1098 XG Amsterdam, The Netherlands}
	
	\maketitle
	\DeclareGraphicsExtensions{.pdf,.png,.jpg}
	\newpage
	
	From the Zeeman to the quantum Hall effect, magnetic fields biasing electronic systems alter their spectrum and imprint helicity on their eigenstates. Electrons travelling along a closed path gain a phase proportional to the enclosed magnetic flux that depends on direction –- evidencing broken time-reversal symmetry $\mathcal{T}$~\cite{peshkin1989aharonov}. Resulting interference phenomena enable unidirectional transport and shift energy levels, leading to topologically nontrivial band structures and chiral conduction channels. Recent years have seen an exploding interest in bringing such geometrical phases~\cite{Cohen2019} and the resulting synthetic magnetism to bosonic systems in photonics, acoustics, and cold atoms, to explore nonreciprocal functionality~\cite{Sliwa2015,ruesink2016nonreciprocity,fang2017generalized,MercierdeLepinay2020} and various topological insulators~\cite{goldman2016topological,Huber2016,Ozawa2019}. 
	
	In a parallel, largely unconnected development, researchers turned to non-Hermitian systems~\cite{moiseyev2011non} such as parity-time ($\mathcal{PT}$) symmetric systems~\cite{bender1998real,Ruter2010,Hodaei2017}, featuring dynamical phase transitions linked to spectral singularities such as exceptional points~\cite{Mirieaar7709}. Here, controlled gain and loss are the resources that lead to unique eigenmode symmetries and tuning of \textit{complex} eigenfrequencies $\epsilon$. Bosonic systems form the natural realm for these phenomena, with lasing and self-oscillation ubiquitous in photonics and mechanics. In particular, bosonic \emph{squeezing} is described by Hamiltonians that do not conserve excitation number, and can induce effective non-Hermitian dynamics and distinct phases with either stable, decaying, or unboundedly growing dynamics~\cite{Flynn2020}. 
	
	Very recently, the combination of topology and non-Hermiticity has attracted strong interest~\cite{coulais2021topology,Bergholtz2019}.  Tailoring gain and loss in topological insulators showed lasing into protected states~\cite{st2017lasing,Bandres2018,Hu2021} and topological phase transitions~\cite{zeuner2015observation}. In principle, one could expect states with symmetries, dynamics, and spectra that are altogether different from Hermitian chiral systems~\cite{Lieu2018,Gong2018}. Indeed, various unique non-Hermitian topological phases have been predicted, with associated phenomena including chirally-amplified and unstable edge modes~\cite{barnett2013edge,Peano2016a,Peano2016d}, quadrature-dependent chiral transport~\cite{McDonald2018,Wanjura2020} and anomalous bulk-boundary correspondence accompanied by extreme sensitivity to boundary conditions~\cite{Bergholtz2019}, as recently observed~\cite{ghatak2020observation,Helbig2020,Weidemann2020,Wang2021}. So far, the rich resources of squeezing interactions and geometrical phases have however remained experimentally unexplored in this context.
	
	Here we demonstrate Aharonov-Bohm (AB) interference and chirality of nanomechanical states in multi-resonator networks where both $\mathcal{T}$-breaking geometrical phases and non-Hermiticity are induced through radiation pressure. On the one hand, optomechanical interactions are widely used to establish laser-controlled mechanical amplification and damping, through dynamical backaction or parametric driving~\cite{Aspelmeyer2013}. On the other hand, optomechanical control allowed synthetic magnetism for photons~\cite{ruesink2016nonreciprocity,fang2017generalized} and phonons~\cite{xu2018nonreciprocal, Mathew2020}, since suitable laser drives can stimulate frequency-converting transitions. We combine both here, using optomechanical particle-conserving and squeezing interactions to create non-Hermitian dynamics without dissipation~\cite{Koutserimpas2017,Wang2019} and uncover new geometrical phases. With the extreme precision with which light can actuate and detect nanomechanical motion, we reveal the unique effects of this merger on chiral transport, dynamical phases, and squeezing –- and actively control them in space and time.
	
	\begin{figure*}[!ht]
		\includegraphics[width=1\linewidth]{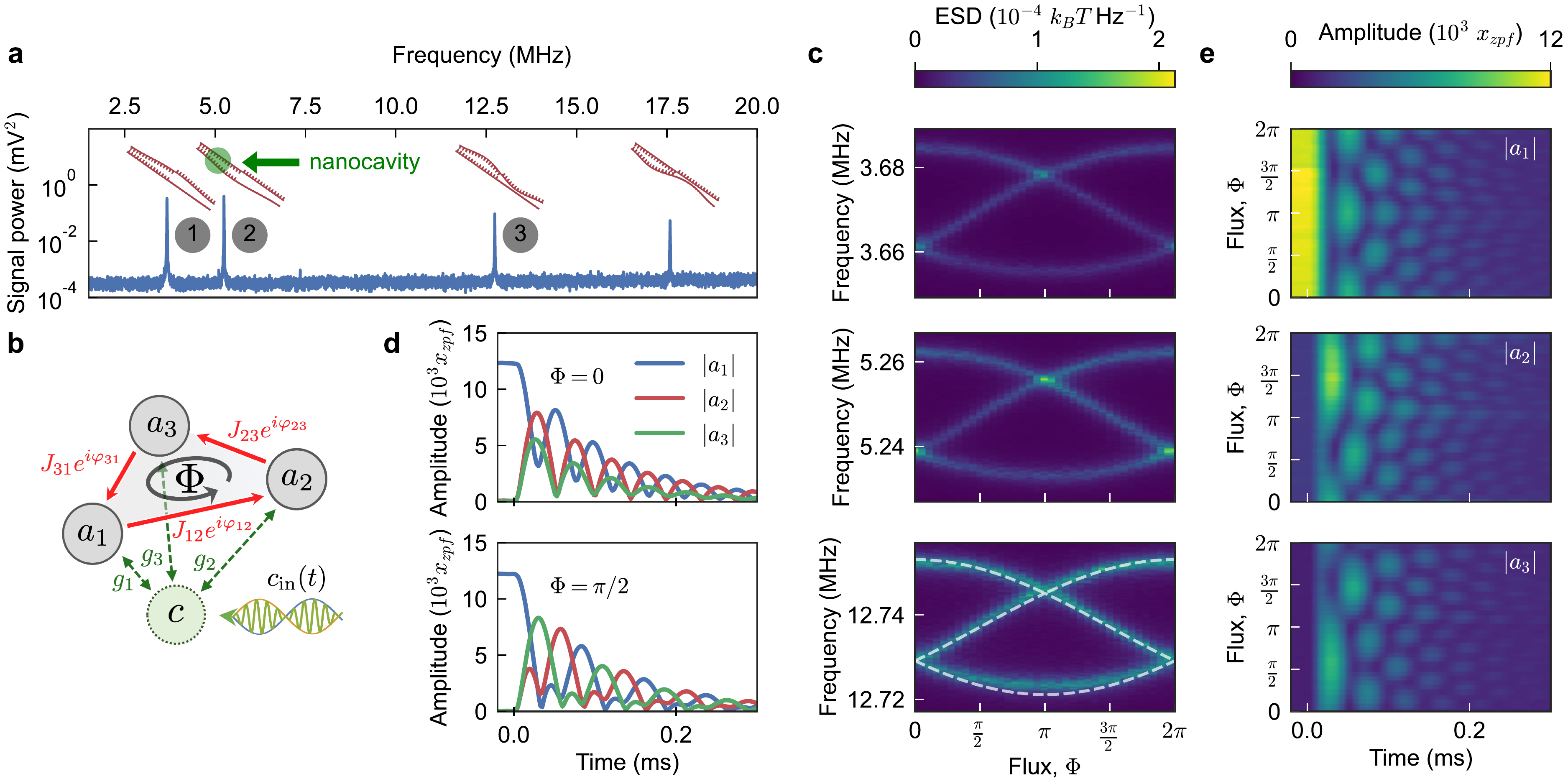}
		\caption{\textbf{Aharonov-Bohm interference in a Hermitian nano-optomechanical network.}  \textbf{a} Thermomechanical fluctuation spectrum of the sliced photonic crystal nanobeam, imprinted on a laser reflected from a nanocavity with linewidth $\kappa/(2\pi)=320$ GHz. Resonances correspond to mechanical flexural modes at frequencies $\omega_i/(2\pi)$ = \{3.7, 5.3, 12.8\} MHz with loss rates $\gamma_i/(2\pi) \approx 1 - 3$ kHz. \textbf{b} The modulated cavity field $c$ couples three resonators in a loop with rates $J_{ij}/(2\pi) = 8$ kHz and Peierls phases $\varphi_{ij}$, adding up to flux $\Phi$. \textbf{c} Thermomechanical noise spectra imprinted on the detection laser around each resonator's sideband versus flux. Hybridised Floquet modes tune with synthetic flux. \textbf{d} Time evolution of resonator amplitudes $|\langle a_i\rangle|\equiv|a_i|$ for $\mathcal{T}$ unbroken ($\Phi = 0$) and broken ($\Phi = \pi/2$). Resonator $a_1$ is coherently driven until $t = 0$ ms, when excitation is stopped and couplings are established. \textbf{e} Time evolution of resonator amplitudes for varying flux, showing crossover from helical to non-helical transport through an intermediate regime with generally aperiodic dynamics, and reversal of chirality with flux sign ($\Phi\mapsto -\Phi$). ESD, energy spectral density.} \label{fig:1}
	\end{figure*}

	We first induce phononic chirality through time-reversal symmetry breaking in a network with Hermitian closed-system dynamics, henceforth simply called \emph{Hermitian}. We use a sliced photonic crystal nanobeam~\cite{leijssen2017nonlinear} supporting multiple non-degenerate MHz-frequency flexural mechanical modes coupled to the optical field of a nanocavity. Each mode (‘resonator’) $i$ changes the cavity frequency by an amount $g_{0}^{(i)}x_i$ through a displacement $x_i$ (normalised to the zero-point amplitude) and experiences a force $\propto g_{0}^{(i)}n_c$, with $g_{0}^{(i)}$ the vacuum optomechanical coupling rate and $n_c$ the intracavity photon number. Figure \ref{fig:1}a shows the system’s mechanical resonances at distinct frequencies in the thermomechanical noise spectrum, read out as modulations of a probe laser (detuning $\Delta_\text{probe}\approx-2.5\kappa$) reflected from the cavity, with decay rate $\kappa$, at normal incidence.

	While the mechanical resonators have well-separated frequencies $\omega_i$, they are made to interact by temporal modulation of the intensity of a control laser detuned from cavity resonance. Thus, the mechanical spectrum serves as a synthetic dimension~\cite{Ozawa2019}, along which we study mode hybridisation and excitation transport. For optimal laser detuning $\Delta=-\kappa/(2\sqrt{3})$, mechanical displacement modulates the intracavity intensity instantaneously at mechanical timescales ($\kappa\gg\omega_i$). The mixing of a control laser intensity modulation at the difference frequency $\omega_j-\omega_i$ of resonators $i$ and $j$ and the radiation pressure force sideband of resonator $i$ creates a sideband resonant at $\omega_j$. This results in a `cross-mode optical spring effect' \cite{Mathew2020} that induces linear, particle-conserving beamsplitter coupling between the resonators at a rate $J_{ij}=\cm g_ig_j\Delta/(\Delta^2+\kappa^2/4)$, with $g_i = g_{0}^{(i)} \sqrt{\bar{n}_c}$ the optomechanical coupling enhanced by the average cavity population $\bar{n}_c$ and $\cm$ the modulation depth (Methods).

	The three lowest-frequency resonators are coupled in a ring network by simultaneously applying three suitable modulation tones (cf.~\autoref{fig:1}b). Describing the resonators in frames rotating at their resonance frequencies, the Hamiltonian for this `beamsplitter trimer’ (BST) reads
	\begin{align}
	H_{\mathrm{BST}}=\sum_{i=1,j\neq i}^{3}J_{ij}e^{-i\varphi_{ij}} a_i^\dag a_j,&&\varphi_{ji}=-\varphi_{ij}.  
	\label{eq:bst-hamiltonian}
	\end{align}
	This Hamiltonian is phonon-number-preserving, but importantly imprints the modulation phase $\varphi_{ij}$ in a nonreciprocal fashion on the transfer of phonons along links in the loop –- precisely like the Peierls phase imprinted by a magnetic vector potential~\cite{Fang2012,Mathew2020}. The gauge-invariant geometrical phase $\Phi=\varphi_{12}+\varphi_{23}+\varphi_{31}$ around the loop of resonators then represents a synthetic magnetic flux threading the plaquette.

	\begin{figure*}[!ht]
	\centering
		\includegraphics[width=0.8\linewidth]{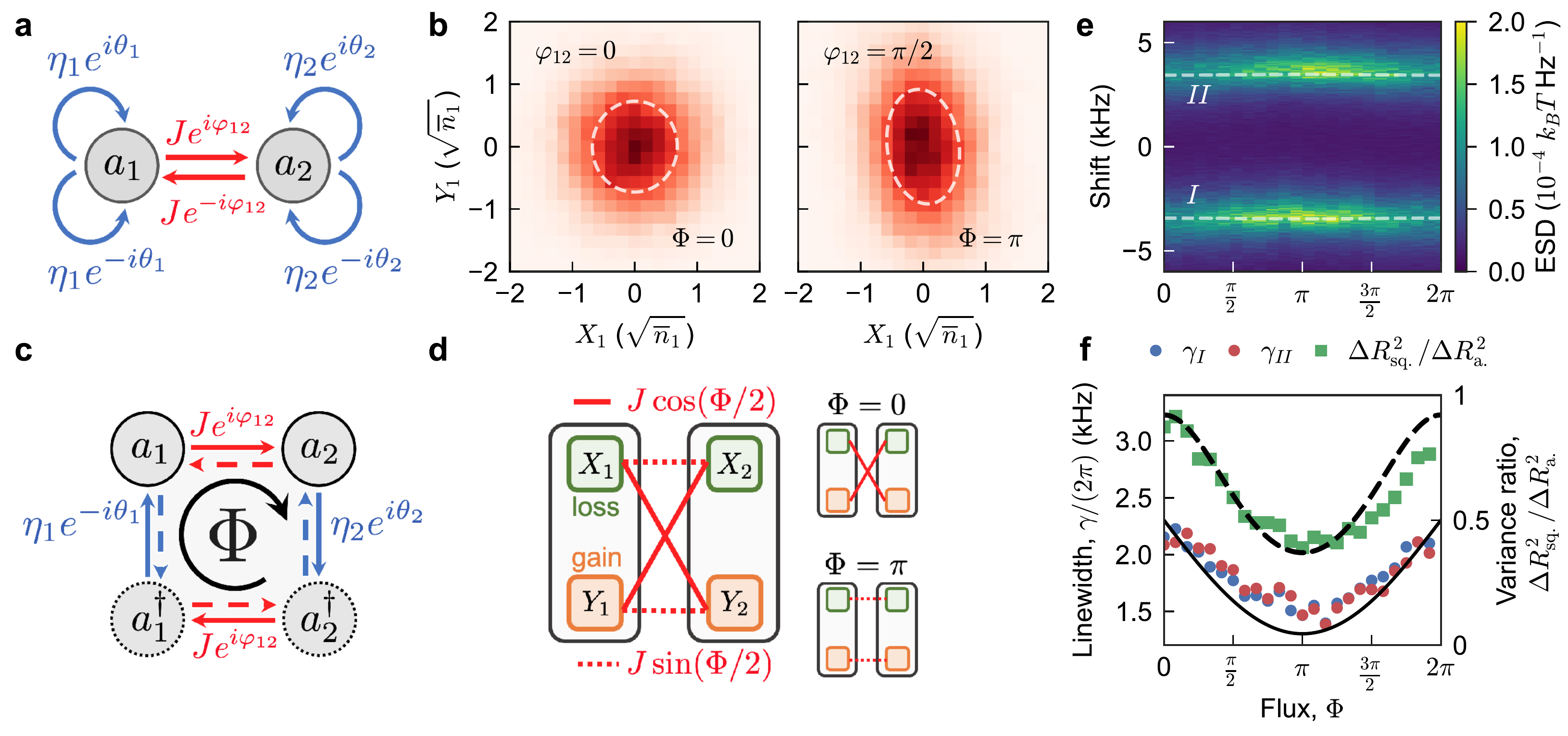}
		\caption{\textbf{AB interference along non-Hermitian squeezing loops}: \textbf{a} The squeezing dimer encompasses two resonators driven at $2\omega_i$ and $\omega_2-\omega_1$. These introduce single-mode squeezing (blue self-loops) and beamsplitter coupling (red). \textbf{b} Histograms of the steady-state phase space distribution of resonator 1 for varying beamsplitter Peierls phase $\varphi_{12}$, showing its effect on thermomechanical squeezing. Dashed ellipses depict the standard deviation of the principal components of the quadrature covariance matrix. Here $\theta_1=\theta_2=\pi/2$. \textbf{c} Graph associated to the Hamiltonian matrix (Methods Eq. 3), unwrapping self-loops in \textbf{a} over particles (annihilated by $a_i$) and holes (annihilated by $a_i^{\dagger}$). The clockwise loop 
		is threaded by synthetic flux $\Phi$, 
		the counterclockwise by $-\Phi$. \textbf{d} Coupling diagram for resonator quadratures, where $\Phi$ controls coupling between squeezed (green) and anti-squeezed (orange) quadratures of the two resonators. \textbf{e} Thermomechanical spectra for the SD around $\omega_1$. \textbf{f} Sweeping flux continuously tunes the fitted apparent resonance linewidths $\gamma_{I,II}$ (blue and red circles), compared to the theoretical loss rate of the lowest-loss eigenfrequency of $\hsd$ (solid curve). Flux-dependent level of squeezing, measured as the ratio of the variances $\Delta R_\text{sq.}^2$ and $\Delta R_\text{a.}^2$ of the quadratures squeezed and antisqueezed along the principal axes of the covariance matrix, respectively, in experiment (green squares) and theory (dashed curve, Supplementary Information~\autoref{sec:flux_squeezing} and~\autoref{sec:SD_as_AB}). Here, $J/(2\pi) = 3.5$ kHz, $\eta_1/(2\pi) = \eta_2/(2\pi) = 0.5$ kHz, and average loss is $\overline{\gamma}/(2\pi) = 2.3$ kHz. ESD, energy spectral density. 
		}\label{fig:2}
	\end{figure*}
	
	Setting equal $J_{ij}=J$, Hamiltonian (\autoref{eq:bst-hamiltonian}) is translationally invariant in a gauge with equal Peierls phases, and therefore diagonal in the momentum basis $\tilde{a}_k=\sum_{j=1}^{3}{e^{i2\pi kj/3}a_j}/\sqrt{3}$ for discrete momenta $k=\{-1,0,1\}$. Through AB interference along the loop, the enclosed flux shifts the eigenfrequencies $\epsilon_k=2J\cos{\left((2\pi k+\Phi)/3\right)}$~\cite{peshkin1989aharonov}. Figure~\ref{fig:1}c reveals these states in the thermomechanical spectra, for each of the resonances splits into a (Floquet) triplet due to strong coupling $J>\gamma_i$, with mechanical damping rates $\gamma_i$. This demonstration of nanomechanical flux-tuning is e.g. paralleled in the conductance of Josephson junctions~\cite{peshkin1989aharonov} and spectra of quantum rings under magnetic fields~\cite{Fuhrer2001}.  
	
	The flux-tuning manifests AB interference over a given sense of rotation --- the mechanism ultimately responsible for chirality of quantum Hall edge states~\cite{Ozawa2019} and nonreciprocal dynamics~\cite{Roushan2017}. Figures \ref{fig:1}d,e show the evolution of a mechanical excitation, initialised in resonator 1 through resonantly modulated radiation pressure. At time $t=0$ ms, its driving is switched off and the modulation implementing \autoref{eq:bst-hamiltonian} is switched on. For $\Phi\in\{0,\pi\}$,
	the BST is time-reversal symmetric (the Hamiltonian matrix $H$ obeys $H=H^\ast$ in some gauge, Methods) and energy simultaneously hops to both other resonators. For any other flux, breaking $\mathcal{T}$ lifts the degeneracy between modes with opposite inter-resonator phase lag ($\tilde{a}_{\pm 1}$ for $\Phi=0$), enabling chiral energy transport. For $\Phi=\pi/2$ ($\Phi=-\pi/2$), this circulates the loop in a clockwise (counterclockwise) fashion, with intermode exchange time $\tau_e=2\pi/(3J\sqrt3)$. 
	
	We thus demonstrated a chiral phononic circulator~\cite{Habraken2012} using light-induced nanomechanical beamsplitter interactions, with scaling potential to topological lattices~\cite{Mathew2020}. Still, vastly richer phenomenology is uncovered by introducing squeezing interactions in the nodes and links of the network. We implement single-mode ($i=j$) or two-mode ($i\neq j$) squeezing by modulating the radiation pressure at a sum-frequency $\omega_i+\omega_j$. The Hamiltonian reads (Methods)
	\begin{equation}\label{eq:squeezing_int}
		H^{\mathrm{sq}}=\sum_{i,j} \frac{\eta_{ij}}{2}(e^{i\theta_{ij}}a_ia_j+e^{-i\theta_{ij}}a_i^{\dagger}a_j^{\dagger}),
	\end{equation}
	with interaction strength $\eta_{ij}=\cm g_ig_j\Delta/(\Delta^2+\kappa^2/4)$ and modulation phase $\theta_{ij}$ now imprinted on the creation or annihilation of phonon pairs. Squeezing angles $\theta_{ij}$ form a powerful control resource, as the Peierls phases $\varphi_{ij}$ before. Indeed, spatially controlled squeezing -- providing anomalous pairing terms -- enables topological bosonic states unparalleled by their fermionic (topological superconductor) counterparts and is essential for proposed topological amplifiers~\cite{Peano2016d}.

	
	We first consider a `squeezing dimer' (SD,~\autoref{fig:2}a) consisting of two resonators, each experiencing single-mode squeezing through modulation at $2\omega_1$, $2\omega_2$ and mutually coupled through a drive at $\omega_2 - \omega_1$ (Hamiltonian $H_\text{SD} = \eta_1 e^{i\theta_1} a_1^2/2 + \eta_2 e^{i\theta_2} a_2^2/2 + J e^{i\varphi_{12}} a_2^\dagger a_1 + \text{H.c.}$). Remarkably, we find that the \textit{level} of squeezing of thermal fluctuations is not only determined by the magnitude of the interactions $\eta_i$, $J$, but also by their phases $\theta_i$, $\varphi_{12}$. Figure~\ref{fig:2}b shows experimental phase-space distributions for $\eta_1 = \eta_2 = \eta$, defining quadratures such that, for $J=0$, $X_i=(a_i+a_i^{\dagger})/\sqrt{2}$ ($Y_i=i(a_i^{\dagger}-a_i)/\sqrt{2}$) are squeezed (anti-squeezed), i.e. $\theta_1=\theta_2=\pi/2$. With beam-splitter coupling $J \gg \eta$, we observe that single-mode squeezing is maximal when $\varphi_{12}=\pi/2$, but essentially disappears if $\varphi_{12}\in\{0, \pi\}$. 
	
	We now show that this observation can be associated with a \textit{non-Hermitian} version of AB interference. Even though the coupled-mode picture \autoref{fig:2}a shows no plaquette, we can recognise a loop along which excitations experience a geometric phase when we combine graph representation~\cite{Ranzani2015} with Bogoliubov-de Gennes (BdG) formalism~\cite{Flynn2020}. The latter treats $a_i$ and $a_i^{\dagger}$ as separate degrees of freedom -- ‘particles’ and ‘holes’ -- and squeezing (\autoref{eq:squeezing_int}) as particle-hole conversion. Crucially, this representation (\autoref{fig:2}c) reveals for SD a conjugate pair of superimposed loops in particle-hole space, threaded by gauge-invariant fluxes $\Phi = 2\varphi_{12} - \theta_1 + \theta_2$ and $-\Phi$. 
    As these fluxes change interference conditions in the loop, they control the connection between quadratures in the two resonators: While for $\Phi=\pi$ the squeezed quadratures are connected and squeezing is maximal, for $\Phi=0$ the squeezed quadrature $X_1$ is connected to the anti-squeezed quadrature $Y_2$ and vice versa, cancelling the overall squeezing (\autoref{fig:2}d, Methods).
	
	This geometric phase again impacts the normal mode frequencies, which are now generally complex. These correspond to the eigenvalues of the BdG dynamical matrix $\hsd$, which defines the equations of motion $i\dot{\vec{\alpha}}=\hsd \vec{\alpha}$, where $\vec{\alpha}=(a_1,a_2,a_1^{\dagger},a_2^{\dagger})$ (Supplementary Information~\autoref{sec:SD_as_AB}), while the associated vectors form a $\Sigma_z$-orthonormal eigenbasis ($\Sigma_z=\mathrm{diag}(\mathbb{1},-\mathbb{1})$). Even without dissipation ($\gamma_i=0$), $\hsd$ is necessarily non-Hermitian, preserving only $\Sigma_z$-pseudo-Hermiticity ($\hsd^\dagger = \Sigma_z \hsd\Sigma_z$) to satisfy bosonic commutation relations~\cite{Flynn2020}. 
	AB-like interference in the BdG loop thus acquires a non-Hermitian character, where now frequency \textit{and} linewidth evolve with flux. Indeed, the thermomechanical spectra in the strongly coupled, dynamically stable regime ($J > \eta$, $2\eta < \gamma_i$, ~\autoref{fig:2}e,f), show that $\Phi$ strongly tunes linewidth and thermal amplitude of the hybridised eigenmodes, in unison with squeezing. The squeezed and antisqueezed partners that we recognised for $\Phi=\pi$ in~\autoref{fig:2}d correspond to a broad and narrow resonance, respectively~\cite{Huber2020}, with the latter dominating the spectrum (Methods).

\begin{figure}
		\includegraphics[width=1\linewidth]{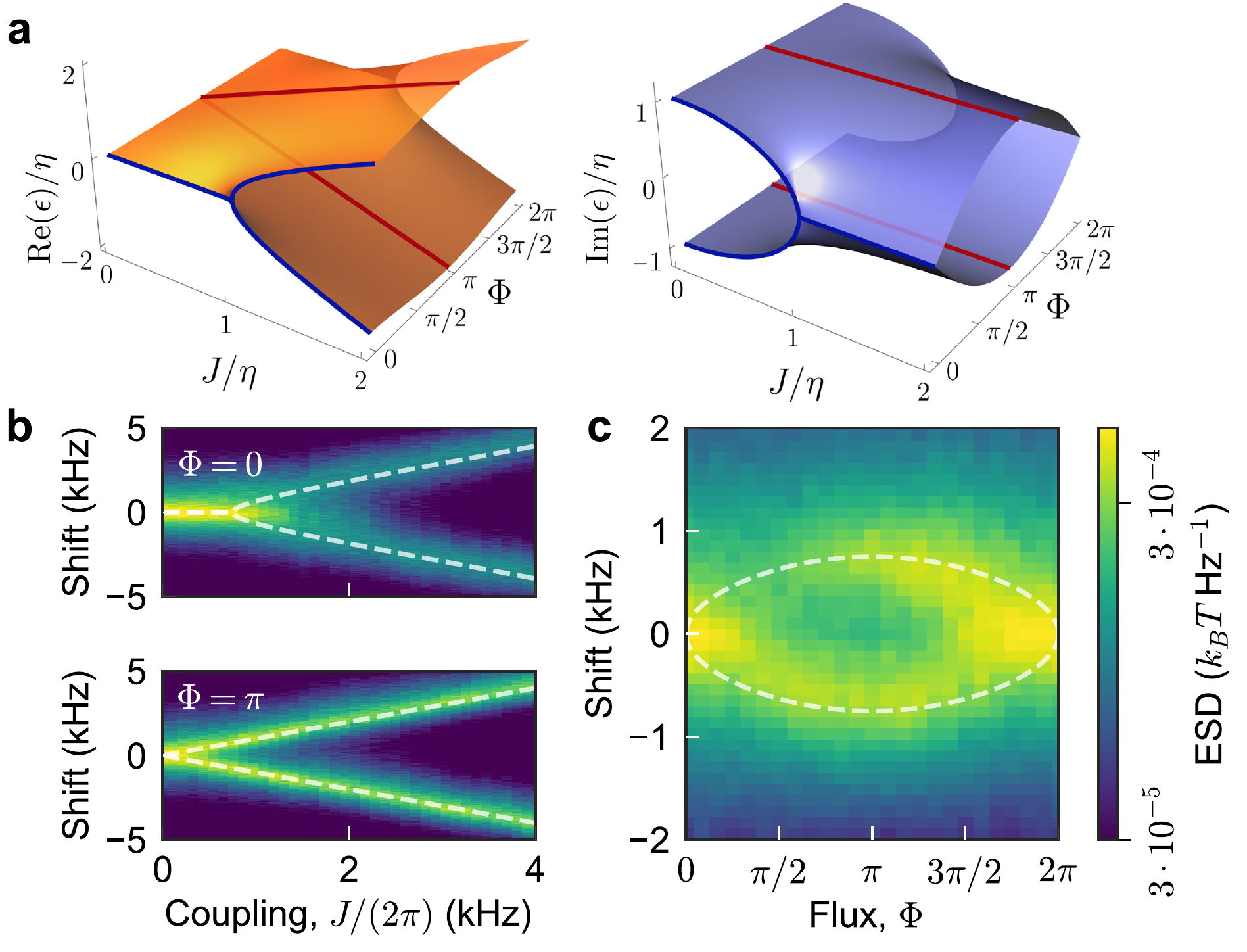}
		\caption{ \textbf{Flux-control of non-Hermitian dynamical phases.} \textbf{a} Complex eigenfrequency surfaces of the SD in $J-\Phi$ space for $\gamma_i=0$, tuned by the non-Hermitian AB effect acting on its beamsplitter and squeezing links. The surfaces are two-fold degenerate except for $\Phi\in\{0,2\pi\}$ and $\eta=J$, where $\mathcal{PT}$ symmetry breaks spontaneously and the eigenspectrum coalesces into two $2^{\mathrm{nd}}$ order EPs. \textbf{b} Fingerprints of complex degeneracies in the thermomechanical spectra for resonator 1 at $\eta/(2\pi)=0.75$ kHz and varying $J$. Nonzero flux breaks $\mathcal{P}_{X_iY_j}\mathcal{T}$ symmetry explicitly, precluding EPs. \textbf{c} Flux-tuned spectra for resonator 1 when $J/(2\pi)\approx\eta/(2\pi)=0.75$ kHz, showing mode coalescence at the EP at $\Phi\in\{0,2\pi\}$. For \textbf{b} and \textbf{c}, theory eigenvalues $\Re(\epsilon)$ are shown as dashed lines.} \label{fig:3}
	\end{figure}

	The complex eigenvalues define surfaces in $J/\eta -\Phi$ space, with varying degeneracy -- and symmetry of  $\hsd$ -- indicating distinct dynamical phases. Their physical properties, linked to $\mathcal{PT}$-symmetric systems, are readily appreciated by studying the dynamical matrix in the quadrature basis $\hsd^{XY}$. 
	For $\Phi=0$, $\mathcal{H}_\text{SD}^{XY}$ respects $\mathcal{P}_{X_iY_j}\mathcal{T}$-symmetry for the two degenerate ``quadrature dimers'' $X_i Y_{j\neq i}$ (\autoref{fig:2}d), where $\mathcal{P}_{X_iY_j}$ exchanges $X_i \leftrightarrow Y_{j}$. We thus demonstrate $\mathcal{PT}$-symmetric physics by means of squeezing dynamics, instead of coupling to dissipative baths~\cite{Koutserimpas2017,Wang2019}. In consequence, the SD features a pair of complex eigensurfaces, two-fold degenerate in their real and imaginary parts (\autoref{fig:3}a). The only effect of non-zero but equal dissipation rates is a uniform displacement of $\Im(\epsilon)$~\cite{Ornigotti2014}.

		\begin{figure*}[!ht]
		\includegraphics[width=\linewidth]{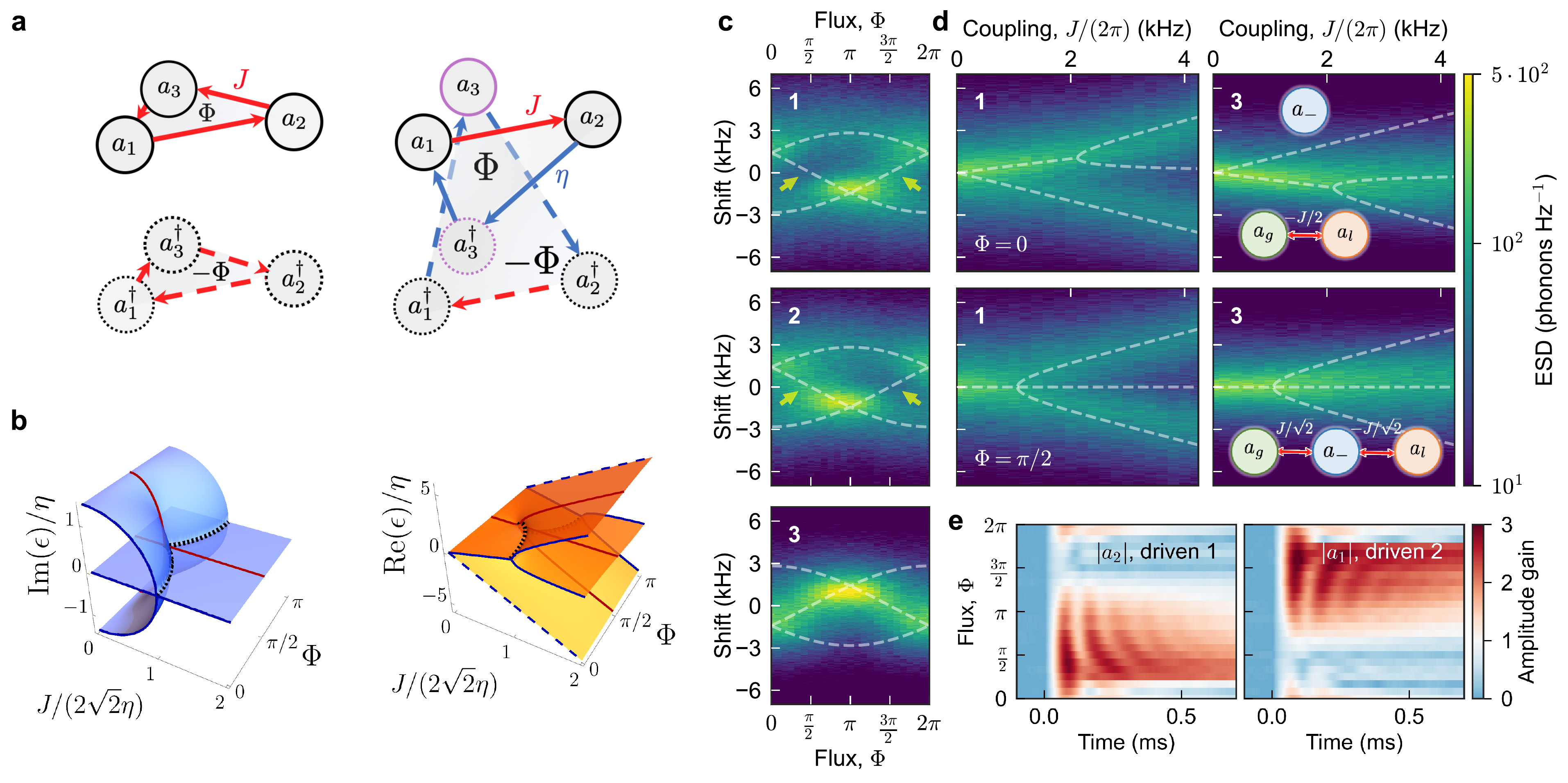}
		\caption{\textbf{Chirality in a non-Hermitian network.} \textbf{a} Sketch of the networks in particle-hole space corresponding to the BST (left) and SCT (right), manifesting their topological resemblance: The diagrams involve disjoint loops with no $a_i- a_i^{\dagger}$ connections. \textbf{b} Complex eigensurfaces for the SCT ($\gamma_i=\gamma$) depicted from $\Phi=0$ to $\Phi=\pi$ for clarity. Imaginary parts are referenced to $\gamma$. \textbf{c} Thermomechanical spectra of the three resonators (label denoted in the plot) for $\eta/(2\pi) = 1$ kHz, $J = 2\sqrt{2}\eta$. Feedback is employed to equalise mechanical loss rates $\gamma_i/(2\pi) = \gamma/(2\pi) = 4$ kHz. The sideband of the `conjugated’ resonator 3 is reflected in frequency compared to the other two. Localisation of eigenstates is observed, including 1-2 asymmetry indicated by arrows. Theoretical eigenfrequencies are shown as dashed lines.   \textbf{d} Spectra for resonators 1 and 3 for $\eta/(2\pi)=0.75$ kHz for a trivial flux $\Phi=0$, and in the maximally chiral case $\Phi=\pi/2$. The breaking of inversion symmetry for resonators 1 and 2 morphs a $2^\text{nd}$ order EP into a $3^\text{rd}$ order one, where $\pglt$ is spontaneously broken (see text). Insets show the effective $\pglt$ dimer/trimer structure for both flux values. \textbf{e} Ratio between instantaneous and initial coherent amplitudes (normalized to phonon number), in the unstable and nonlinear regime $\eta/(2\pi)=J/(2\pi)=5$ kHz, without feedback (mechanical loss rates $\gamma_i/(2\pi) = \{2.5, 1.6, 4.1\}$ kHz). Resonator 1 (left) or 2 (right) is driven for $t<0$, and couplings are established when $t>0$. This induces amplified transport to the other resonator and self-oscillation bounded by nonlinear dynamics. The amplification between sites (1,2) is strongly nonreciprocal with direction $1\rightarrow 2$ for $\Phi=\pi/2$ and $2\rightarrow 1$ for $\Phi=3\pi/2$, showing the chirality of the unstable eigenstates.}
		\label{fig:4}
	\end{figure*}
	
	The thermomechanical spectra in~\autoref{fig:3}b evidence the distinct dynamics in different regions. Along $\Phi=0$, we recognise behaviour analogous to the conventional $\mathcal{PT}$-dimer~\cite{Ruter2010}: Eigenmodes (hosted by quadrature dimers) respect $\mathcal{PT}$ symmetry for $J>\eta$, with equal linewidths and frequency splitting increasing with $J$. For $J<\eta$,  $\mathcal{PT}$ symmetry is spontaneously broken, with real frequency of all SD eigenmodes degenerate and independent of $J$, while their linewidths split.  At $J = \eta$, these phases are separated by a degenerate pair of second-order exceptional points (EPs) (one per quadrature dimer) where $\mathcal{H}_\text{SD}$ becomes defective. 
	Finite fluxes break the $\mathcal{P}_{X_iY_j}\mathcal{T}$ symmetry of $\mathcal{H}^{XY}_\text{SD}$ explicitly, thereby preventing the occurrence of EPs and a symmetry-broken phase for any value of $J$ or $\eta$, as observed in~\autoref{fig:3}b (bottom). 
	The effect of varying flux on the complex spectra is striking for $J\approx \eta$ (\autoref{fig:3}c), where we find strong tuning of both frequency \emph{and} linewidth, with the eigenmodes coalescing at the degenerate EPs $\Phi\in\{0,2\pi\}$.

	The behaviour of SD is intrinsically quadrature-dependent, as the paths in quasiparticle space link conjugated elements $a_i$ and $a_i^\dagger$ either directly or indirectly. The response to any real excitation (a superposition of $a_i$'s and $a_i^\dagger$'s) then depends on the particle-hole phase difference, i.e. the excited quadrature. Another example is phase-dependent amplification in the bosonic Kitaev chain (without synthetic flux). \cite{McDonald2018,Flynn2020}	One can, however, imagine the creation of loops that do not contain such links, where we expect that nonreciprocity and chirality are quadrature-independent (Methods). In fact, the Hermitian BST represents a trivial example, comprising two \emph{disjoint} loops connecting all particles and all holes, respectively (\autoref{fig:4}a). 

	We find a non-Hermitian system featuring disjoint loops by `conjugating' one resonator in the BST, i.e. swapping the role of particle $a_3$ with its hole $a_3^\dagger$. We implement this `singly conjugated trimer' (SCT) by modulating at $\omega_2 - \omega_1, \omega_1+\omega_3$ and $\omega_2+\omega_3$. The latter tones induce two-mode squeezing, specifically $H_\text{SCT} = J e^{i\varphi_{12}} a_2^\dagger a_1 + \eta_{23} e^{i\theta_{23}} a_3 a_2 + \eta_{13} e^{-i\theta_{13}} a_1^{\dagger} a_3^{\dagger} + \text{H.c.}$, and loops threaded by fluxes $\Phi =\varphi_{12}+\theta_{23}-\theta_{13}$ and $-\Phi$ (\autoref{fig:4}a).

	
	The disjoint loop topology of the quasiparticle network implies \emph{block-diagonality} of the  BdG dynamical matrix $\hsct$, with blocks corresponding to single loops and related by negative conjugation. The unique interplay of AB interference and non-Hermiticity in SCT leads to stability transitions between dynamical phases unmatched by BST. These are recognised in the complex eigenvalues of a single block of $\hsct$. Figure~\ref{fig:4}b shows these as surfaces in $J/\eta - \Phi$ space for $\eta_{13} = \eta_{32} = \eta$ and equal dissipation $\gamma=0$. We identify a stable phase with real eigenfrequencies and an unstable phase with three distinct imaginary parts.

	Interestingly, for $J=2\sqrt{2}\eta$ eigenvalues of a single loop of $\hsct$ are real and coincide in magnitude with those of a homogeneous BST ($J_{ij}=J$) for all $\Phi$. The thermal spectra in \autoref{fig:4}c show, however, that the frequency components around $\omega_3$ associated with the `conjugated' resonator (3) appear reflected, since particles (holes) evolve with positive (negative) frequencies in the non-rotating frame. 
	Moreover, we observe asymmetries between resonators 1 and 2 in the thermal amplitude of the middle band at $\Phi\in\{\pi/2,3\pi/2\}$. This asymmetry cannot occur in the BST as long as $J_{13}=J_{23}$, and must be due to the combination of chirality, which is maximal at these fluxes, and particle-non-conserving squeezing interactions. Indeed, theory shows it persists even for zero temperature (Supplementary Information~\autoref{sec:flux_a}). 
    The asymmetric, flux-controlled localisation of fluctuations links to chiral oscillations in incoherently pumped $\mathcal{PT}$-symmetric trimers~\cite{Downing2020} and suggests the SCT functions like a nonreciprocal amplifier~\cite{Sliwa2015,ruesink2016nonreciprocity,MercierdeLepinay2020,Peano2016d} for phonons.

    We see from the eigensurfaces (\autoref{fig:4}b) that the system transitions to an unstable phase if $J/\eta$ is reduced, but now for any flux $\Phi$, at an exceptional \emph{contour} (black dotted line) in parameter space. To associate this with a $\mathcal{PT}$ symmetry, we consider the eigenmode basis for $J=0$. Specified in the $\theta_{ij}=0$ gauge, these are the `gainy' and `lossy' modes $a_{g,l}=(a_+ \pm ia_3^\dag)/\sqrt{2}$ with $\epsilon_{g,l}=\pm\sqrt{2}i\eta$, and the `neutral' mode $a_-$ with $\epsilon_-=0$, where $a_\pm = (a_1 \pm a_2)/\sqrt{2}$. A finite beamsplitter interaction $J>0$ couples these three modes, with flux-dependent effective couplings. For $\Phi\in\{0,\pi\}$, the gainy and lossy modes form a $\pglt$-symmetric dimer, where $\mathcal{P}_{gl}$ exchanges $a_g\leftrightarrow a_l$, while the neutral mode is isolated at $\epsilon_-=-J$ ($\Phi=0$) or $\epsilon_-=J$ ($\Phi=\pi$). Figure~\ref{fig:4}d(top) shows the spectral signature of the second order EP at $J=2\sqrt{2}\eta$ that, for increasing $J$, indicates the transition from a state where the dimer's eigenstates spontaneously break $\pglt$ symmetry to a $\pglt$-symmetric, stable phase. However, for $\Phi\in\{\pi/2,3\pi/2\}$, the three modes $a_l$, $a_-$ and $a_g$ are coupled in a loss-neutral-gain chain configuration. Interestingly, this trimer features a stability transition where $\pglt$ is spontaneously broken at a \emph{third-order} EP at $J=\sqrt{2}\eta$, as observed in \autoref{fig:4}d(bottom). Indeed, the presence of a higher-order EP is mandated by the eigensurface topology (\autoref{fig:4}b).

    The fact that finite fluxes explicitly break the mirror symmetry $\mathcal{P}_{12}$ greatly impacts specifically the $\pglt$-broken phase. In a three-site chain (gainy-neutral-lossy resonators), $\mathcal{PT}$-broken states delocalise non-uniformly over the central and a boundary site~\cite{Hodaei2017}. If $\Phi=\pm\pi/2$, the eigenmodes thus involve non-uniform combinations of $a_-$ and boundary sites $a_g$ or $a_l$. As a result, gain is biased towards the bare oscillator $a_1$ ($\Phi=\pi/2$) or $a_2$ ($\Phi=-\pi/2$) as the third-order EP point is crossed. This flux-tunable chiral gain becomes strikingly visible in the transient dynamics of the SCT in the unstable regime. In~\autoref{fig:4}e, the interaction is switched on at $t=0$~ms with squeezing gain exceeding mechanical dissipation. An initial excitation in resonator 1 (2) is amplified coherently -- above initial amplitudes -- towards 2 (1) if the flux is set to $\Phi=\pi/2$ ($\Phi=3\pi/2$), and attenuated quickly in the opposite direction. In contrast, for $\Phi=0$, gain distributes evenly in resonators 1 and 2 and dynamics are reciprocal. In fact, linear analysis breaks down as the system crosses the instability ($\Im(\epsilon)>0$) threshold, and we see that optomechanically-induced Duffing nonlinearities saturate the amplitudes and lead to coherent self-oscillation, even for excitations of only a few times the thermal amplitude. Indeed, this points the way to investigating strongly nonlinear systems with broken Hermiticity and time-reversal symmetry.

	In conclusion, we observed chiral, non-Hermitian phonon dynamics in nano-optomechanical networks with fully-controlled beamsplitter and squeezing interactions. Through a powerful diagrammatic framework, we uncovered new geometrical phases acting on excitations in particle-hole space that control $\mathcal{PT}$ symmetry through a non-Hermitian Aharonov-Bohm effect. The resulting phenomena of tunable squeezing, (higher-order) exceptional points and nonreciprocal amplification point to applications in nanomechanical sensing~\cite{Lau2018}, signal processing~\cite{Peano2016d}, and classical Ising machines~\cite{Mahboob2016}. But these mechanisms will have equally powerful consequences in other bosonic domains, from photonics to cold atoms.
	While the effects are probed here with thermal and coherent excitations, they persist down to the quantum domain, and may form the essential ingredients of the exploration of new linear and nonlinear non-Hermitian topological phases.

   \section*{Methods}
   
   \subsection*{Derivation of the effective Hamiltonian}\label{sec:adiab_el}
   We construct a comprehensive theoretical model for the optomechanically-mediated nanomechanical interactions in our platform. A cavity mode with frequency $\omega_c$ and photon loss rate $\kappa$ is coupled to an ensemble of nanobeam mechanical modes (\autoref{fig:1}a) with frequencies $\omega_i$ (index $i\in\{1,2\cdots,N\}$) with vacuum optomechanical coupling rates $g_{0}^{(i)}$, according to the Hamiltonian
   \begin{subequations}
   \begin{equation}
   \tilde{H}_s=\sum_{i}\omega_{i}\tilde{a}_i^{\dagger}\tilde{a}_i-\Delta c^{\dagger}c-\sum_{i}g_{0}^{(i)}c^{\dagger}c(\tilde{a}_i+\tilde{a}_i^{\dagger})\label{eq:H}.
   \end{equation}
   Here, mechanical annihilation operators in the lab frame are denoted by $\tilde{a}_i$ and we set $\hbar=1$. The cavity field annihilation operator $c$ is expressed in the rotating frame of a control field at frequency $\omega_L$ detuned by $\Delta=\omega_L-\omega_c$ from the cavity resonance. We operate in the regime of large detuning and bandwidth ($\Delta,\kappa\gg\omega_i$). With cavity in-coupling rate $\kin$, a control field with slowly-varying amplitude $C_{\mathrm{in}}(t)$ addresses the intracavity photon population instantaneously, displacing the cavity mode by a steady-state amplitude approximated by the $g_{0}^{(i)}=0$ solution
   \begin{equation}\label{eq:ss_dyn}
   \bar{c}(t)\approx\sqrt{\kin}\chi_c C_\mathrm{in}(t),
   \end{equation}
   with bare cavity susceptibility value $\chi_c=
   (\frac{\kappa}{2}-i\Delta)^{-1}$.
   
   We linearise the cavity amplitude around the solution~\autoref{eq:ss_dyn} by writing $c(t)\rightarrow \bar{c}(t)+\delta c(t)$ and neglecting terms $\mathcal{O}((\delta c)^2)$, assuming small cavity fluctuations $\delta c(t)$. Neglecting constant terms, the linearised Hamiltonian $\tilde{H}_s=\tilde{H}_0+\tilde{V}^{\text{rp}}+\tilde{V}^{\text{d}}$ contains the mean-field Hamiltonian $\tilde{H}_0$, the radiation-pressure interaction $\tilde{V}^{\text{rp}}$ and the drive term $\tilde{V}^{\text{d}}$, reading
   \begin{align}\label{eq:decomp_linear}
   \tilde{H}_0	= & 	-\Delta \delta c^{\dagger}\delta c  +\sum_{i}\big[\omega_{i}\tilde{a}_i^{\dagger}\tilde{a}_i\hspace{-0.5mm}-\hspace{-0.5mm}|\bar{c}(t)|^{2}g_{0}^{(i)}(\tilde{a}_i+\tilde{a}_i^{\dagger})\big],\hspace{-2mm}\\
   \tilde{V}^{\mathrm{rp}}    =&   - \left(\bar{c}(t)\delta c^{\dagger}+\bar{c}(t)^*\delta c\right)\left(\sum_{i}g_{0}^{(i)}(\tilde{a}_i+\tilde{a}_i^{\dagger})\right),\label{eq:int1}\\
   \tilde{V}^\mathrm{d}    =&    i\sqrt{\kin}\left(C_\mathrm{in}(t)\delta c^{\dagger}-C^*_\mathrm{in}(t)\delta c\right)\label{eq:int2}.
   \end{align}

   Subsequently, fast-evolving fluctuations $\delta c$ are adiabatically eliminated to find an effective phononic Hamiltonian. To apply the approach in~\cite{Reiter2012}, we express the interactions in \autoref{eq:int1} and~\autoref{eq:int2} in terms of operators $a_i=e^{i\omega_it}\tilde{a}_i$ in frames rotating at $\omega_i$, accessed via a unitary transformation $U_F=e^{it\sum_i\omega_i\tilde{a}^\dagger_i\tilde{a}_i}$. This yields the Fourier components
   \begin{equation}\label{eq:F_exp_pertb}
   U_F(\tilde{V}^{\mathrm{rp}}+\tilde{V}^\mathrm{d})U_F^{\dagger}=\sum_f\sum_{q=\pm}v_q^{f}e^{-i \omega_f t},
   \end{equation}
   and the transformed Hamiltonian $H_s=U_F\tilde{H}_sU_F^\dagger-\sum_i\omega_i\tilde{a}_i^{\dagger}\tilde{a}_i$. Here $f$ indexes each of the resulting operator terms $v_q^{f}$ in the rotating frame with associated frequency $\omega_f$, whereas the index $q=\{\pm\}$ splits perturbations that create/destroy excitations in the `excited' subspace (photonic terms $\sim\delta c^{\dagger}$/$\sim\delta c$). The effective, `ground state' (phononic) Hamiltonian $H_\mathrm{eff}=H_{g}+H_{\mathrm{eff}}^{\mathrm{int}}$ includes a displacement term $H_{g}=-|\bar{c}(t)|^{2}(\sum_{i}g_{0}^{(i)}(a_ie^{-i\omega_it}+\mathrm{H.c.}))$ and the effective interaction Hamiltonian 
   \begin{equation}\label{eq:H_eff_general}
   H_\mathrm{eff}^{\mathrm{int}}=\frac{1}{2}\sum_{f,f'}v_-^{f'}e^{i \omega_{f'} t}\left(\frac{\delta c^{\dagger}\delta c}{\Delta+i\frac{\kappa}{2}-\omega_f}+\mathrm{H.c.}\right)v^{f}_{+}e^{-i\omega_f t}.
   \end{equation}
   Here we identify that the frequencies $\omega_f$ in~\autoref{eq:F_exp_pertb} are  $\mathcal{O}(\omega_i)$, implying negligible frequency variations $\Delta -\omega_f\simeq\Delta$, at the same level as in~\autoref{eq:ss_dyn}.  Inserting~\autoref{eq:ss_dyn},~\autoref{eq:int1} and \autoref{eq:int2} into~\autoref{eq:H_eff_general}, $H_\mathrm{eff}^{\mathrm{int}}$ approximates to
   \begin{align}\label{eq:before_aver}
   H_{\mathrm{eff}}^{\mathrm{int}}\simeq&\Delta\kin|\chi_{c}|^{4}|C_{\mathrm{in}}(t)|^{2}\left[\left(\sum_{i}g_{0}^{(i)}(a_{i}e^{-i\omega_{i}t}+\mathrm{H.c.})\right)^{2}\right.\nonumber\\
   &+\left.2\Delta\sum_{i}g_{0}^{(i)}(a_{i}e^{-i\omega_{i}t}+\mathrm{H.c.})\right].
   \end{align} 
   We introduce modulation of the control field intensity using multiple harmonic driving tones $l$, i.e., $|C_\mathrm{in}(t)|^2=\left| \bar{C}_\text{in} \right|^2 \left(1 + \sum_l \cm^{(l)}\cos\left(\om^{(l)} t+\phim^{(l)}\right) \right)$ with frequencies $\om^{(l)}$, modulation depths $\cm^{(l)}$ and phases $\phim^{(l)}$. From \autoref{eq:ss_dyn}, the homogeneous intracavity intensity responds linearly as
   $n_c(t)\approx\left|\bar{c}(t)\right|^2= \bar{n}_c \left(1 + \sum_l \cm^{(l)}\cos\left(\om^{(l)} t+\phim^{(l)}\right) \right)$, where $\bar{n}_c = \kin|\chi_c|^2|\bar{C}_\text{in}|^2$ is the average photon number.
   
   Assuming dynamical modulations are not resonant with any vibrational mode ($\omega_\mathrm{m}\neq\omega_i$), displacement terms $\sim\sum_{i,l}\cos(\omega_\text{m}^{(l)}t+\phi_\text{m})(a_ie^{-i\omega_{i}t}+\mathrm{H.c.})$ in~\autoref{eq:before_aver} average to zero under the Rotating Wave Approximation (RWA). Assuming moderate couplings compared with natural oscillation frequencies, the RWA only retains the \textit{co-rotating} terms, which evolve slowly when expressed in terms of the rotating frame operators $a_i$. With these assumptions, the relevant contributions in \autoref{eq:before_aver} read $H_{\mathrm{eff}}^{\mathrm{int}}\simeq\sum_{i,j,l}H_\mathrm{eff}^{(i,j,l)}$  with $i,j\in\{1,2,\cdots N\}$ and
   \begin{align}\label{eq:time_dep_terms}
   H_{\mathrm{eff}}^{(i,j,l)}=g(t)(a_{i}e^{-i\omega_{i}t}+\mathrm{H.c.})(a_{j}e^{-i\omega_{j}t}+\mathrm{H.c.}),
   \end{align}
   with $g(t)=\Delta|\chi_{c}|^{2}g_{0}^{(i)}g_{0}^{(j)}n_{c}(t)$.
   The static component of $n_c(t)$ is responsible for an optical shift of the mechanical spring constant by $\omega_i\mapsto\omega_i+\delta \omega_i$ that is reabsorbed in the definition of $\omega_i$, where $\delta\omega_i = 2g_i^2\Delta/(\Delta^2+\kappa^2/4)$ and $g_i = g_0^{(i)} \sqrt{\bar{n}_c}$ denotes the cavity-enhanced optomechanical coupling rate~\cite{Aspelmeyer2013}. Crucially, the time-dependent part in~\autoref{eq:time_dep_terms} corresponds to mechanical interactions which can be selected by suitably resonant modulation tones, while imprinting $\phim^{(l)}$ as a \textit{Peierls} phase on the interaction~\cite{Mathew2020}. Within a subsequent RWA, the remaining interaction terms in \autoref{eq:time_dep_terms} correspond to the modulation frequencies $\om^{(l)}$ either approaching a \textit{i)} frequency sum $\Sigma\omega^{\braket{ij}}=\omega_i+\omega_j$ or a \textit{ii)} frequency difference $\Delta\omega^{\braket{ij}}=\omega_i-\omega_j$, with $i,j\in\{1,2\cdots,N\}$. Under previous assumptions, the Hamiltonian~\autoref{eq:H} finally approximates in the rotating frame to 
   \end{subequations}
   \begin{subequations}
   \begin{align}\label{eq:Ht-resonance-conditions}
   H_{\text{eff}}&\simeq\sum_{\mathclap{\om^{(l)}\approx\Delta\omega^{\braket{ij}}}}J_{ij}a_{i}^{\dagger}a_{j}e^{-i((\omega_{\mathrm{m}}^{(l)}-\Delta\omega^{\langle ij\rangle})t+\varphi_{ij})}+\text{H.c.}\nonumber\\&+\sum_{\mathclap{\om^{(l)}\approx\Sigma\omega^{\braket{ij}}}}\eta_{ij}a_{i}^{\dagger}a_{j}^{\dagger}e^{-i((\omega_{\mathrm{m}}^{(l)}-\Sigma\omega^{\langle ij\rangle})t+\theta_{ij})}/2+\text{H.c.},
   \end{align}
   where the sums run over the tones $l$ and indices $\braket{i,j}$ that satisfy the specified resonance condition. Note that only a single pair of indices $\braket{i,j}$ satisfies resonance with a difference frequency $\Delta\omega^{\braket{ij}}$, while resonance with a sum frequency $\Sigma\omega^{\braket{ij}}$ is satisfied by both $\braket{i,j}$ and $\braket{j,i}$.
   
   The hopping (squeezing) amplitudes, denoted $J_{ij}$ ($\eta_{ij}$), are proportional to the modulation depth $\cm^{(l)}$ of the corresponding drive tone $l$~\cite{Mahboob2014},\cite{Mathew2020} and read
   \begin{align}
   \left\{J_{ij},\eta_{ij}\right\} = \cm^{(l)} \frac{g_i g_j \Delta}{(\Delta^2+\kappa^2/4)} = \cm^{(l)}\frac{\sqrt{\delta\omega_i\delta\omega_j}}{2}.
   \end{align}
   
   Similarly, the hopping (squeezing) phases, denoted $\varphi_{ij}$ ($\theta_{ij}$), are equal to the corresponding modulation phase $\phim^{(l)}$. The RWA is valid for moderate coupling strengths $J_{ij},\eta_{ij}\ll\omega_i$ (in the experiment, $J_{ij}/\omega_i,\eta_{ij}/\omega_i\sim 10^{-3}-10^{-2}$), and moderate detuning of the control tones, as well as no commensurable frequency scales ($\omega_i\pm\omega_j\neq\omega_k$ for all modes $i,j,k$). 
   
   Besides moderate effective coupling, the RWA relies on the assumption that the modulated drive is quasi-resonant with each relevant process. In the large detuning limit and for large parametric drive, significant deviations are expected~\cite{Leuch2016}. Parametric resonators are more naturally treated in this case in terms of the natural amplitudes $x$~\cite{CalvaneseStrinati2019,Bello2019} or employing quadratures in a generalised rotating frame~\cite{Guckenheimer1984}.
   For modulation frequencies resonant with
   $\Delta\omega^{\braket{ij}},\Sigma\omega^{\braket{ij}}$, \autoref{eq:Ht-resonance-conditions} is exactly time-independent. In this limit, we encode the beam-splitter interactions that conserve the phonon number $n_\text{ph}=\sum_{i=1}^Na_i^{\dagger}a_i$ in the elements $\mathcal{A}_{ij} = J_{ij} e^{-i\varphi_{ij}}$, $\mathcal{A}_{ji} = \mathcal{A}_{ij}^{*}$ of the Hermitian \emph{hopping matrix} $\mathcal{A}$. Subsequently, we define the symmetric \emph{squeezing matrix} $\mathcal{B}$ that encodes the particle-non-conserving squeezing interactions in its elements $\mathcal{B}_{ij} = \eta_{ij} e^{i\theta_{ij}}$, $\mathcal{B}_{ji} = \mathcal{B}_{ij}$.  \autoref{eq:Ht-resonance-conditions} then writes succinctly as the general quadratic form
   \begin{equation}\label{eq:Ht}
   H_{\mathrm{eff}}\simeq\sum_{i,j}a_{i}^{\dagger}\mathcal{A}_{ij}a_{j}+\frac{1}{2}(a_{i}^{\dagger}\mathcal{B}_{ij}a_{j}^{\dagger}+a_{i}\mathcal{B}^*_{ij}a_{j}).
   \end{equation}
   \end{subequations}
   \subsection*{Bogoliubov-de-Gennes framework and symmetries}
   
   The time-independent Hamiltonian~\autoref{eq:Ht} allows for a straightforward application of the toolbox of quadratic bosonic Hamiltonians. After defining the Nambu-like vector $\vec{\alpha}=(\vec{a},\vec{a}^\dagger)^{T}$, with $\vec{a}=(a_{1},\cdots,a_{N})$, the effective Hamiltonian in the rotating frame reads
   \begin{align}
   H_{\mathrm{eff}}=\frac{1}{2}\vec{\alpha}^{\dagger}H\vec{\alpha},&&
   H= \left(\begin{array}{cc}
   \mathcal{A} & \mathcal{B}\\
   \mathcal{B}^* & \mathcal{A}^*
   \end{array}\right).\label{eq:Ht_alt}
   \end{align} 
   To faithfully model the ubiquitous mechanical dissipation and thermal fluctuations in the experiment, we introduce coupling to $N$ independent environmental baths in a Heisenberg-Langevin formalism~\cite{gardiner2004quantum}. The corresponding equation of motion for mechanical modes, namely $\dot{\vec{\alpha}}(t)=-i\mathcal{M}\vec{\alpha}(t)+\vec{\alpha}_{\mathrm{in}}(t)$, depends on the open-system dynamical matrix $\mathcal{M}=\hbdg-i\frac{\Gamma}{2}$, containing the dissipation matrix $\Gamma=\mathrm{diag}(\gamma_1,\cdots,\gamma_N,\gamma_1,\cdots,\gamma_N)$, and the Bogoliubov-de-Gennes matrix~\cite{blaizot1986quantum,Rossignoli2005}
   \begin{align}\label{eq:HBDG}
   \hbdg=\Sigma_zH=\left(\begin{array}{cc}
   \mathcal{A} & \mathcal{B}\\
   -\mathcal{B}^{*} & -\mathcal{A}^{*}
   \end{array}\right),
   \end{align}
   where $\Sigma_z=\sigma_z\otimes\mathbb{1}=[\vec{\alpha},\vec{\alpha}^{\dagger}]$ encodes bosonic commutation relations. Cavity-mediated corrections to mechanical dissipation ($\gamma_i\kappa/(\Delta^2+\kappa^2)\ll 1$)\cite{Reiter2012} will be neglected. The rotating source terms $\vec{\alpha}_\mathrm{in}=(a_\mathrm{in},a_\mathrm{in}^\dagger)^T$ represent baths with Bose occupations $\bar{n}_i\simeq k_\mathrm{B}T/\omega_i$.  These fulfil the same Markovian correlations as their lab-frame counterparts, i.e. $\braket{\vec{\alpha}_\mathrm{in}(t)\vec{\alpha}^{\dagger}_\mathrm{in}(t')}=\mathcal{D}\delta(t-t')$ with diffusion matrix $ 		\mathcal{D}=\mathrm{diag}(\gamma_1(\bar{n}_1+1)\cdots,\gamma_1\bar{n}_1\cdots)$~\cite{Aranas2017}.
   
   Treating creation and annihilation operators, $a_i$ and $a_i^{\dagger}$, as separate entities in Hamiltonian and BdG matrices shows closed dynamics in particle-hole space.	When squeezing interactions -- which inter-convert particles and holes -- are absent ($\mathcal{B}=0$), the dynamics of $a_i$ and $a_i^\dagger$ are independent, and simply governed by the Hermitian matrices $\mathcal{A}$ and $-\mathcal{A}^*$ respectively. On top of this, if the loss matrix is proportional to the identity, namely $\Gamma=\gamma\mathbb{1}$, the dynamics can be simply mapped to the closed system via a rigid displacement of the imaginary parts of eigenvalues by $\gamma/2$. Therefore, whenever $\mathcal{B}=0$ is zero, we say that the mechanical modes undergo \emph{Hermitian} dynamics.
   
   However, even for $\Gamma = 0$, $\mathcal{M}$ and $\hbdg$ are non-Hermitian if squeezing is present ($\mathcal{B}\neq0$). This allows $\hbdg$ to host eigenvectors with potentially \emph{complex} eigenfrequencies $\epsilon$, indicating their oscillatory (real $\epsilon$), exponential (imaginary $\epsilon$) or combined (complex $\epsilon$) evolution. We say that the time evolution of the amplitudes $\vec{\alpha}(t)$, readily obtained from the spectral decomposition of $\hbdg$ (see \cite{Flynn2020} and Supplementary Information~\autoref{sec:BdGForm}), manifests \emph{non-Hermitian dynamics}.
   
   Such decomposition reveals that, depending on the system parameters, mechanical modes feature distinct dynamical phases corresponding to dissimilar eigenpairs of the  BdG dynamical matrix $\hbdg$, characterised by different partial degeneracies in the real and imaginary parts of its eigenfrequencies. For example, purely oscillatory eigenstates indicate a stable phase, while positive imaginary eigenfrequencies indicate an unstable phase. These dynamical phases can also be conveniently classified by comparing the symmetries of $\hbdg$ and its eigenstates, which can be embedded into generalised parity-time ($\gpt$) symmetries~\cite{Flynn2020}. Dynamical phase transitions occur precisely at regions in parameter space where the eigenvectors break a $\gpt$ symmetry of $\hbdg$ spontaneously. Recall the symmetries of the dynamical matrix $\mathcal{M}$ that include energy dissipation directly follow those of $\hbdg$ if $\gamma_i=\gamma$ after the appropriate offset of imaginary parts, so we can refer indistinctly to the symmetries of $\hbdg$ or $\mathcal{M}$ in this case.
   
   Such phase boundaries are characterised by non-Hermitian singularities known as exceptional points,\cite{heiss2012physics} where eigenvalues and eigenvectors simultaneously coalesce, leading to a defective eigenvector subspace, or splitting of eigenvalues off the real axis without loss of diagonalisability of $\hbdg$. Coalescences can be readily found in the studied systems by analytical diagonalisation (see Supplementary Information~\autoref{sec:SD_as_AB},~\autoref{sec:ECS},~\autoref{sec:loop_eig},~\autoref{sec:quad_dep}). Even when such an analytic approach becomes impractical, the defectiveness of $\hbdg$ can be assessed from the condition number $\mathrm{cond}(V^{-1})$ for the inverse of the numerical eigenvector matrix $V$, which acquires larger values when $\hbdg$ is close to non-diagonalisability~\cite{moiseyev2011non}.
   
   It must be noted the system-dependent $\gpt$ symmetries coexist with another two built-in symmetries of the bosonic $\hbdg$ that simply reflect the redundancies introduced in the splitting of $a_i$ and $a_i^{\dagger}$.  Namely,
   \begin{enumerate}
   \item  \textit{Charge-conjugation} that arises from mutual adjointness of creation and annihilation operators: $\mathcal{C}\hbdg\mathcal{C}=-\hbdg$, where $\mathcal{C}=(\sigma_x\otimes \mathbb{1})\mathcal{K}$ with complex conjugation $\mathcal{K}$ and $x$ Pauli matrix $\sigma_x$.
   \item \textit{$\Sigma_z$-Pseudo-Hermiticity}, with ensures bosonic commutators:  $\Sigma_z	\hbdg\Sigma_z=	\hbdg^{\dagger}$ .
   \end{enumerate}
   
   These symmetries are not necessarily fulfilled in the open-system dynamical matrix $\mathcal{M}$: dissipation breaks $\mathcal{C}$ and $\Sigma_z$. Nevertheless, if loss rates are symmetrical ($\gamma_i\equiv\gamma$), both charge-conjugation and pseudo-Hermiticity can be restored in the the dynamically-offset basis $\bar{\alpha}(t) = e^{-\frac{\gamma}{2} t}\vec{\alpha}(t)$,\cite{Ornigotti2014}\cite{li2020} which effectively maps $\mathcal{M}\mapsto\hbdg$. This crucial fact allows our systems where losses are engineered to be equal to be catalogued by the very same symmetries as $\hbdg$.
   
   \begin{figure}
   \centering
   \includegraphics[width=0.8\linewidth]{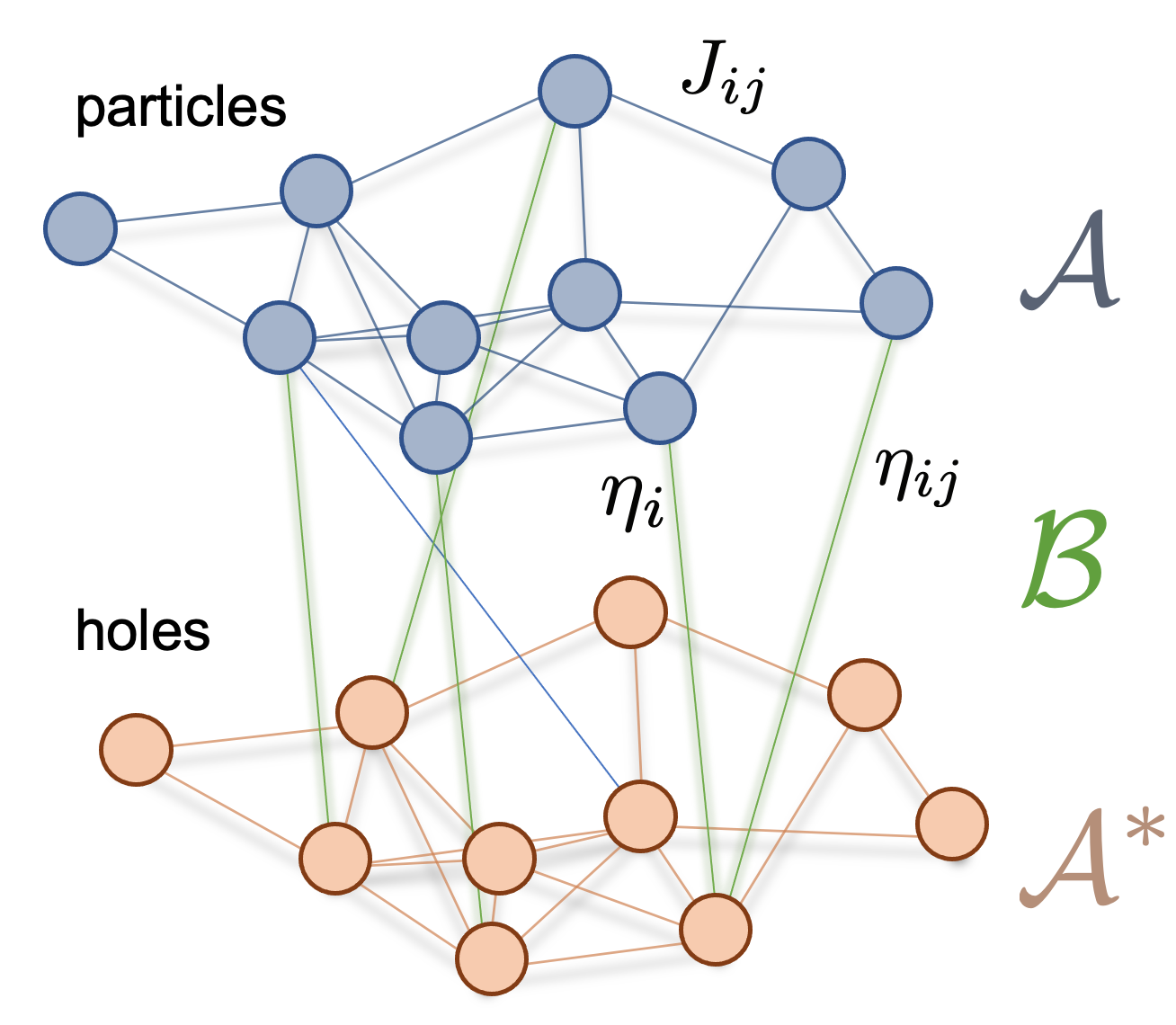}
   \caption{\textbf{Network graph representation of general quadratic Hamiltonians.} Schematic of an arbitrary dynamical matrix $\hbdg$, acting on a Nambu-like vector $\vec{\alpha}=(a_1,a_2,\cdots,a_N,a_1^{\dagger},a_2^{\dagger},\cdots,a_N^{\dagger})$. Particle annihilation (hole creation) operators, $a_i$, are represented by blue nodes, whereas hole annihilation (particle creation) operators are represented by orange nodes. $\hbdg$ includes excitation-conserving interactions (matrix $\mathcal{A}$), which link particle operators (e.g. terms $\mathcal{A}_{ij}a_i^{\dagger}a_j$) and hole operators (e.g. terms $\mathcal{A}^*_{ji}a_ja_i^{\dagger}$). Squeezing interactions (with complex amplitude matrix $\mathcal{B}$) contain pairs $\mathcal{B}_{ij}a_i^{\dagger}a_j^{\dagger}$ which can be visualized to either annihilate two particles $i,j$ or to annihilate a particle in $i$ an create hole in $j$, hence the connection between particle and hole networks (green). Mutatis mutandis, terms $\mathcal{B}_{ij}^*a_i a_j$ can be similarly visualized.\label{fig:graph_rep}} 
   \end{figure}
   
   \subsection*{Graph representation of quadratic bosonic Hamiltonians}
   We introduce a convenient graphical representation for the Hamiltonian in~\autoref{eq:Ht_alt}
   uncovering the different forms of loops and nontrivial $U(1)$ gauge fields in particle-hole space. For it we consider $\mathcal{A}$ and $\mathcal{A}^*$ as the adjacency matrices for network graphs $\mathcal{G}_{a}$  and $\mathcal{G}_{a^{\dagger}}$, disposed in two layers where nodes correspond to $a_i$ and $a_i^{\dagger}$ operators respectively (see \autoref{fig:graph_rep}). In this two-layer network representation, particle-conserving systems, where $\mathcal{B}=0$, feature disjoint networks $\mathcal{G}_{a}$ and $\mathcal{G}_{a^{\dagger}}$, mapped into each other via particle-hole conjugation $\mathcal{C}$.
   
   Systems with parametric gain, where $\mathcal{B}\neq0$, have links that couple $\mathcal{G}_{a}$ to $\mathcal{G}_{a^{\dagger}}$ through $\mathcal{B}$ and back via $\mathcal{B}^*$. We adopt this graph representation representing $H$ in the main text, but note that a similar representation follows for the  BdG dynamical matrix in~\autoref{eq:HBDG}, where the adjacency matrices for network graphs $\mathcal{G}_{a}$  and $\mathcal{G}_{a^{\dagger}}$ are $\mathcal{A}$ and $-\mathcal{A}^*$, connected with each other through non-Hermitian couplings $\mathcal{B}$ and $-\mathcal{B}^*$. This in particular reveals that $\mathcal{B}\neq0$ unlocks loops along which dynamics are \textit{non-Hermitian}. Both network graphs representing $H$ and $\hbdg$ equivalently allow the recognition of loops in enlarged particle-hole space, with quantified geometrical phases that only differ in trivial phase factors of $\pi$. 
   
   \subsection*{Disjoint graphs and quadrature-independent transport}\label{sec:disjoint_loops}
   
   Quadrature-independent transport is found in networks that feature
   \textit{disjoint} graphs (e.g. loops), which do not contain (indirect) links between particles $a_i$ and their corresponding holes $a_i^{\dagger}$ (\autoref{fig:4}). 
   This \textit{sublattice} symmetry implies that a set of nodes $\vec{\alpha}_\mathcal{L}=(a_i,\cdots,a^{\dagger}_j)$ in an independent graph is governed by an uncoupled block in $\hbdg$. For $M$ disjoint graphs $\mathcal{L}_1,\mathcal{L}_2,\cdots,\mathcal{L}_M$, we can block-diagonalize $\hbdg$ by permuting the modes of $\vec{\alpha}$ into each of the graphs via transformations $G$:
   $\vec{\alpha}\mapsto G\vec{\alpha}=(
   \vec{\alpha}_{\mathcal{L}_{1}} , \vec{\alpha}_{\mathcal{L}_{1}}^{\dagger}, \ldots , \vec{\alpha}_{\mathcal{L}_{M}} , \vec{\alpha}_{\mathcal{L}_{M}}^{\dagger})^{T}.
   $
   As a result, the transformed  BdG dynamical matrix $\hbdg\mapsto \hbdg'= G \hbdg G$ reads
   \begin{align}
   \hbdg'=\mathrm{diag}(\mathcal{L}_1,- \mathcal{L}_{1}^*,\cdots\hspace{-1mm},\mathcal{L}_M,-\mathcal{L}_{M}^*). \hspace{-1mm}
   \end{align}
   The dynamical matrices for each pair of conjugated graphs $\mathcal{L}_i$ and $-\mathcal{L}_{i}^*$ have eigenvectors related by charge conjugation $\mathcal{C}$. This fact ensures an even number of graphs in the system reflecting the doubling of degrees of freedom introduced by the BdG particle-hole description. In the dynamical evolution, particles and their corresponding hole excitations never mix as they propagate through the graphs.
   
   These properties have consequences in dynamics. Without loss, the dynamics in graph $\mathcal{L}_i$ ($\mathcal{L}^*_i$) follow from $i\partial_t\vec{\alpha}_{\mathcal{L}_{i}}=\mathcal{L}_i\vec{\alpha}_{\mathcal{L}_{1}}$ ($i\partial_t\vec{\alpha}_{\mathcal{L}_{i}}^{\dagger}=-\mathcal{L}_i^*\vec{\alpha}_{\mathcal{L}_{i}}^{\dagger}$). From the formal solution of these equations, the time evolution of bosonic populations $\vec{n}=(a_1^\dagger a_1,\cdots,a_N^\dagger a_N)$ obeys
   \begin{align}\label{eq:E_evo}
   \vec{n}(t)=
   &e^{i\mathcal{L}_i^*t}\vec{n}(0)e^{-i\mathcal{L}_it},
   \end{align}
   i.e. their evolution depends on the initial energies but not the relative phases of $a_i(0)$ and $a_i^{\dagger}(0)$ (or quadrature of resonator $i$). 
   
   \subsection*{Dynamical matrices of the studied examples}
   A Hermitian system where we study $\mathcal{T}$-breaking synthetic fluxes is the \textit{beam-splitter trimer} (BST). We calibrate modulation depths in the experiment to set $J_{ij}=J$. In a gauge where $\varphi_{i,(i\hspace{0.5mm}\mathrm{mod}\hspace{0.5mm}3)+1}=\Phi/3$, it is governed by a circulant hopping matrix 
   \begin{align}\label{eq:BST_A}
   \mathcal{A}=J\left(\begin{array}{ccc}
   0 & e^{-i\Phi/3} & e^{i\Phi/3}\\
   e^{i\Phi/3} & 0 & e^{-i\Phi/3}\\
   e^{-i\Phi/3} & e^{i\Phi/3} & 0
   \end{array}\right).
   \end{align}
   Here $\mathcal{T}$ is explicitly broken by a non-trivial flux $\Phi \neq 0, \pi$, for which there is no $U(1)$ gauge transformation rendering the Hamiltonian matrix in~\autoref{eq:Ht_alt} real~\cite{Koch2010}.
   
   The two examples of non-Hermitian networks that we study are the minimal instance of a loop in particle-hole space, namely the  \textit{squeezing dimer} (SD), governed by respective hopping and squeezing matrices
   \begin{align}\label{eq:SD_building_blocks}
   \mathcal{A}=J\left(\begin{array}{ccc}
   0 & e^{-i\varphi_{12}} \\
   e^{i\varphi_{12}} & 0\\
   \end{array}\right),&&	\mathcal{B}=\eta\left(\begin{array}{ccc}
   e^{-i\theta_1} & 0\\
   0 & e^{-i\theta_2}
   \end{array}\right),
   \end{align}
   which incorporates beam-splitter and single-mode squeezing interactions, and the \textit{singly conjugated trimer }(SCT), encompassing a beam-splitter and a pair of two-mode squeezing links:
   \begin{subequations}
   \begin{align}
   \mathcal{A}=&J\hspace{-1mm}\left(\begin{array}{ccc}
   0 & e^{-i\varphi_{12}} & 0\\
   e^{i\varphi_{12}} & 0 & 0\\
   0 & 0 & 0
   \end{array}\right),\\
   \mathcal{B}=&\eta\hspace{-1mm}\left(\begin{array}{ccc}
   0 & 0 & e^{-i\theta_{13}}\\
   0 & 0 & e^{-i\theta_{23}}\\
   e^{-i\theta_{13}} & e^{-i\theta_{23}} & 0
   \end{array}\right).\hspace{-2mm}
   \end{align}
   \end{subequations}
   The effective parity-time symmetries for these non-Hermitian examples are detailed in the subsequent section ``Gain-loss bases and effective $\mathcal{PT}$ symmetries''.
   
   \subsection*{Non-Hermitian Aharonov-Bohm effect}
   
   The SD shown in \autoref{fig:2} presents the minimal instance of a plaquette in particle-hole space permeated by a nontrivial flux, and illustrates the contrast between the Hermitian and non-Hermitian Aharonov-Bohm (AB) effects. We describe how the latter is manifested in the ``energy'' eigenbasis, with generally complex eigenvalues, and in the flux-dependenent coupling of gainy/lossy quadratures.
   
   In a Hermitian four-mode loop with flux distributed evenly over its links, the Fourier modes $\tilde{a}_k=\sum_{j=1}^4 a_i e^{2\pi ikj/4}/2$ ($k=\{-2,-1,0,1\}$) are its (uncoupled) eigenmodes. Their (multimode) interference with nontrivial Peierls phases produces a flux-dependent, real spectrum (AB effect). However, particles and holes are in-equivalent entities in the network-graph of the SD, breaking cyclic-permutation invariance $\alpha_j\mapsto\alpha_{j+1}$. This translates into the fact that the naively defined ``Fourier'' modes for such a loop, $\tilde{\alpha}_k=\sum_{j=1}^4\alpha_j e^{2\pi ikj/4}/2$, do not respect bosonic commutation relations. This violation of pseudo-Hermiticity -- or equivalently the fact that $\Sigma_z$ and the Fourier matrix do not commute -- implies $\Sigma_z$ is not diagonal in the Fourier basis, but instead couples the $\tilde{\alpha}_k$ in the  BdG dynamical matrix \autoref{eq:HBDG} that describes the system's evolution. We show in Supplementary Information~\autoref{sec:eig_therm} that the corresponding coupling matrix is non-Hermitian. In a scenario where an effective flux threads the plaquette, this implies a \emph{non-Hermitian} AB effect, where interference effects and non-Hermitian coupling coexist. This results in eigenvectors with flux-dependent, complex eigenfrequencies. For arbitrary flux, eigenfrequencies come in the quartet $\{\epsilon,\epsilon^*,-\epsilon,-\epsilon^*\}$,\cite{Flynn2020} with
   \begin{align}
   \epsilon=\sqrt{\eta^2-J^2+2iJ\eta\sin\Phi}.
   \end{align}

   To understand that the non-Hermitian AB effect can  induce a flux-dependent coupling between quadratures, which implies redistribution of gain and squeezing in the dimer, we note that the dynamics of the SD are governed by two superimposed loops $\mathcal{L}$ and $\mathcal{L}^*$ in particle-hole space. These are related by conjugation and represent clockwise and counterclockwise propagation of excitations. We choose the gauge $\theta_i=\pi/2$, for which the local quadratures $X_i=(a_i + a_i^\dagger)/\sqrt{2}$ ($Y_i=i(a_i^{\dagger}-a_i)/\sqrt{2}$) experience loss (gain) in the beam-splitter-uncoupled limit ($J=0$). The flux in this gauge is simply given by $\Phi=2\varphi_{12}$. 
   The resonator quadratures are delocalised in particle-hole space and their interactions can be decomposed in terms of particle-hole conversions along the two loops, i.e. $\mathcal{H}_{\mathrm{SD}}=\mathcal{H}_{\mathrm{SD}}^\mathcal{L}+\mathcal{H}_{\mathrm{SD}}^{\mathcal{L}^*}$ (loop order $\{a_1,a_2,a_2^{\dagger},a_1^{\dagger}\}$), with 
   \begin{subequations}
   \begin{align}
   \mathcal{H}_{\mathrm{SD}}^\mathcal{L}=\left(\begin{array}{cccc}
   0 & \bar{J} & 0 & 0\\
   0 & 0 & -i\eta & 0\\
   0 & 0 & 0 & -\bar{J}\\
   -i\eta & 0 & 0 & 0
   \end{array}\right), &&\hspace{-1mm}
   \mathcal{H}_{\mathrm{SD}}^{\mathcal{L}^*}=(\Sigma_z\mathcal{H}_{\mathrm{SD}}^\mathcal{L}\Sigma_z)^{\dagger},
   \end{align}
   \end{subequations}
   where ($\bar{J}=Je^{-\frac{i \Phi }{2}}$).
   The mapping into the quadrature basis $Q$ (order
   $\{X_1,X_2,Y_1,Y_2\}$) renders the loop matrices into
   \begin{align}
   Q\mathcal{H}_{\mathrm{SD}}^\mathcal{L}Q^{\dagger}=&\frac{1}{2}\left(\begin{array}{cccc}
   -i\eta & \bar{J} & \eta & i\bar{J}\\
   -\bar{J} & -i\eta & i\bar{J} & -\eta\\
   \eta & -i\bar{J} & i\eta & \bar{J}\\
   -i\bar{J} & -\eta & -\bar{J} & i\eta
   \end{array}\right),\\
   Q\mathcal{H}_{\mathrm{SD}}^{\mathcal{L}^*}Q^{\dagger}=&\frac{1}{2}\left(\begin{array}{cccc}
   -i\eta & -\bar{J}^{*} & -\eta & i\bar{J}^{*}\\
   \bar{J}^{*} & -i\eta & i\bar{J}^{*} & \eta\\
   -\eta & -i\bar{J}^{*} & i\eta & -\bar{J}^{*}\\
   -i\bar{J}^{*} & \eta & \bar{J}^{*} & i\eta
   \end{array}\right).
   \end{align}
   
   In the quadrature basis, the  BdG dynamical matrix $\mathcal{H}^{XY}_\mathrm{SD}=Q\mathcal{H}_{\mathrm{SD}}^\mathcal{L}Q^{\dagger}+Q\mathcal{H}_{\mathrm{SD}}^{\mathcal{L}^*}Q^{\dagger}$ reads
   \begin{align}\label{eq:h_XY_sd}
   \mathcal{H}_{\mathrm{SD}}^{XY}=\left(\begin{array}{cccc}
   -i\eta & -iJ_{\parallel} & 0 & iJ_{\perp}\\
   iJ_{\parallel} & -i\eta &  iJ_{\perp} & 0\\
   0 & -iJ_{\perp} &  i\eta & -iJ_{\parallel}\\
   -iJ_{\perp} & 0  & iJ_{\parallel} & i\eta\\
   \end{array}\right),
   \end{align}
   where the combination of clockwise and counter-clockwise processes with nontrivial Peierls phases leads to the flux-dependent couplings  $J_{\parallel}=J\sin(\frac{\Phi}{2})$ and $J_{\perp}=J\cos(\frac{\Phi}{2})$ between quadratures.
   
    \begin{figure*}
   \centering
   \includegraphics[width=1\linewidth]{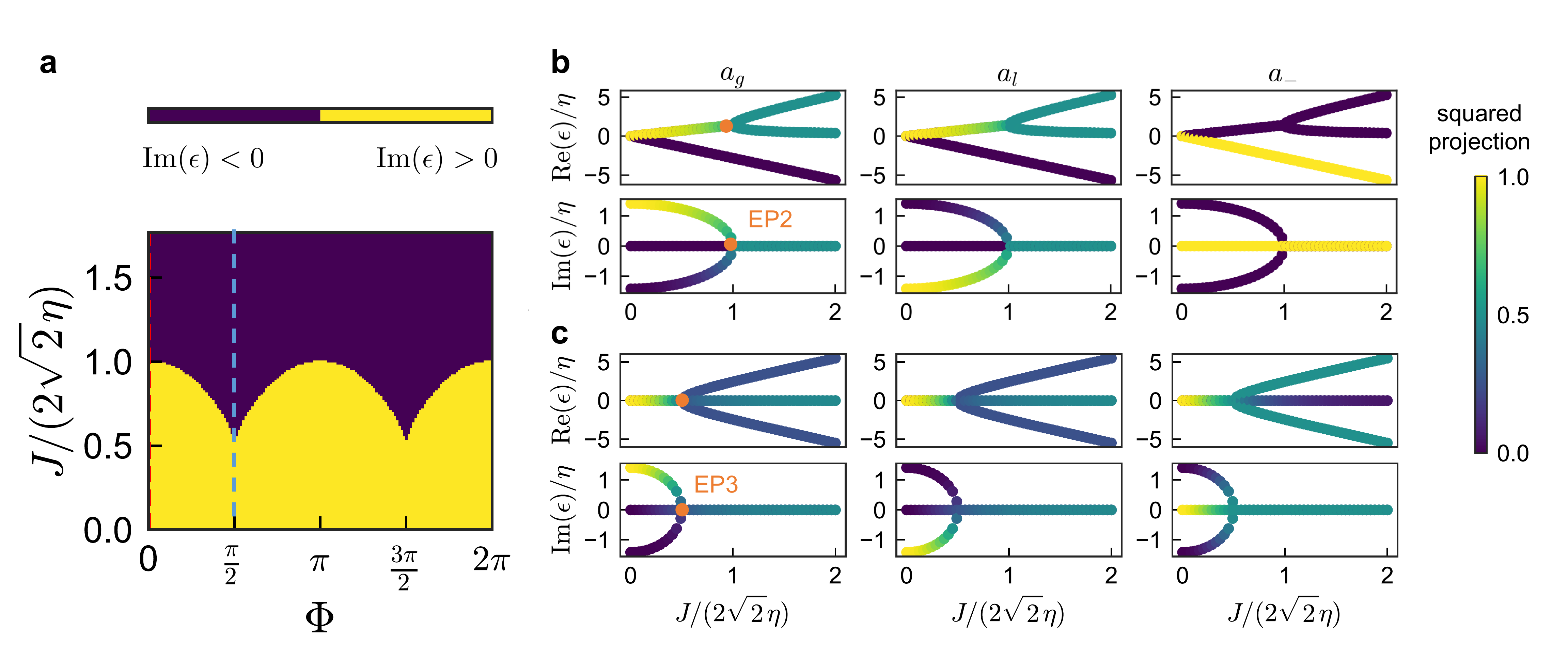}
   \caption{\textbf{Calculated eigenstates of the loop $a_1,a_2,a_3^{\dagger}$ in the SCT studied in \autoref{fig:4}.} \textbf{a} Phase diagram for the imaginary part of the eigenfrequencies, showing the stability-to-instability boundary in $\xi-\Phi$ space, where $\xi=J/(2\sqrt{2}\eta)$ and $\gamma_i=0$. Such boundary is associated with a $2^{\mathrm{nd}}$ order exceptional contour. \textbf{b} Cuts of the eigenfrequency Riemann surfaces along $\Phi=0$, shown as a red dashed trajectory in the phase diagram, as a function of the ratio $\xi=J/(2\sqrt{2}\eta)$. The squared weights of the $J=0$ eigenstates in the corresponding eigenvectors are shown in the colorscale. The weights are calculated from the symplectic projections ($\Sigma_z$ product) on the gainy/lossy combinations $a_g,a_l$ and the passive mode $a_-$.  A second order exceptional point (denoted EP2), found for $J=2\sqrt{2}\eta$, is highlighted. As $J<2\sqrt{2}\eta$, $\mathcal{P}_{gl}\mathcal{T}$ symmetry is spontaneously broken, inducing eigenstate localisation. The antisymmetric 1-2 mode $a_-$ is detached from this mechanism and remains uncoupled.	Real and imaginary parts are re-scaled by $\eta$. \textbf{c} Similar data along the cut $\Phi=\pi/2$ (corresponding to the blue dashed line in \textbf{a}, which shows the third-order exceptional point (EP3, at $J=\sqrt{2}\eta$). The $\mathcal{P}_{gl}\mathcal{T}$ symmetry broken states are now hybrid combinations of $a_g, a_-$ and $a_l, a_-$ modes. Such combinations break $\mathcal{P}_{12}\mathcal{T}$ as well, as explained in the main text.\label{fig:ED_3}}
   \end{figure*}
   
   \subsection*{Gain-loss bases and effective $\mathcal{PT}$ symmetries}\label{sec:gain_loss_basis}
   
   Adequate bases for the SD and the SCT can be determined for which one easily recognises an inversion plane that separates gain and loss at either side, and therefore potentially a parity-time ($\mathcal{PT}$) symmetry.
   In the case of the SD, a $\mathcal{PT}$ symmetry is found using the local quadratures $\{X_i,Y_i\}$. In the basis $\left\{X_1,Y_2,X_2,Y_1\right\}$ and in a gauge with parametric driving phases $\theta_i=\pi/2$, the  BdG dynamical matrix \autoref{eq:h_XY_sd} is block-diagonal for $\Phi=0$ and reads $\hs^{XY}=\mathrm{diag}(\mathcal{H}^{X_1,Y_2},\mathcal{H}^{X_2,Y_1})$ with the blocks
   \begin{subequations}
   \begin{align}\label{eq:SD_subb}
   \mathcal{H}^{X_{1},Y_{2}}=i\left(\begin{array}{cc}
   -\eta & J\\
   -J & \eta
   \end{array}\right)=\mathcal{H}^{X_{2},Y_{1}}.
   \end{align}
   \end{subequations}
   governing the dynamics of the independent ``quadrature dimers'' $X_1 Y_2$ and $X_2 Y_1$.
   
   Each of the blocks is $\mathcal{P}_{X_iY_j}\mathcal{T}$ symmetric, with parity symmetries $\mathcal{P}_{X_iY_j}:X_i\leftrightarrow Y_j$. The eigenfrequencies for each block, $\epsilon^{X_iY_j}= \pm\sqrt{\eta^2-J^2}$, 
   are real within the $\mathcal{P}_{X_iY_j}\mathcal{T}$-symmetric region $J>\eta$, in which the corresponding eigenstates respect the symmetry of the dynamical matrix. This is no longer true if $J\leq \eta$, where $\mathcal{P}_{X_iY_j}\mathcal{T}$ is spontaneously broken, with a second order EP at $J=\eta$ indicating the transition.  
   
   The recognition of this parity-time symmetry allows explaining why non-zero fluxes imply complex, non-real eigenvalues and the disappearance of the EP: they induce coupling between the sub-blocks~\autoref{eq:SD_subb} and the \textit{explicit} breaking of $\mathcal{P}_{X_iY_j}\mathcal{T}$. This dynamical phase transition along $\Phi\geq0$ from real to complex eigenvalues can equivalently be characterised in terms of \textit{spontaneous} breaking of a generalised $\mathcal{PT}$ ($\gpt$) symmetry without loss of diagonalisability~\cite{Flynn2020}.  An extended theoretical analysis shows that asymmetries in SD cause small shifts in the location of degeneracies in the experimental regime. For instance, if $\Phi=\pi$, asymmetry in the loss rates transforms the degeneracy at $J=0$ into an EP at $J=|\gamma_2-\gamma_1|/2$, overshadowed by dissipation $\gamma_i$. This case is a particular instance of breaking of $\mathcal{PT}$ symmetries fulfilled for arbitrary fluxes and the expansion of exceptional points into contours in parameter space (see Supplementary Information~\autoref{sec:ECS}). This can explain why in \autoref{fig:3}b (top) the experimentally observed peak at zero shift extends for slightly higher $J$ than expected in the idealised theory.
   
   Similarly, the dynamical phases of the SCT can be classified by $G\mathcal{PT}$ symmetries, implied from the $\Sigma_z$-pseudo-Hermiticity of bosonic dynamical matrices. It is therefore paramount that the $\Sigma_z$-pseudo-Hermiticity of $\hbdg$ is fulfilled, modulo a constant displacement in the imaginary parts ($a_i\mapsto\bar{a}_i=a_ie^{-\gamma t/2}$). In experiment, we achieve this by applying feedback control to modify the resonator damping rates to be equal. The $G\mathcal{PT}$ symmetry is again straightforwardly recognised in the  basis of the eigenmodes for vanishing beam-splitter coupling ($J=0$), which corresponds to a basis where a mirror plane separates gain and loss in the system. 
   The SCT's dynamics can be integrated using a single block of $\hbdg'$, for example the block acting on $\{a_1,a_2,a_3^\dagger\}$ (gauge $\theta_{23}=\theta_{13}=0$, where the flux simply reads $\Phi=\varphi_{12}$),
   \begin{equation}\label{eq:H_sct_loop}
   \mathcal{L}=\left(\begin{array}{ccc}
   0 & Je^{-i\Phi} & \eta\\
   Je^{i\Phi} & 0 & \eta\\
   -\eta & -\eta & 0
   \end{array}\right).
   \end{equation}
   Now we switch to the eigenbasis of~\autoref{eq:H_sct_loop} for $J=0$, via the unitary transformation,
   \begin{align}
   U_{gl}=\frac{1}{\sqrt{2}}\left(\begin{array}{ccc}
   \frac{i}{\sqrt{2}} & -\frac{i}{\sqrt{2}} & -1\\
   \frac{i}{\sqrt{2}} & -\frac{i}{\sqrt{2}} & 1\\
   1 & 1 & 0
   \end{array}\right).
   \end{align}
   The corresponding eigenvectors (column vectors of $U_{gl}$) are denoted as $a_l=(a_3^{\dagger}+ia_+)/\sqrt{2}$ (where $a_+=(a_1+a_2)/\sqrt{2}$ is the symmetric superposition of resonator 1 and 2 states), $a_g=(a_3^{\dagger}-ia_+)/\sqrt{2}$ and $a_-=(a_2-a_1)/\sqrt{2}$. The effective sites $a_l,a_-$ and $a_g$ regroup gain and loss in the system. Adopting the order $\{a_l,a_g,a_-\}$, we decompose the transformed matrix $\mathcal{L}_{gl}=U_{gl}^{\dagger}\mathcal{L}U_{gl}=\Xi+\Theta$ into the contribution for $J=0$
   \begin{equation}
   \Xi\equiv U_{gl}^{\dagger}\mathcal{L}U_{gl}|_{J=0}=\mathrm{diag}(-i\sqrt{2}\eta,i\sqrt{2}\eta,0),
   \end{equation}
   and the effective frequency shifts and interactions of modes $a_g,a_l,a_-$,
   \begin{align}
   \Theta=\left(
   \begin{array}{ccc}
   \frac{1}{2} J \cos (\Phi )  & -\frac{1}{2} J \cos (\Phi ) & \frac{J \sin (\Phi )}{\sqrt{2}} \\
   -\frac{1}{2} J \cos (\Phi ) & \frac{1}{2} J \cos (\Phi )  & \frac{J \sin (\Phi )}{\sqrt{2}} \\
   -\frac{J \sin (\Phi )}{\sqrt{2}} & \frac{J \sin (\Phi )}{\sqrt{2}} & -J \cos (\Phi ) \\
   \end{array}
   \right).
   \end{align}
   In this basis we recognise that $\mathcal{L}_{gl}$ respects $\mathcal{P}_{gl}\mathcal{T}$ symmetry, where the parity operation $\mathcal{P}_{gl}:a_g\leftrightarrow a_l$ swaps the effective gain and loss sites, and $\mathcal{T}:i\mapsto-i,\Phi\mapsto-\Phi$.   In particular,  for zero flux, the dynamical matrix reads
   \begin{subequations}
   \begin{align}\label{eq:f0}
   \mathcal{L}_{gl}|_{\Phi=0}=\left(
   \begin{array}{ccc}
   \frac{J}{2}- i \sqrt{2} \eta & -\frac{J}{2} & 0 \\
   -\frac{J}{2} & \frac{J}{2}+ i \sqrt{2} \eta & 0 \\
   0 & 0 & -J \\
   \end{array}
   \right),
   \end{align}
   and shows that mode $a_-$ is uncoupled from the remaining $\mathcal{P}_{gl}\mathcal{T}$-symmetric $2\times 2$ effective dynamical matrix for $a_g$ and $a_l$ (see inset in \autoref{fig:4}d).
   
    \begin{figure*}
  	\centering
  	\includegraphics[width=\linewidth]{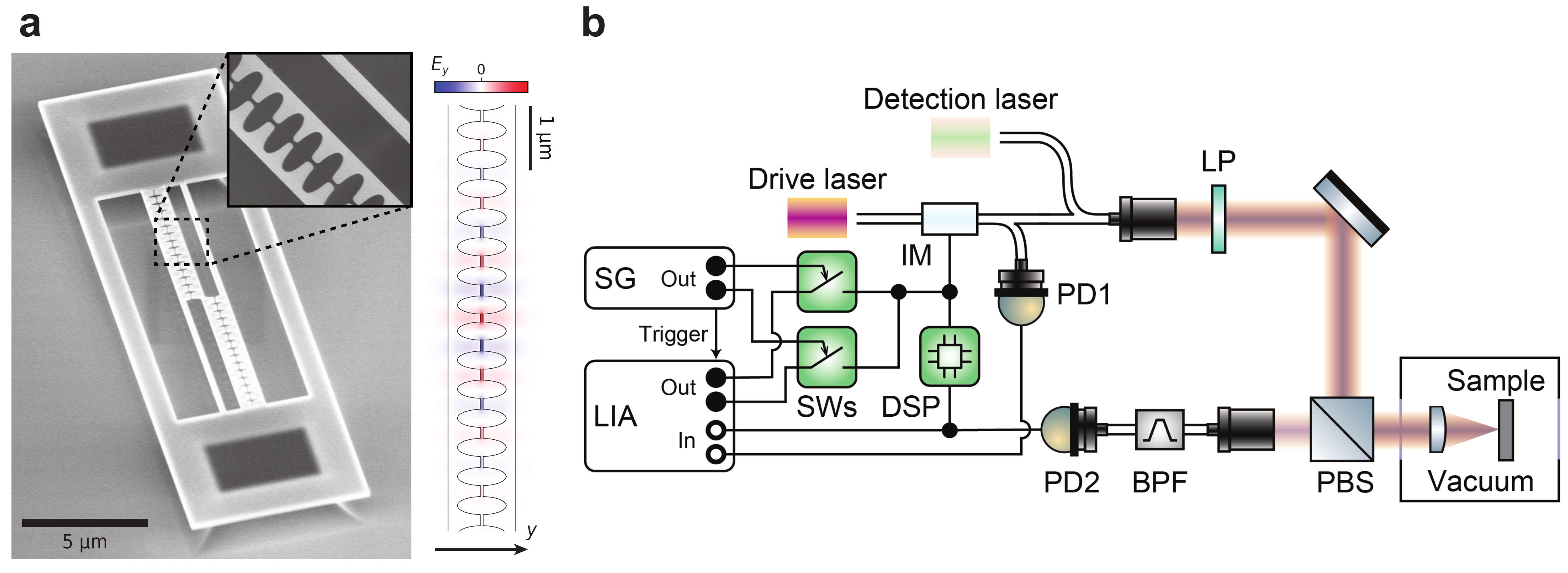}
  	\caption{\textbf{Experimental setup. }\textbf{a} Electron micrograph (left; tilt $45\degree$, inset; top view) showing a device as used in our experiments. In the top silicon device layer (thickness $220$ nm), three suspended beams are defined with teeth separated by a narrow slit ($\sim 50$ nm). Between each outer beam and the central beam, a photonic crystal cavity is defined that hosts an optical mode (right; simulated electric field $y$-component $E_y$). The mode's energy is strongly confined to the narrow slits, inducing large parametric interaction with flexural mechanical resonances of the two beams. The cavity's off-centre position ensures coupling to both even and odd resonances. In the presented experiments, we only use one of the two cavities. The widths of the outer beams' straight sections are intentionally made unequal, such that the mechanical resonances of all beams are detuned. The top layer is supported by pedestals etched out in the buried silicon oxide layer. \textbf{b} Schematic of the experimental set-up. IM, intensity modulator; LP, linear polarizer; PBS, polarizing beamsplitter; BPF, optical bandpass filter; PD1, PD2, photodiode; DSP, digital signal processor; SWs, microwave switches; LIA, ultrahigh-frequency lock-in amplifier; SG, signal generator. The LIA ports serve to (Out) drive the IM through an amplification stage (not shown) and to (In) analyse intensity modulations of the drive laser (for calibration) and detection laser. For time-resolved measurements, the SG is programmed to (Out) actuate the drive signal switches and trigger the LIA acquisition. The DSP optionally generates a feedback signal to modify resonator damping rates. \label{fig:device}}
  \end{figure*}
   
   Conversely, a linear trimer structure follows at $\Phi=\pm \pi/2$, where the dynamical matrix reads
   \begin{align}\label{eq:fpi/2}
   \mathcal{L}_{gl}|_{\Phi=\pm\pi/2}=\left(
   \begin{array}{ccc}
   -i \sqrt{2} \eta  & 0 & \mp\frac{J}{\sqrt{2}} \\
   0 & i \sqrt{2} \eta  & \pm\frac{J}{\sqrt{2}} \\
   \mp\frac{J}{\sqrt{2}} & \pm\frac{J}{\sqrt{2}} & 0 \\
   \end{array}
   \right).
   \end{align}
   \end{subequations}
   Here we note the explicit morphing of the effective dimer into a three-mode chain configuration when changing the flux in the SCT from $0,\pi$ to $\pm \pi/2$ (see main text). In addition, $\mathcal{P}_{gl}\mathcal{T} \mathcal{L}_{gl}|_{\Phi=\pm\pi/2}=\mathcal{L}_{gl}|_{\Phi=\pm\pi/2}$, noting $\mathcal{T}\Theta|_{\Phi=\pm\pi/2}=\Theta|_{\Phi=\mp\pi/2}$.  From \autoref{eq:f0} and \autoref{eq:fpi/2}, we can directly observe flux affects the nature of the arising EPs, which can be either second or third order. Note that while finite synthetic fluxes retain $\pglt$ of $\mathcal{L}$, they break the mirror symmetry $\mathcal{P}_{12}$, affecting the localisation transition above the EP (see main text,~\autoref{fig:ED_3}). The full expressions for the eigenspectra that illustrate this behaviour can be found in the Supplementary Information
  ~\autoref{sec:loop_eig}.

   \subsection*{Subdominant linewidths in thermal spectra for the squeezing dimer}
   
   In the thermomechanical noise spectra of the SD in \autoref{fig:2}e,f, we expect narrow and broad, frequency-degenerate, resonances. This is shown by the ideal SD ($\gamma_i=\gamma$), whose spectrum is obtained in a closed form using the relationship (Supplementary Information~\autoref{sec:eig_therm})
   \begin{equation}\label{eq:noise_sp}
   \mathcal{S}(\omega)=\langle\vec{\alpha}^{\dagger}(\omega)\vec{\alpha}(\omega)\rangle=\chi_m^{\dagger}(\omega)\mathcal{D}\chi_m(\omega),
   \end{equation}
   with mechanical susceptibility matrix  $\chi_m(\omega)=i/(\omega\mathbb{1}-\hs)$ (see Supplementary Information~\autoref{sec:phase_sp_rep}).
   The noise spectrum of resonator $i\in(1,2)$ is given by the diagonal element $\mathcal{S}_{ii}(\omega)$. An explicit calculation for the SD shows that even in the simplified limit of equal resonator bath occupations $\bar{n}_i=\bar{n}$, the spectrum consists of 4 superimposed Lorentzian responses located at the real parts of the eigenfrequencies of $\hs$.
   
   The full expression is omitted for simplicity, but we present the result for the relevant limit $\Phi=\pi$, where two pairs of resonances split by $2J$ and
   \begin{align}\label{eq:noise_sp_SD}
   \mathcal{S}_{ii}(\omega)\propto\gamma\sum_{\Omega=\pm J}(&\frac{\bar{n}+1}{(\gamma+2\eta)^{2}+4(\omega-\Omega)^{2}}+\nonumber\\
   &\frac{\bar{n}}{(\gamma-2\eta)^{2}+4(\omega-\Omega)^{2}}).
   \end{align}
   From~\autoref{eq:noise_sp_SD}, it is apparent that the spectral weight in the rotating frame at $\pm J$ in the stable regime ($\gamma>2\eta$) is concentrated in a dominant, narrow resonance with linewidth $\gamma-2\eta$, on top of an additional, heavily damped contribution with linewidth $\gamma+2\eta$.

   \subsection*{Design and fabrication}
   The device, shown in~\autoref{fig:device}a, was designed as a sliced photonic crystal nanobeam with two beam halves of different mass to create non-degenerate mechanical modes. The cavity was defined away from the beams' centres to optically access flexural modes with even as well as odd symmetries. Devices were fabricated from a silicon-on-insulator substrate, with a 220 nm device layer and \SI{3}{\micro\meter} buried oxide layer (BOX). A 50 nm layer of diluted hydrogen silsesquioxane resist (1:2 in methyl isobutyl ketone) was spin-coated, and electron-beam lithography (Raith Voyager) was used to write patterns on the sample. After developing in tetramethylammonium hydroxide, an anisotropic etch of the exposed device layer was done using inductively coupled plasma–reactive ion etching with HBr and O$_2$ gases. The nanobeams were suspended in a wet etch of the underlying BOX layer with hydrofluoric acid followed by critical point drying.
   
   \begin{figure*}
	\centering
	\includegraphics[width=\linewidth]{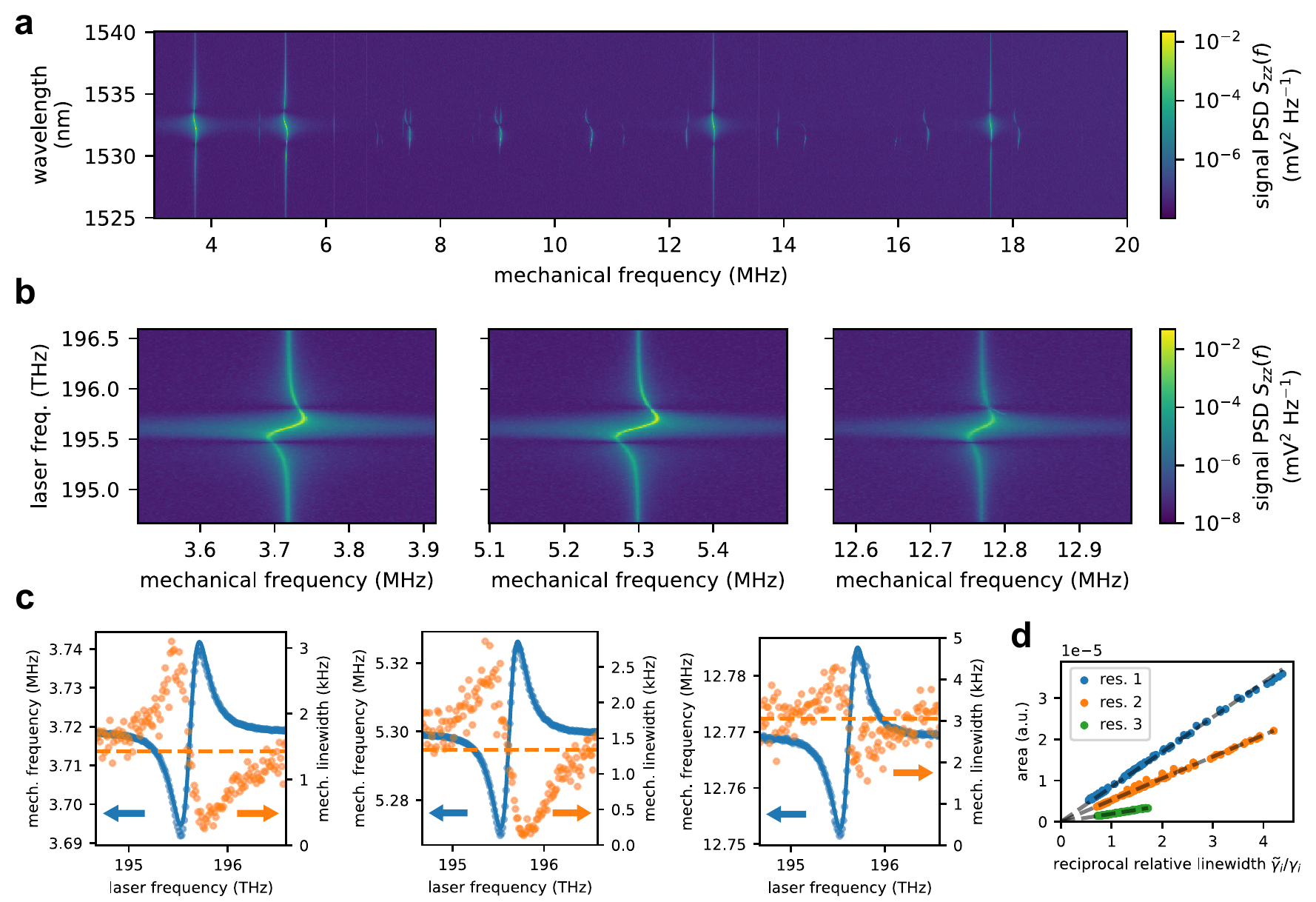}
	\caption{\textbf{Optical spring shift and opto-thermal backaction.} \textbf{a} Thermomechanical noise spectra of the first few mechanical modes imprinted on an unmodulated single drive/detection laser, as the laser's frequency ($\omega_L$) is swept across the cavity resonance. The four most intense peaks around frequencies $\omega_i/2\pi \approx \{3.7, 5.3, 12.8, 17.6\}$ MHz correspond to flexural modes (labelled $i$) of the individual beam halves and show frequency tuning characteristic to the optical spring effect, while the other modes represent non-linearly transduced harmonics of those modes. \textbf{b} Zoomed-in thermomechanical noise spectra of the first three resonators. \textbf{c} From the spectra in \textbf{b}, resonance frequencies $\omega_i$ (blue circles) and linewidths $\gamma_i$ (orange circles) are extracted. The resonance frequencies are fitted using the standard optical spring model (solid blue). Across all resonators, we find agreement in the fitted cavity resonance $\omega_c/2\pi = 195.62$ THz and linewidth $\kappa/2\pi = 320$ GHz (Q factor $Q\approx600$). The small sideband resolution $\omega_i/\kappa \approx 10^{-5}$ suggests very little change in linewidth due to dynamical cavity backaction (dashed orange). The linewidth modulations we observe suggest the presence of an opto-thermal retardation effect~\cite{hauer2019dueling}. \textbf{d} Drive laser frequency sweep while now using a separate, fixed frequency, far-detuned detection laser. The fixed transduction of mechanical motion onto this detection laser allows a comparison of resonance peak area $A_i(\omega_L)$, versus linewidth $\gamma_i(\omega_L)$ as the drive laser frequency $\omega_L$ is varied. The resonance peak area of mode $i$ is proportional to the variance $\langle X_i^2 \rangle$ of its displacement $X_i$, which is proportional to its temperature $T_i$. Dynamical backaction modifies the effective mode temperature through $T_i = T_0 \left( \tilde{\gamma}_i/\gamma_i \right)$ \cite{Aspelmeyer2013}, where $T_0$ is the initial temperature and $\tilde{\gamma}_i$ is the mode's intrinsic linewidth, determined by switching off the drive laser. Our data is well explained by linear fits of $A_i(\omega_L)$ versus $\tilde{\gamma}_i/\gamma_i(\omega_L)$ (dashed), confirming the effective temperature model.\label{fig:optical_spring}}
\end{figure*}
   
   \subsection*{Experimental setup}
  
A schematic of the experimental setup is presented in~\autoref{fig:device}b. The sample was placed, with the devices rotated by $45\degree$ relative to the vertical polarisation of the incoming light, in a vacuum chamber at room temperature at a pressure of $\sim 2 \times 10^{-6}$~mbar. 
A tunable laser (Toptica CTL 1500) connected through a Thorlabs LN81S-FC intensity modulator (IM) was used as the drive laser. A small part of the modulated drive laser light was split using a fibre-based beam splitter and fed onto a fibre-coupled fast photodetector (New Focus 1811, DC-coupled) to monitor the drive signal. A second laser (New Focus TLB-6328 or Toptica CTL 1550) far detuned from the cavity resonance ($\omega_\text{det} - \omega_c \approx -2.5\kappa$) was used as the detection laser. The lasers were combined on a fibre-based beam combiner and launched using a fibre collimator into the free-space setup.

Control signals were generated by a Zurich Instruments UHFLI lock-in amplifier. One output of the lock-in amplifier carried signals to generate interactions, while the other output carried coherent excitation signals. Both outputs were routed through individual radio-frequency (RF) switches (Mini-Circuits ZYSWA-2-50DR+), combined, amplified (Mini-Circuits ZHL-32A+ with 9 dB attenuation) and connected to the RF port of the IM to drive and modulate the nanobeam mechanics. For time-resolved experiments, a synchronised two-channel signal generator (Siglent SDG1062X) was used to generate pulses to actuate both RF switches and trigger the lock-in amplifier acquisition. 

Reflected detection laser light that interacted with the cavity was filtered using a cross-polarised detection scheme, fibre coupled, separated from the drive laser using a tunable bandpass filter (DiCon), and detected on a fast, low-noise photodetector (New Focus 1811, AC-coupled). Intensity modulations of the detection laser encoding resonator displacements were analysed using the lock-in amplifier. 

To generate a feedback signal, the electronic displacement signal was split and filtered using a digital signal processor (DSP, RedPitaya STEMlab 125-14) that implemented a configurable electronic bandpass filter with tunable gain and phase shift (using the PyRPL suite). The output of the DSP was combined with the control signals just before the RF amplifier.

\subsection*{Experimental procedure}
\subsubsection*{Resonator characterisation}
The intrinsic, optically unmodified resonator frequencies $\tilde{\omega}_i$ and linewidths $\tilde{\gamma}_i$ were obtained by switching off the drive laser and recording a thermomechanical spectrum with the detection laser. A power sweep of the detection laser verified that the detection laser did not induce a noticeable optical shift in frequency or linewidth.

To compensate for variations in incoupling and outcoupling efficiency, caused by position drift of the sample stage, the following reference procedure was performed immediately before every experiment: A thermomechanical spectrum was taken to obtain the spring-shifted resonator frequencies $\omega_i$, linewidths $\gamma_i$, and root-mean-square (rms) displacement voltage levels $z_{\text{rms},i}$. From the rms level, the displacement voltage corresponding to a single phonon was calculated using $z_{\text{ph},i}^2 = z_{\text{rms},i}^2/\left(\bar{n}_i \tilde{\gamma}_i / \gamma_i \right)$, where $\bar{n}_i = k_B T / \hbar \omega_i$ is the occupation of the resonator's phonon bath at room temperature $T = 295$ K. The ratio $\tilde{\gamma}_i / \gamma_i$ compensates for thermo-optically induced dynamical backaction\cite{hauer2019dueling} that changes the effective bath temperatures\cite{Aspelmeyer2013} (see~\autoref{fig:optical_spring}).

\begin{figure}
	\includegraphics[width=1\linewidth]{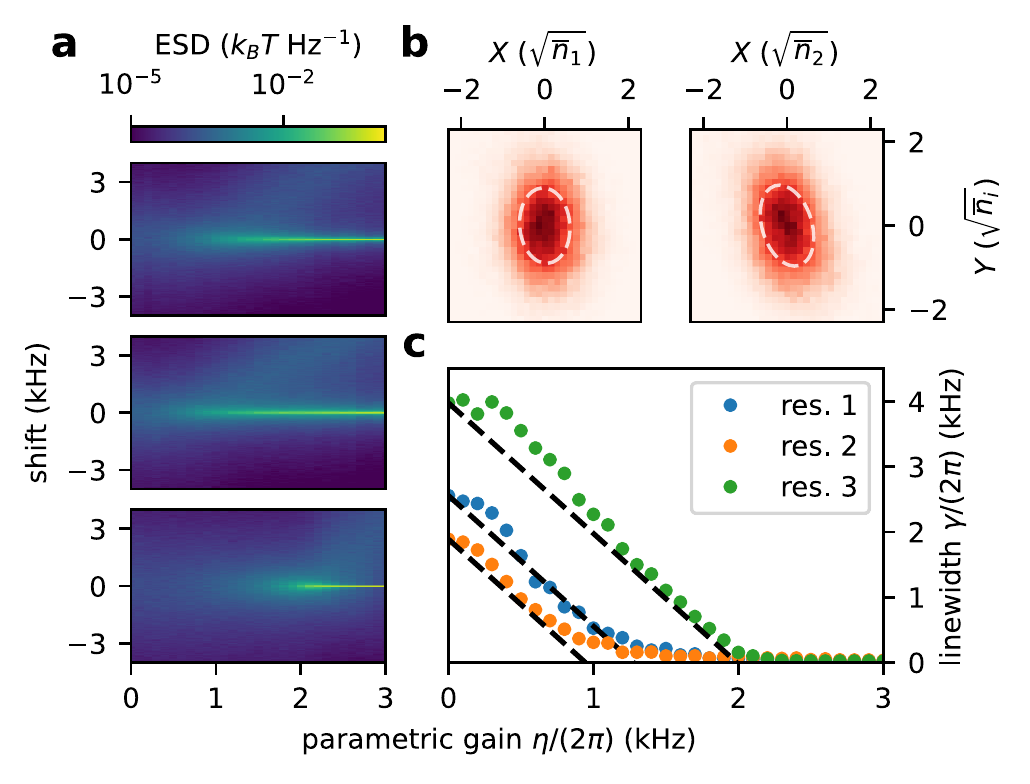}
	\caption{\textbf{Single-mode squeezing and linewidth modulation by parametric driving.} \textbf{a} Parametric gain induced by a single-mode squeezing interaction observed in thermomechanical spectra. Each row corresponds to a separate experiment where resonator $i$ (top: 1, middle: 2, bottom: 3) is subjected to a single-mode squeezing interaction of strength $\eta$. As $\eta$ is increased, the resonance transitions from the broad intrinsic linewidth to a narrow parametric resonance. \textbf{b} The phase-space distribution of the thermal fluctuations of resonator $i$ (left: 1, right: 2) subject to a single-mode squeezing interaction of strength $\eta/(2\pi) = 0.5$ kHz with squeezing angle $\theta = \pi/2$ reveals a squeezed thermal state. The squeezed (antisqueezed) quadrature $X$ ($Y$), measured in units of the thermal equilibrium amplitude $\sqrt{\bar{n}_i}$, are referenced using the propagation delay (Methods). The principal components of the quadrature covariance matrix (standard deviations depicted by dashed ellipses) show slight residual phase offsets, estimated from a full sweep of $\theta$ (not shown) at $7\degree$ and $13\degree$ respectively, which are corrected for in all relevant experiments. \textbf{c} Fitted Lorentzian full-width at half-maximum linewidths of the resonances show in a). Even though a superposition of two degenerate resonances is expected -- a broadened resonance of the antisqueezed quadrature and a narrowed resonance of the squeezed quadrature -- only a single one can be successfully fitted in each spectrum. This reflects the fact that the highly populated narrowed resonance dominates the broadened resonance. As the parametric gain $\eta$ is increased, each resonator's squeezed quadrature linewidth is expected to decrease by $\Delta\gamma = -2\eta$ (dashed lines), until parametric threshold is reached at $\eta = \gamma_i/2$, where $\gamma_i$ is the intrinsic linewidth of resonator $i$. The fitted linewidths follow the expected trend quite closely for intermediate $\eta$, while for lower $\eta$ the narrow resonance is presumably not yet fully dominant and for larger $\eta$ high-amplitude non-linear effects are prominent.
		\label{fig:sm-squeezing}}
\end{figure}

\subsubsection*{Calibration of control signals}
To find the linear operation point of the IM, a sinusoidal modulation voltage was applied while sweeping its amplitude and monitoring the modulated drive laser. The IM bias voltage was varied to minimise the variation in DC transmission as a function of modulation amplitude. To compensate for frequency-dependent transmission in the RF chain, the relation between control signal voltage amplitude $V_\text{m}$ and modulation depth $c_\text{m}$ was measured individually for every tone using the DC-coupled modulation monitor detector.

For the BST experiments shown in \autoref{fig:1}, the linear relation between modulation amplitude $V_\text{m}$ and the beam-splitter coupling $J_{ij}$ induced by sinusoidal modulation at $\omega_\text{m} = \omega_i - \omega_j$, $i\neq j$ was established by sweeping $V_\text{m}$, recording thermomechanical spectra of resonators $i$ and $j$ and fitting the frequency splitting of the hybridised modes.

\begin{figure*}
	\centering
	\includegraphics[width=\linewidth]{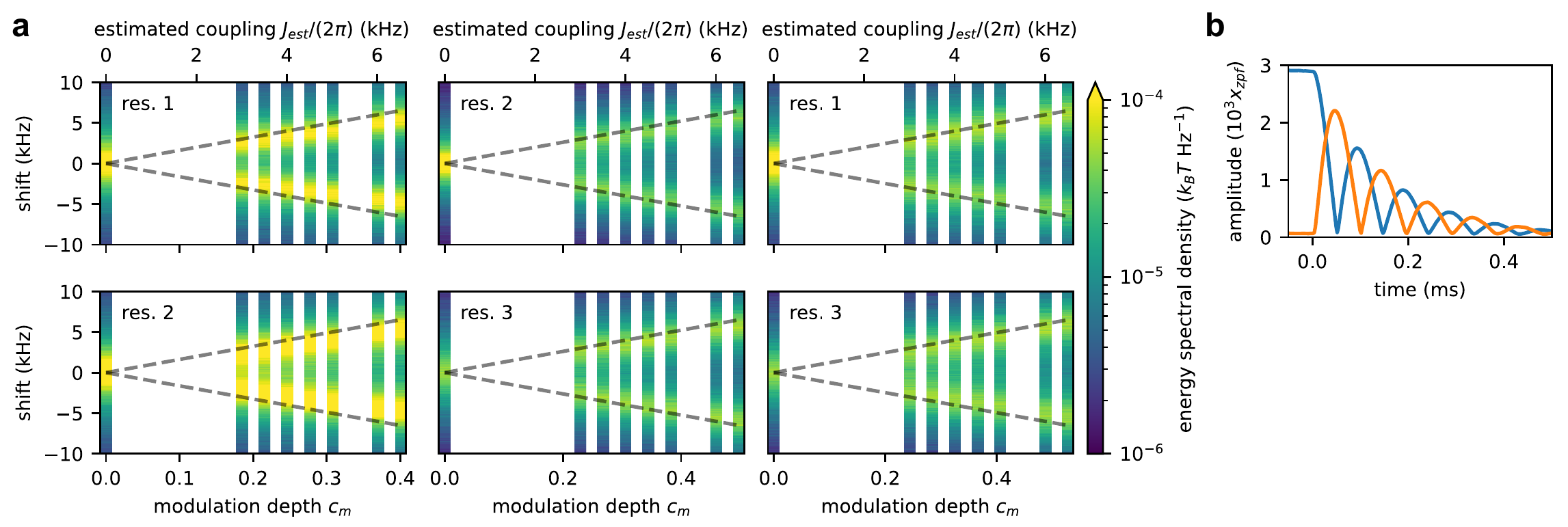}
	\caption{\textbf{Estimation of beam-splitter interaction strengths.} \textbf{a} Mode splitting induced by a beam-splitter interaction observed in thermomechanical spectra. Each column corresponds to a beam-splitter interaction induced between a pair of resonators $i \leftrightarrow j$ (left: $1 \leftrightarrow 2$, middle: $2 \leftrightarrow 3$, right: $1 \leftrightarrow 3$) by a single drive laser modulation at frequency $\Delta\omega_{ij} = \omega_i - \omega_j$, where $\omega_{i,j}$ is the frequency of resonator $i$, $j$. Thermomechanical spectra (top row: resonator $i$, bottom row: resonator $j$) are recorded for increasing modulation depth $c_\text{m}$. The linear relation $J_\text{est} = c_\text{m} \sqrt{\delta\omega_i \delta\omega_j} / 2$ is used to estimate the coupling strength $J_\text{est}$ (top axis) from $c_\text{m}$, where $\delta\omega_{i,j}$ is the optical spring shift of mode $i$, $j$. The estimated mode splitting (dashed) is slightly larger than observed, presumably due to frequency-dependent transduction (at DC and $\Delta\omega_{ij}$) in the measurement of $c_\text{m}$. The difference is quantified by extracting Lorentzian peak frequencies from the spectra and subsequently fitting those linearly against modulation depth, and results in an observed mode splitting slope that is $78\%$, $90\%$ and $90\%$ of the estimated slope respectively. The average estimation offset of $86\%$ is applied to all (beam-splitter and squeezing) interaction strength calculations in our experiments. \textbf{b} Time evolution of the coherent amplitude (in units of their zero point fluctuations) of a pair of resonators (1, blue and 2, orange) coupled through a beam-splitter interaction (strength $J/2\pi = 5$ kHz). Resonator $1$ is initially (time $t < 0$) driven to a high amplitude steady state by a coherent drive laser modulation. At $t=0$, the drive is switched off and the interaction is switched on. Rabi oscillations induced by the coupling interaction are observed, where energy is transferred back and forth between the resonators until the coherent energy in the resonators is dissipated.\label{fig:beam_splitter_calib}}
\end{figure*}

Spectral estimation of the strength of a squeezing interaction is less precise due to the spectral superposition of gain and loss (see~\autoref{fig:sm-squeezing}). Therefore, in the other experiments, the squeezing and beam-splitter interaction strengths $\eta_{ij},J_{ij}$ induced by a sinusoidal drive laser modulation at frequency $\omega_\text{m} = \omega_i \pm \omega_j$ (for $i\neq j$ or $i=j$) and modulation depth $c_\text{m}$ were obtained using the relation $\eta_{ij},J_{ij} = c_\text{m} \sqrt{\delta\omega_i \delta\omega_j} / 2$, where $\delta\omega_i = \omega_i - \tilde{\omega}_i$ is the optical spring shift of resonator $i$. Note that $\delta\omega_i$ and $\delta\omega_j$ always have the same sign. Using this relation avoids the need to know the photon-phonon coupling rates $g_{0i}$ and cavity incoupling efficiency precisely. To verify, the effective beam-splitter interaction strength obtained above was compared to the frequency splitting observed in thermomechanical spectra for a sweep of the modulation depth $c_\text{m}$ (see~\autoref{fig:beam_splitter_calib}). From this, a difference between calculated and actual interaction strength of about $10\%$ was obtained, presumably due to a difference in the modulation detector sensitivity at DC. This difference was applied as a correction factor to all calculated interaction strengths. 

For the SD experiments (\autoref{fig:2} and \autoref{fig:3}), an additional, linear correction on the scaling of the beam-splitter coupling $J$ was obtained by fitting the linear frequency splitting for $\Phi=\pi$ (as shown in \autoref{fig:3}b) as a function of $c_\text{m}$.

In the BST experiments, the flux offset $\Phi_0 = \varphi_{23} + \varphi_{31}$ was obtained by extracting eigenfrequencies from thermomechanical spectra as a function of $\varphi_{12}$ and fitting those to the eigenfrequencies $\epsilon_k = 2J\cos((2\pi k + \Phi)/3)$ of the Hamiltonian $H_\text{BST}$ in Eq.~(1) indexed by $k = \{-1,0,1\}$, where $\Phi = \Phi_0 + \varphi_{12}$. In the other experiments, to circumvent spectral estimation of the flux and to facilitate the analysis of (anti)squeezed quadratures, the phases of the control tones are referred to an effective time origin internal to the lock-in amplifier, which allows to define a deterministic gauge in which the modulation phases are set. This method was verified by applying it to the BST and comparing it to the flux offset fitting method outlined above.

\begin{figure*}
	\centering
	\includegraphics[width=0.8\linewidth]{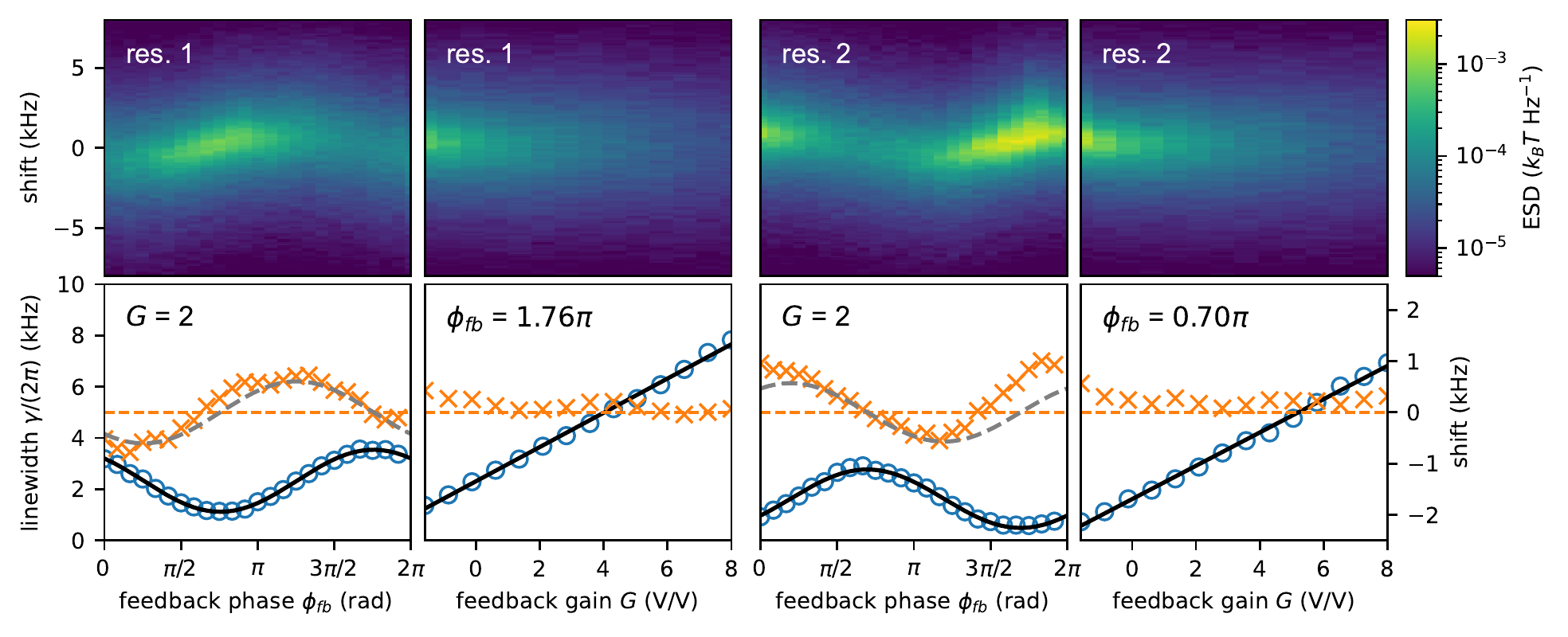}
	\caption{\textbf{Damping rate adjustment by feedback.} Resonator thermomechanical spectra (top row) and fitted full-width half-maximum linewidths (bottom row) adjusted by feeding back electronically filtered and phase-shifted resonator displacement signals onto the drive laser modulation (left two columns, resonator 1; right two columns, resonator 2). The resonator linewidth (circles) and frequency shift (crosses) vary sinusoidally with the feedback phase $\phi_\text{fb}$ (odd columns). By fitting the linewidth variation (solid black), the optimal phase shift to increase the damping rate is selected. The frequency variation (dashed grey) expected from the fitted linewidth modulation, relative to the resonator frequency with feedback off (dashed orange), lags by $\pi/2$ radians. For the optimal feedback phase shift, an increase in linewidth is observed for increasing gain $G$, while the resonator frequency remains unaffected (even columns). The slope of the linear fit (solid black) can be used when setting a resonator's linewidth to a desired value.\label{fig:si:feedback}}
\end{figure*}

To realize the modulation of dissipation rates in the SCT experiments, a feedback signal was obtained by filtering the electronic displacement signal around each resonator's frequency $\omega_i$ in parallel (second-order filter half-width at half-maximum $78$ kHz), applying individual gains and phase shifts, and digitally combining the filtered signals. For each mode, the optimal feedback phase shift was found by taking thermomechanical spectra using fixed feedback gain for a full sweep of the phase shift, fitting the extracted linewidths with a sinusoidal variation and selecting the shift with the most significant change in linewidth (see
~\autoref{fig:si:feedback}). Subsequently, for the optimal phase shift, thermomechanical spectra were taken for various settings of the feedback gain and a linear relation was fitted between gain and extracted linewidths.

\subsubsection*{Analysis of the displacement signal}
The electronic displacement signal was demodulated in parallel at each resonator's frequency $\omega_i$ using electronic local oscillators internal to the lock-in amplifier that are referenced to the same time origin as the control tones. For each resonator, the demodulated in-phase ($I_i$) and quadrature ($Q_i$) components were filtered (third-order low-pass filter, $3$ dB bandwidth $50$ kHz) and combined into a complex amplitude $z_i(t) = I_i(t) + iQ_i(t)$ that is formally equivalent to the resonator amplitude in the rotating frame. The complex amplitudes of all resonators involved were acquired simultaneously, at a rate between $50$ and $500$ kSa/s, depending on the experiment. These complex time traces were normalized using the signal levels obtained in the reference procedure described earlier and were either \textit{i)} analysed directly to yield phase-space distributions; \textit{ii)} averaged coherently, i.e. $\langle z_i(t) \rangle$; 
or \textit{iii)} Fourier transformed (Hann windowing function), squared and averaged to yield energy spectral densities (ESD). In the last case, the low-pass filter was compensated for by dividing spectral densities by the filter frequency response. Time-resolved experiments were averaged over 1000 runs.

\begin{figure*}
	\centering
	\includegraphics[width=0.75\linewidth]{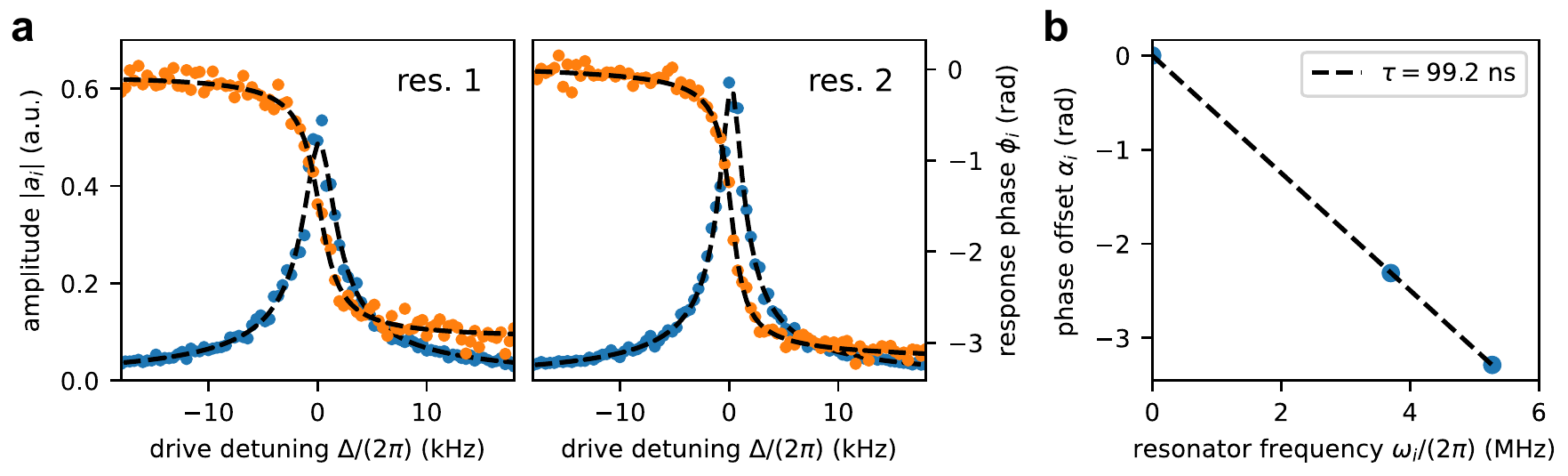}
	\caption{\textbf{Resonator coherent response.} \textbf{a)} Amplitude $|a_i|$ (blue, left axis) and phase $\phi_i$ (orange, right axis) of the complex response $a_i(\Delta) = e^{i(\phi_i(\Delta)+\alpha_i)} |a_i(\Delta)|$ of resonators $1$ and $2$ (resonance frequencies $\omega_i$) to a drive laser modulation at a frequency $\omega_d$ close to resonance (drive detuning $\Delta = \omega_d - \omega_i$). $\alpha_i$ is the phase offset due to signal delay through the set-up. A Lorentzian response $a_i = e^{i\alpha_i} A_i \frac{\gamma_i/2}{i\gamma_i/2 - \Delta}$ is fitted to the data (dashed). \textbf{b)} Phase offset $\alpha_i$ versus resonance frequency $\omega_i/(2\pi)$. A linear fit (dashed) of $\alpha_i = -\omega_i \tau$ implies a signal delay $\tau = 99.2$ ns.\label{fig:resonator_phase_offset}
	}
	
	
\end{figure*}

The total signal delay through the setup, from the LIA control outputs via the sample to the LIA input, was determined by driving each of the resonators and measuring the coherent response (see ~\autoref{fig:resonator_phase_offset}). The phase offset $\alpha_i$ between drive tone and coherent response of resonator $i$ was extracted and fitted linearly against the resonator frequencies $\omega_i$. The fitted delay was used to relate the quadratures of the demodulated amplitudes $z_i(t)$ to those defined by the control tones. This relation was verified for resonators 1 and 2 by turning on a single-mode squeezing interaction, recording a thermomechanical time trace, constructing a phase space distribution and fitting the angle of the squeezed and anti-squeezed principal quadrature axes (see~\autoref{fig:sm-squeezing}). Slight offsets on the order of $10\degree$, possibly stemming from dispersion between signals at $2\omega_i$ and $\omega_i$, were found and subsequently corrected for.

\bibliographystyle{apsrev4-2}
\bibliography{refs_all_cleaned}

\subsection*{Acknowledgements}
The authors thank Clara Wanjura, Andreas Nunnenkamp, and Matteo Brunelli for useful discussions, and Marc Serra-Garcia, Said Rodriguez, Femius Koenderink, and Oded Zilberberg for critical reading of the manuscript. This work is part of the research programme of the Netherlands Organisation for Scientific Research (NWO). The authors acknowledge support from the the European Research Council (ERC starting grant no. 759644-TOPP) and the European Union’s Horizon 2020 research and innovation programme under grant agreement no. 732894 (FET Proactive HOT). J. d. P. acknowledges financial support from the ETH Fellowship program (grant no. 20-2 FEL-66).

\clearpage
\pagebreak
\renewcommand{\theequation}{S\arabic{equation}}
\renewcommand{\thefigure}{S\arabic{figure}}
\setcounter{equation}{0}
\setcounter{figure}{0}
\setcounter{table}{0}
\setcounter{page}{1}

\vspace{0.9cm}

\onecolumngrid
\part{\Large Supplementary Information: Non-Hermitian chiral phononics through optomechanically-induced squeezing}
\twocolumngrid

\setcounter{secnumdepth}{2} 
\section{Further theoretical details}
\subsection{Bogoliubov modes and their dynamics}\label{sec:BdGForm}

Here we extend on the formalism for nanomechanical dynamics exposed in Methods. We focus on interpreting the eigenmodes of the non-Hermitian BdG dynamical matrix and its links with the unitary dynamics of Hermitian systems. A closed, linear bosonic system is governed by the the Heisenberg equations $i\dot{\vec{\alpha}}=\hbdg\vec{\alpha}$, with the dynamical Bogoliubov-de-Gennes (BdG) matrix defined as~\cite{blaizot1986quantum_,Flynn2020_},
\begin{equation}\label{eq:HBDG}
	\hbdg=\Sigma_zH=\left(\begin{array}{cc}
		\mathcal{A} & \mathcal{B}\\
		-\mathcal{B}^{*} & -\mathcal{A}^{*}
	\end{array}\right).
\end{equation}

The quasiparticles $\psi_n$ or eigenoperators of the effective Hamiltonian $H_\text{eff}$, as well as the solution of $\vec{\alpha}(t)$ can be expanded in terms of the eigenstates of $\hbdg$, $\ket{\psi_n}$ ($\hbdg\ket{\psi_n}=\epsilon_n\ket{\psi_n}$), in a similar fashion to Hermitian systems.  Nevertheless, $\hbdg$ is no longer diagonalisable via a unitary transformation~\cite{blaizot1986quantum_,Rossignoli2005_,Flynn2020_}, which breaks bosonic commutation rules.\footnote{This
	stems from the different character of $\mathcal{A}$ (Hermitian) and $\mathcal{B}$ (symmetric), and is ultimately caused by the different effects of unitary transformations: $\mathcal{A}\rightarrow U\mathcal{A}U^{\dagger}$, $\mathcal{B}\rightarrow U \mathcal{B}U^T$).} Instead the normal modes of $\mathcal{H}_\mathrm{eff}$, defined by $\vec{\psi}=T^{-1}\vec{\alpha}$ can only be found from a paraunitary canonical transformation $T$, namely $T^{-1}=\Sigma_zT^{\dagger}\Sigma_z$ where $\Sigma_z=\sigma_z\otimes\mathbb{1}_N=[\vec{\alpha},\vec{\alpha}^{\dagger}]$.  $\hbdg$ is diagonalizable in a complete eigenbasis with respect to the $\Sigma_z$ inner product (also denoted symplectic product). This non-unitary diagonalization also links to the existence of complex eigenvalues. These eigenvalues reflect redundancies caused by internal symmetries of $\hbdg$, described in Methods, in particular:
\begin{enumerate}
	\item  \textit{Charge-conjugation} $\mathcal{C}$ implies that if $\ket{\psi_n}$ is an eigenvector of $\hbdg$ with eigenvalue $\epsilon_{n}$ ($n\in (1,\cdots,2N)$ where $N$ is the number of modes), then $\mathcal{C}\ket{\psi_n}$ is an eigenvector of $\hbdg$ with eigenvalue $-\epsilon_n^*$. 
	\item \textit{$\Sigma_z$-Pseudo-Hermiticity} signifies that if $\ket{\psi_n}$ is an eigenvector of $\hbdg$ with eigenvalue $\epsilon_{n}$ ($n\in (1,\cdots,2N)$), then $\Sigma_z\ket{\psi_n}$ is an eigenvector of $\hbdg^\dagger$ with eigenvalue $\epsilon_n^*$. 
\end{enumerate}
In general, the eigenvalues of $\hbdg$ thus come in quartets $\{\epsilon_n,\epsilon_n^*,-\epsilon_n,-\epsilon_n^*\}$, of which some elements may be equal, e.g. when $\epsilon_n$ is real or imaginary. The redundancy in the BdG description introduced by the above symmetries plays a role in steady-state quantities, as detailed in~\autoref{sec:eig_therm}. 

Denoted by $\ket{\psi_n},\ket{\psi_{m*}}$, the eigenvectors corresponding to eigenvalues $\epsilon_n$ and $\epsilon_m^*\neq \epsilon_n^*$ are $\Sigma_z$-orthonormal: $\braket{\psi_{n*}|\Sigma_z|\psi_m}=\delta_{nm}$, while the usual norm vanishes $\braket{\psi_{n}|\Sigma_z|\psi_m}=0$. This basis allows for the spectral decomposition of $	\hbdg=\sum_{n=1}^{2N}\epsilon_{n}|\psi_{n}\rangle\langle\psi_{n*}|\Sigma_{z}$, and the expansion of \textit{Bogoliubov quasiparticles} as  $
\psi^{\dagger}_n=\vec{\alpha}^{\dagger}\Sigma_z\ket{\psi_n}$ and $\psi_{n*}=\bra{\psi_{n*}}\Sigma_z\vec{\alpha}$, containing superpositions of $a_i$ and $a_i^{\dagger}$. The Bogoliubov modes fulfil \textit{pseudo-canonical} commutation relations
\begin{align}
	[\psi_n,\psi^{\dagger}_{m*}]=\delta_{nm},&&[\psi_n,\psi_m]=0,&&[\psi_{n*},\psi_{m*}]=0,
\end{align}
and allow the expansion of the effective Hamiltonian as $H_\mathrm{eff}=\frac{1}{2}\sum_{n=1}^{2N}\epsilon_n\psi^{\dagger}_n\psi_{n*}$.\footnote{Note however that for $\epsilon_n\in\mathbb{C}$, the operators $\psi^{\dagger}_n$ and $\psi_{n*}$ are not related by Hermitian conjugation.}

The eigenspectrum of $\hbdg$ relates with time-dynamics of physical quantities. Akin to Hermitian systems, we can project the time evolution of the operator $b(0)=\bra{v^0}\vec{\alpha}$ onto the eigenbasis of $\hbdg$ as $b(0)=\sum_n\braket{v^0|\psi_n}\bra{\psi_{n*}}\Sigma_z\vec{\alpha}$ and apply the non-Hermitian evolution operator $\mathcal{U}=e^{-i\hbdg t}$ to $b(0)$:
\begin{align}\label{eq:expansion_time}
	b(t)=\sum_{n=1}^{2N}e^{-i\epsilon_{n}t}\bra{v^{0}}\psi_{n}\rangle\langle\psi_{n*}|\Sigma_{z}\vec{\alpha},
	\intertext{or recognising the eigenoperators $\psi_n,\psi_{n*}$}
	b(t)=\sum_{n=1}^{2N}e^{-i\epsilon_{n}t}\bra{v^{0}}\psi_{n}\rangle\psi_{n*}.
\end{align}
The expansion~\autoref{eq:expansion_time} is similar to the result for unitary dynamics, except for the potentially complex phase evolution of each of the eigencomponents, which is given by the eigenvalues.

\subsection{Bound-state Hermitian and non-Hermitian Aharonov-Bohm effect}\label{sec:bound_state}
Here we provide further mathematical background for the comparison between Hermitian and non-Hermitian AB effects in Methods. We use as an example a single loop with $N$ nodes, which can be particle-like or include both particles and hole nodes. In a Hermitian chain ($\mathcal{B}=0$), with coupling amplitudes $J$ and periodic boundary conditions, the Hamiltonian matrix $H$ (see Methods) is diagonal in the Fourier basis
$a_{k}=\sum_{j=1}^{N}a_{j}e^{2\pi ijk/N}/\sqrt{N}$ with circular wavenumbers  $k$.\footnote{$k\in\left\{ -[N/2],\cdots,[N/2]\right\}$
	for $N$ odd, or $k\in\left\{ -[N/2],\cdots,[N/2]-1\right\}$ for $N$ even, where $[]$ denotes the integer part function.} Noting $\sum_{j=1}^{N}e^{2\pi ij(k-k')/N}=N\delta_{k,k'}$ and choosing a gauge where all Peierls phases are equally distributed, $\varphi_{ij}=\Phi/N$, the Hamiltonian of the ring is given by
\begin{align}\label{eq:H_ring}
	H_\mathrm{ring}=&J\sum_{j=1}^{N}a_{j}^{\dagger}a_{j+1}e^{i\Phi/N}+\mathrm{H.c.}.\nonumber\\
	\intertext{This is transformed to the Fourier basis as}
	H_\mathrm{ring}=&\frac{J}{N}\sum_{k,k'}a_{k}^{\dagger}a_{k'}e^{2\pi ij(k-k')/N}e^{2\pi i((k+\Phi/(2\pi))/N)}+\mathrm{H.c.}=\nonumber\\
	=&2J\sum_{k}\cos\left((2\pi k+\Phi)/N\right)a_{k}^{\dagger}a_{k}.
\end{align}
Aharonov-Bohm interference is manifest in the second line of~\autoref{eq:H_ring}, where the phases $\varphi_{ij}$ displace the wavenumber $k$, after being combined via $\sum_{k'}$. 

We seek a generalisation of this idea to loops that involve particles and holes.	In the BdG formalism, a Hermitian loop decomposes into a pair of particle-hole related disjoint loops (Methods). Equation~\ref{eq:HBDG} is thus Peierls-phase dependent -- from now on explicitly stated with a curly bracket notation -- through the Hamiltonian matrix $H(\{\varphi_{ij}\})=\mathrm{diag}(\mathcal{A}(\{\varphi_{ij}\}),\mathcal{A}^{*}(\{\varphi_{ij}\}))$. Fourier decomposition is equivalent to the block diagonal unitary transformation $\vec{\alpha}=\mathcal{U}_H\vec{\alpha}$ with $\mathcal{U}_H=\mathrm{diag}(U,U^{*})$, where $U_{kj}=e^{2\pi ijk/N}/\sqrt{N}$ preserves bosonic commutators ($[\mathcal{U}_{\mathrm{H}},\Sigma_{z}]=0$). The BdG matrix transforms as
\begin{align}
	\hbdg(k)=
	\Sigma_{z}\mathrm{diag}(U\mathcal{A}(\{\varphi_{ij}\})U^{\dagger},U^{\dagger}\mathcal{A}^{*}(\{\varphi_{ij}\})U),
\end{align}
with a diagonal matrix at the r.h.s., given $\mathcal{A}$ is circulant~\cite{gray2006toeplitz_}. Interference including nontrivial Peierls phases now
enters within each the blocks of $\hbdg(k)$.

For loops involving particles and holes, we define Fourier
modes $\mathcal{U}_{\mathrm{NH}}$, with $\vec{\alpha}=\mathcal{U}_\mathrm{NH}\vec{\alpha}$ and $(U_\mathrm{NH})_{kj}=e^{2\pi ijk/N}/\sqrt{N}$  that diagonalize the (Hermitian) Hamiltonian
matrix $H$:
\begin{equation}
	H(\{\varphi_{ij}\},\{\theta_{ij}\})\mapsto\mathcal{U}_{\mathrm{NH}}H(\{\varphi_{ij}\},\{\theta_{ij}\})\mathcal{U}_{\mathrm{NH}}^{\dagger}.
\end{equation}

Importantly,  $\mathcal{U}_{\mathrm{NH}}$ no longer respects bosonic commutation relations ($[\mathcal{U}_{\mathrm{NH}},\Sigma_{z}]\neq0$), and thus does not diagonalise $\hbdg$. Instead, $\hbdg\mapsto\hbdg'=\mathcal{U}_{\mathrm{NH}}\hbdg\mathcal{U}_{\mathrm{NH}}^{\dagger}$ with
\begin{equation}
	\hbdg'(\{\varphi_{ij}\},\{\theta_{ij}\})=\mathcal{V}\Lambda(\{\varphi_{ij}\},\{\theta_{ij}\}),
\end{equation}
where $\Lambda(\{\varphi_{ij}\},\{\theta_{ij}\})=\mathcal{U}_{\mathrm{NH}}H\mathcal{U}_{\mathrm{NH}}^{\dagger}$
is a real diagonal matrix by construction, which contains the eigenvalues of the analogous Hermitian loop, see~\autoref{eq:H_ring}. This matrix contains the outcome of interference
of Fourier waves with nontrivial Peierls phases. But on top of this
effect, the Peierls-phase-independent term $\mathcal{V}=\mathcal{U}_{\mathrm{NH}}\Sigma_{z}\mathcal{U}_{\mathrm{NH}}^{\dagger}$ is in general a non-Hermitian matrix -- being a product of non-commuting Hermitian matrices -- that couples Fourier states with different $k$. This non-Hermitian interaction of Fourier modes with nontrivial phases, on top of their interference, embodies the non-Hermitian AB effect. Note that if all holes were replaced by particles ($\Sigma_z\rightarrow\mathbb{1}$), then a trivial coupling matrix follows $\mathcal{V}\rightarrow\mathbb{1}$, given the fact that $\mathcal{U}_\mathrm{NH}$ is unitary.

\subsection{Eigenmodes and thermomechanical spectra}\label{sec:eig_therm}

In our experiment, we employ thermomechanical noise spectra to probe the effective phononic density of states. Here we mathematically justify how that statement is still valid in a system with parametric interactions within the stable regime (i.e. a steady-state is well defined). Extending the discussion in~\autoref{sec:BdGForm}, we show how the internal symmetries of $\hbdg$ imply that the noise spectra can be estimated by only determining half of the eigenmodes of $\hbdg$, despite the doubling of degrees of freedom. 

Thermomechanical spectra are calculated from the Heisenberg-Langevin equations for the system in the rotating frame (see Methods). Their solution in the frequency domain reads $\vec{\alpha}(\omega)=\chi_m(\omega)\vec{\alpha}_{\mathrm{in}}(\omega)$ with $\vec{\alpha}=(\vec{a},\vec{a}^{\dagger})$ and susceptibility matrix
\begin{equation}
	\chi_m(\omega)=
	\frac{i}{\omega\mathbb{1}-(\hbdg-i\frac{\Gamma}{2})} =\left(\begin{array}{cc}
		\chi_{m,\vec{a}\vec{a}}(\omega) & \chi_{m,\vec{a}\vec{a}^{\dagger}}(\omega)\\
		\chi_{m,\vec{a}^{\dagger}\vec{a}}(\omega) & \chi_{m,\vec{a}^{\dagger}\vec{a}^{\dagger}}(\omega)\end{array}\right),
\end{equation}
which we split into diagonal and off-diagonal response blocks. The steady-state fluctuation spectra read~\cite{Meystre2007_} 
\begin{equation}\label{eq:noise_sp}
	\mathcal{S}(\omega)=\langle\vec{\alpha}^{\dagger}(\omega)\vec{\alpha}(\omega)\rangle=\chi_m^{\dagger}(\omega)\mathcal{D}\chi_m(\omega),
\end{equation}	
with diffusion matrix $ 		\mathcal{D}=\mathrm{diag}(\gamma_1(\bar{n}_1+1)\cdots,\gamma_1\bar{n}_1\cdots)$.

We deduce further properties of the output via eigen-expansion of $\hbdg$. In particular, for $\mathcal{B}=0$, eigenmodes of the system do not mix $a_i$ and $a_i^{\dagger}$, implying vanishing off-diagonal blocks $\chi_{m,\vec{a}\vec{a}}(\omega),\chi_{m,\vec{a}^{\dagger}\vec{a}^{\dagger}}(\omega)= 0$, and
\begin{equation}\label{eq:chi_cons}
	\chi_{m,\vec{a}\vec{a}}(\omega)=i(\omega\mathbb{1}-(\mathcal{A}-i\Gamma/2))^{-1}.
\end{equation}
Here $\Gamma$ now denotes the un-duplicated loss matrix $\Gamma=\mathrm{diag}(\gamma_1,\cdots,\gamma_N)$.
Expanding~\autoref{eq:chi_cons} in terms of the eigenmodes of $\mathcal{A}-i\Gamma/2$ shows that the noise spectrum in~\autoref{eq:noise_sp} probes the density of the states, with poles at the eigenvalues of $\mathcal{A}-i\Gamma/2$ and eigenvectors providing the weights of each eigenmode, e.g. for the BST network ($k=\{-1,0,1\}$, $N=3$)
\begin{align}
	\chi_{m,a_ja_j}(\omega)=\frac{1}{N}\sum_{k}\frac{e^{2\pi ijk/3}}{(\frac{\gamma}{2}-i(\omega-\omega_{k}(\Phi)))}.
\end{align}
Here we defined the eigenfrequencies $\omega_{k}(\Phi)=2J\cos((2\pi k+\Phi)/3)$. Note that the rotating picture permits a treatment entirely analogous to time-independent systems. The inverse of the rotating frame transformation must be applied to relate the results above to the experimental outcome. To illustrate the result qualitatively, we consider the dynamics of BST for $\gamma=0$ in the lab frame. Expanding $\tilde{a}_j(t)$ in terms of the eigenmodes $a_k$,
\begin{equation}\label{eq:lab_frame}
	\tilde{a}_j(t)=\frac{1}{\sqrt{3}}\sum_{k=\{-1,0,1\}} e^{i2\pi jk/3}e^{-i(\omega_j+\omega_k(\Phi))t}a_k(0).
\end{equation}
According to to~\autoref{eq:lab_frame}, the noise spectra in the lab frame show sidebands at the natural resonator frequencies $\omega_j$, which are surrounded by side peaks with spectral weight corresponding to the rotating eigenmodes.

We now deduce general properties of the noise spectra in parametrically driven scenarios.
The eigenvalues of $\hbdg$ are real in Hermitian systems ($\mathcal{B}=0$), but also in generalised $\mathcal{PT}$-symmetric regimes of non-Hermitian systems ($\mathcal{B}\neq0$)~\cite{Flynn2020_}. Off-diagonal contributions $\chi_{m,\vec{a}\vec{a}},\chi_{m,\vec{a}^{\dagger}\vec{a}^{\dagger}}$ are in this case nonzero, with a redundant information content due to particle-hole symmetry: $\chi_{m,\vec{a}^{\dagger}\vec{a}^{\dagger}}(\omega)=-\chi^{*}_{m,\vec{a}\vec{a}}(-\omega)$. When eigenvalues $\epsilon_n\in\mathbb{R}$, they can be divided into two groups $\{\epsilon_n,-\epsilon_n\}$ with corresponding eigenvectors $\ket{\psi_n}$ (particle-like\footnote{If $\ket{\psi^\mathcal{L}}$ is eigenvector of $\hbdg$, then $\Sigma_z\ket{\psi_n^\mathcal{L}}=\ket{\psi_n}$.}) and $\ket{\tilde{\psi}_n}=\mathcal{C}\ket{\psi_n}$ (hole-like\footnote{If $\ket{\psi_n^\mathcal{L}}$ is eigenvector of $\hbdg$, then  $\Sigma_z\mathcal{C}\ket{\psi^\mathcal{L}n}=-\mathcal{C}\ket{\psi_n}$.}).  These states have positive and negative $\Sigma_z$-norms respectively and are orthogonal, i.e. $
\braket{\psi_n|\Sigma_z|\psi_m}=\delta_{nm}$, $\braket{\tilde{\psi}_n|\Sigma_z|\tilde{\psi}_m}=-\delta_{nm}$, and $\braket{\psi_n|\Sigma_z|\tilde{\psi}_m}=0$ \cite{blaizot1986quantum_}. Inserting the identity $\mathbb{1}=\sum_n(|\psi_{n}\rangle\langle\psi_{n}|-|\tilde{\psi}_{n}\rangle\langle\tilde{\psi}_{n}|)\Sigma_{z}$, the susceptibility matrix mimics this splitting
\begin{equation}\label{eq:susc_expand}
	\chi_m(\omega)=i\sum_{n=1}^{N}\frac{|\psi_{n}\rangle\langle\psi_{n}|\Sigma_{z}}{\omega-\epsilon_{n}}-\frac{|\tilde{\psi}_{n}\rangle\langle\tilde{\psi}_{n}|\Sigma_{z}}{\omega+\epsilon_{n}}.
\end{equation}
From~\autoref{eq:susc_expand} we deduce that the resonator susceptibility at positive (negative) frequency sidebands involves only particle-like (hole-like) eigenstates. This splitting also extends to the case of of disjoint loops encompassing a subset of particles and holes each (e.g. SCT, Methods), where $\ket{\psi_n},\ket{\tilde{\psi}_n}$ would denote eigenstates hosted by each loop. 

The noise spectrum of each physical resonator is given by $\mathcal{S}_{ii}(\omega)=\bra{e_i}\mathcal{S}\ket{e_i}$ for $i\in(1,\cdots,N)$. This involves the calculation of the projection $\bra{e_i}\chi_m^{\dagger}(\omega)$ where $\ket{e_i}=(0,\cdots,1,\cdots,0)$ is only non-zero at the $i$th position. This state can only overlap with states from a given loop (e.g. either a particle or a hole state), hence either  $\braket{e_i|\Sigma_z|\psi_n}\neq0$ and $\braket{e_i|\Sigma_z|\tilde{\psi}_n}=0$ or vice-versa.
The particle-hole redundancy of the BdG description thus implies that only one of the terms in the r.h.s. of~\autoref{eq:susc_expand} will contribute to the projection $\chi_m(\omega)\ket{e_i}$, implying only half of the eigenstates are required in the computation and have physical content. Following this logic, the noise spectrum for resonator $i$ reads

\begin{subequations}
	\begin{align}
		\mathcal{S}_{ii}(\omega)=&
		&\sum_{k=1}^{2N}\mathcal{D}_{kk}\sum_{n=1}^{N}\frac{|\langle e_{i}|\psi_{n}\rangle|^{2}|\langle\psi_{n}|\Sigma_{z}|e_{k}\rangle|^{2}}{|\omega-\epsilon_{n}|^{2}}.
	\end{align}	
	If $\psi_n$ contains $a_i$ (or, equivalently $\ket{\psi_n}\propto\ket{e_i}$) but $\tilde{\psi}_n$ does not contain $a_i$, the spectrum will display positive frequency sidebands of the mechanical resonance 	located at  $+\epsilon_{n}$ with $n\in 1\cdots N$ (within the current rotating frame mechanical frequencies are shifted to $\omega=0$)
	Conversely,  for modes $j\neq i$ where $\psi_n\propto a^{\dagger}_j$ ($\ket{\tilde{\psi}_n}\propto\ket{e_j}$), the eigenmodes that play a role are the $\ket{\tilde{\psi}_n}$ instead, associated with negative frequency sidebands at $-\epsilon_{n}$ in $\mathcal{S}_{jj}(\omega)$:
	\begin{equation}
		\mathcal{S}_{jj}(\omega)=\sum_{k=1}^{N}\mathcal{D}_{kk}\sum_{n=1}^{N}\frac{|\langle e_{j}|\tilde{\psi}_{n}\rangle|^{2}|\langle\tilde{\psi}_{n}|\Sigma_{z}|e_{k}\rangle|^{2}}{|\omega+\epsilon_{n}|^{2}}.
	\end{equation}	
\end{subequations}

\subsection{Phase-space representation}\label{sec:phase_sp_rep}
The BdG formalism in particle hole-space is equivalent to a description in terms of quadratures. The latter is helpful in interpreting the main features of flux-tunable quadrature squeezing in the main text. Here we discuss the representation of nanomechanical steadystates as distributions in phase space. Regrouping quadratures into a vector $\vec{R}=(X_1,X_2,\cdots,Y_1,Y_2,\cdots)^T$, the second moments 
$\mathcal{O}=\braket{\vec{R}\vec{R}^T}$ then obey~\cite{Meystre2007_} 
\begin{equation}\label{eq:2nd_mom}
	\dot{\mathcal{O}}=i\left(\hs^{XY}\mathcal{O}+\mathcal{O}(\hs^{XY})^{T}\right)+2\mathcal{D}^{R}.
\end{equation}
where $\hs^{XY}$ is given in Methods, Eq. 14. Note that first moments evolve according to $\langle \dot{\vec{R}}\rangle=-i\hs^{XY}\langle \vec{R}\rangle+\langle\vec{R}_\text{in}\rangle$.
The diffusion matrix $\mathcal{D}^R$ encodes the Markovian correlations $\braket{R^{i}_{\mathrm{in}}(t),R^{j}_{\mathrm{in}}(t')}=\mathcal{D}^R_{ij}\delta(t-t')$, where
\begin{align}
	\braket{X_\mathrm{in}^{i}(t),X_\mathrm{in}^{j}(t')}=&\braket{Y_\mathrm{in}^{i}(t),Y_\mathrm{in}^{i}(t')}=(\bar{n}_i+\frac{1}{2})\delta(t-t'),\nonumber\\
	\braket{X_\mathrm{in}^{i}(t),Y_\mathrm{in}^{j}(t')}=&-\braket{Y_\mathrm{in}^{i}(t),X_\mathrm{in}^{i}(t')}=\frac{i}{2}\delta(t-t').
\end{align}

The thermal steadystates then follow as Gaussian Wigner function of the eigenvalues of $\vec{R}$~\cite{Weedbrook2012_}, denoted $\vec{r}=(x_1,x_2,\cdots,y_1,y_2,\cdots)^T$ with $\braket{R_i}=0$, namely
\begin{equation}
	W(\vec{r})=\frac{1}{(2\pi)^N\sqrt{\det \sigma}}\exp(-\frac{1}{2}\vec{r}^T\sigma^{-1}\vec{r}),
\end{equation}
with a symmetric covariance matrix 
\begin{equation}\label{eq:cov_matrix}
	\sigma_{ij}=\frac{1}{2}(\frac{1}{2}(\mathcal{O}_{ij}+\mathcal{O}_{ji}) - \braket{R_i}\braket{R_j}),
\end{equation}
whose eigenvectors indicate axes along which (anti)squeezing occurs with magnitude given by the corresponding eigenvalues. Note that, in the absence of coherent driving, $\braket{R_i}=0$. The marginal distributions for resonator $k$
\begin{equation}
	W_{k}(x_k,y_k)=\int\prod_{i\neq k}  \mathrm{d}x_i\mathrm{d}y_iW(x_1,y_1,\cdots,x_N,y_N),
\end{equation}
are Gaussian distributions that	show thermomechanical squeezing, visualized using the standard deviation ellipse defined by its covariance matrix. This is for example shown in \autoref{fig:2} of the main text, with explicit calculations in~\autoref{sec:flux_squeezing}.

\begin{widetext}
	\section{Supporting results}
	\subsection{Flux-tunable thermomechanical squeezing in the SD}\label{sec:flux_squeezing}
	Here we show some analytical results for the steadystate properties of the SD, employing the toolbox of~\autoref{sec:phase_sp_rep}. These results assist the interpretation of the results in \autoref{fig:2}b and \autoref{fig:2}c from the main text.
	
	For an ideal SD with $\gamma_i=\gamma$ ($i\in\{1,2\}$) and equal thermal occupation (e.g. resonant modes) $\bar{n}_i=\bar{n}$, the covariance matrix of the system can be calculated analytically from the solution $\dot{\mathcal{O}}=0$ in~\autoref{eq:2nd_mom} and~\autoref{eq:cov_matrix}. We note that due to thermo-optically induced backaction (Methods), the effective resonator bath occupations $\bar{n}_1 \approx \bar{n}_2$ only differ by a few percent for the SD experiments that we show.
	
	The first limit of interest is $\Phi=\pi$, where the covariance matrix becomes diagonal and independent of $J$. In this case, quadratures $X_i$ see their variances decreased with increasing $\eta$ (squeezing) while variances for $Y_i$ are increased (anti-squeezing):
	\begin{align}
		\sigma(\Phi=\pi)= \left(
		\begin{array}{cccc}
			\frac{\gamma +2 \gamma  \bar{n}}{2 \gamma +4 \eta } & 0 & 0 & 0 \\
			0 & \frac{\gamma +2 \gamma  \bar{n}}{2 \gamma +4 \eta } & 0 & 0 \\
			0 & 0 & \frac{\gamma +2 \gamma  \bar{n}}{2 \gamma -4 \eta } & 0 \\
			0 & 0 & 0 & \frac{\gamma +2 \gamma  \bar{n}}{2 \gamma -4 \eta } \\
		\end{array}
		\right).
	\end{align}
	
	The covariance matrix for $\Phi=0$ reads
	\begin{align}
		\sigma(\Phi=0)= \frac{(\bar{n}+\frac{1}{2})}{\gamma^{2}-4\eta^{2}+4J^{2}}\left(\begin{array}{cccc}
			\gamma(\gamma-2\eta)+4J^{2} & 0 & 0 & J\eta\\
			0 & \gamma(\gamma-2\eta)+4J^{2} & J\eta & 0\\
			0 & J\eta & \gamma(\gamma+2\eta)+4J^{2} & 0\\
			J\eta & 0 & 0 & \gamma(\gamma+2\eta)+4J^{2}
		\end{array}\right).
	\end{align}
	In this case, the cross correlations (indicated by off-diagonal elements) still suggest the existence of a basis of hybrid quadratures where squeezing can be found. This result can be referenced to the covariance matrix in the standard two-mode squeezing case, with Hamiltonian $H_\mathrm{TMS}=i\eta a_1^{\dagger}a_2^{\dagger}+\mathrm{H.c.}$. This Hamiltonian produces anti-squeezing in the variables $X_1+X_2$ and $Y_1-Y_2$ and squeezing in $X_1-X_2$ and $Y_1+Y_2$, with no (single mode) squeezing on $X_i$ or $Y_i$~\cite{Bachor2019_}.  
	
	\begin{figure*}
		\centering
		\includegraphics[width=0.75\linewidth]{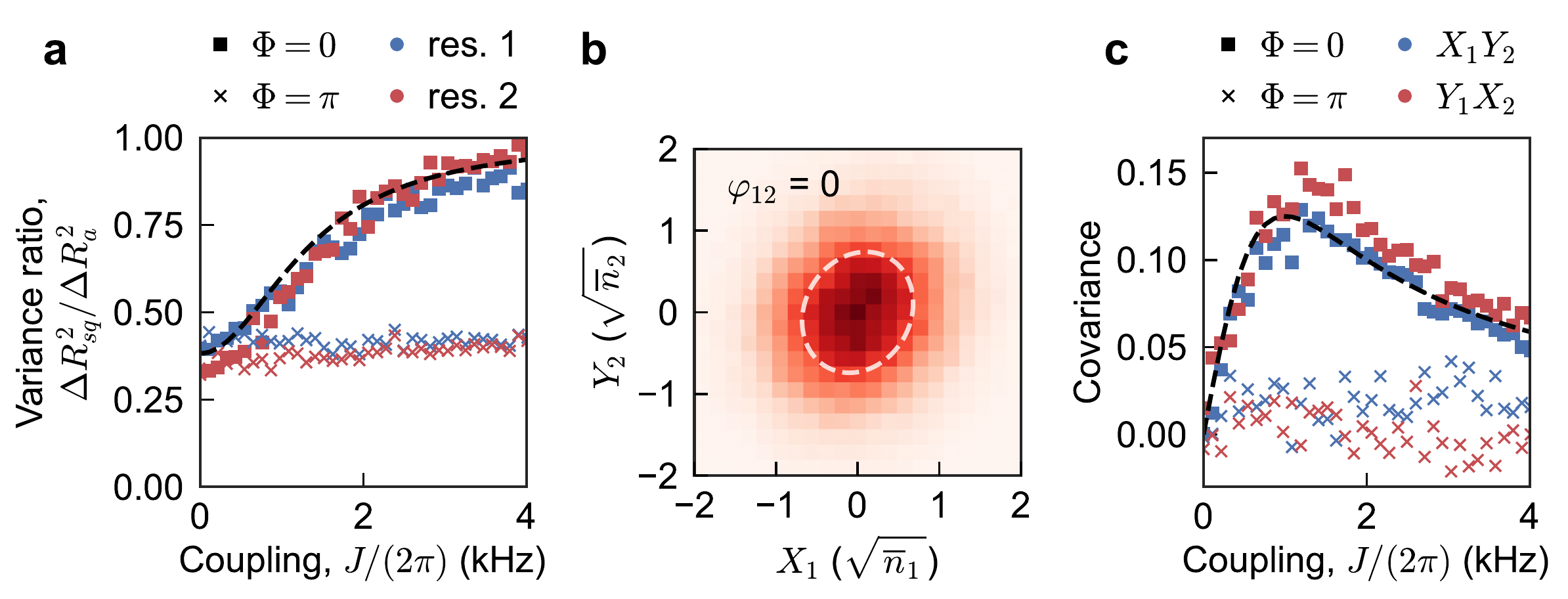}
		\caption{\textbf{Tunable single-mode and effective two-mode squeezing in the squeezing dimer.} \textbf{a} Intra-resonator squeezing as a function of the beam-splitter coupling $J$. Two values $\Phi = 0, \pi$ of the flux are shown for equal single-mode squeezing strengths $\eta_1 = \eta_2 = 0.5$ kHz. The level of single-mode squeezing is expressed by the ratio of the smallest ($\Delta R^2_\text{sq}$) and largest ($\Delta R^2_\text{a}$) eigenvalues of the covariance matrix of the quadrature amplitudes recorded for each resonator. These eigenvalues indicate the amplitude variance along the squeezed and antisqueezed principal quadrature components, respectively. For $\Phi = \pi$, where the squeezed (antisqueezed) quadratures $X_i$ ($Y_i$) of both resonators are coupled (cf. \autoref{fig:2}d), the slight initial imbalance in variance ratio is reduced as $J$ increases while the value of the variance ratio remains low. In contrast, for $\Phi = 0$ -- when the squeezed quadrature $X_i$ in one resonator is coupled to the antisqueezed quadrature $Y_j$ in the other -- we observe cancellation of single-mode squeezing as the variance ratio tends to $1$ with increasing $J$. This agrees well with theory (dashed line), where for simplicity we have assumed equal dissipation rates $\overline{\gamma} = 2.2$ kHz equal to the average of the experimental losses $\gamma_i = \{2.6, 1.9\}$ kHz, as well as equal bath occupations. Due to dynamical (optothermal) backaction, for this particular experiment the effective bath occupations $n_1 \approx n_2$ only differed by a few percent. \textbf{b} Two-mode squeezing observed in the cross-resonator amplitude distribution of quadratures $X_1$ and $Y_2$ for $\Phi = 0$, $J=3.5$ kHz and $\eta_1 = \eta_2 = 0.5$ kHz. The dashed ellipse depicts the standard deviation of the principal components of the quadrature covariance matrix and shows positive correlations between $X_1$ and $Y_2$ (covariance $\cov(X_1,Y_2) = 0.08$). \textbf{c} Covariance of the coupled quadrature pairs $X_1Y_2$ and $Y_1X_2$ as a function of $J$, with $\eta_1 = \eta_2 = 0.5$ kHz. No correlations are found for flux $\Phi = \pi$, when single-mode squeezing is strongest and independent of $J$ (cf. panel a). However, for $\Phi = 0$, positive correlations $\cov(X_1,Y_2), \cov(Y_1,X_2) > 0$ are found when $J$ is increased, as predicted in theory (dashed line). A trade-off between the squeezing axes rotation towards the standard two-mode squeezing limit and the decrease in the overall squeezing level as $J$ is increased leads to a maximum covariance (although not optimal squeezing level for the rotated quadratures) at a coupling $J_\text{opt}$. For the simple theory model with equal dissipation and bath occupation that we use it is given by $J_\text{opt}^2 =  (\gamma^2 - 4 \eta^2) / 4$.\label{fig:TMS_eff}}
	\end{figure*}

	To establish a link with this two-mode squeezing case, we diagonalise $\sigma(\Phi=0)$ to reveal the \emph{rotation} of the principal (squeezing) axes of the covariance matrix. Defining the hybrid quadratures ($\xi=2J/\gamma$)
	\begin{align}
		\left(\begin{array}{c}
			R_{\mathrm{sq.}}^{(1)}\\
			R_{\mathrm{sq.}}^{(2)}\\
			R_{\mathrm{a.}}^{(1)}\\
			R_{\mathrm{a.}}^{(2)}
		\end{array}\right)=\left(
		\begin{array}{cccc}
			-\frac{\sqrt{\frac{1}{\sqrt{\xi ^2+1}}+1}}{\sqrt{2}} & 0 & 0 & \frac{\xi }{\sqrt{2} \sqrt{\xi ^2+\sqrt{\xi ^2+1}+1}} \\
			0 & -\frac{\sqrt{\frac{1}{\sqrt{\xi ^2+1}}+1}}{\sqrt{2}} & \frac{\xi }{\sqrt{2} \sqrt{\xi ^2+\sqrt{\xi ^2+1}+1}} & 0 \\
			\frac{\xi }{\sqrt{2} \sqrt{\xi ^2+\sqrt{\xi ^2+1}+1}} & 0 & 0 & \frac{\xi }{\sqrt{2} \sqrt{\xi ^2-\sqrt{\xi ^2+1}+1}} \\
			0 & \frac{\xi }{\sqrt{2} \sqrt{\xi ^2+\sqrt{\xi ^2+1}+1}} & \frac{\xi }{\sqrt{2} \sqrt{\xi ^2-\sqrt{\xi ^2+1}+1}} & 0 \\
		\end{array}
		\right)\left(\begin{array}{c}
			X_{1}\\
			X_{2}\\
			Y_{1}\\
			Y_{2}
		\end{array}\right),
	\end{align}
	we observe $R_{\mathrm{sq.}}^{(i)}$ are squeezed whereas $R_{\mathrm{a.}}^{(i)}$ are anti-squeezed. The corresponding variances, with $\sigma(\Phi=0)=(\Delta R_\mathrm{sq.}^2,\Delta R_\mathrm{sq.}^2,\Delta R_\mathrm{a.}^2,\Delta R_\mathrm{a.}^2)$, read
	\begin{align}\label{eq:rotated_variances}
		\Delta R_\mathrm{sq.}^2=\frac{\gamma  (2\bar{n}+1) \sqrt{\xi ^2+1}}{2 \gamma  \sqrt{\xi ^2+1}+4 \eta },&&\Delta R_\mathrm{a.}^2=\frac{\gamma  (2\bar{n}+1) \sqrt{\xi ^2+1}}{2 \gamma  \sqrt{\xi ^2+1}-4 \eta }.
	\end{align}
	In the strong coupling limit $\xi\gg 1$, the principal axes rotate to the  
	antisymmetric quadratures $(X_1-Y_2)/\sqrt{2}$ and $(X_2-Y_1)/\sqrt{2}$ (squeezed), besides the symmetric superpositions  $(X_1+Y_2)/\sqrt{2}$ and  $(X_2+Y_2)/\sqrt{2}$ (anti-squeezed).  This rotation can be mapped into the standard case of two mode squeezing $H_\mathrm{TMS}$ after considering the real rotation $Y_2\rightarrow X_2$, $X_2\rightarrow -Y_2$. Note, however, that~\autoref{eq:rotated_variances} indicate the level of squeezing vanishes in this limit, since $\Delta R_\mathrm{sq.}^2\simeq\Delta R_\mathrm{a.}^2$. For flux $\Phi=0$, there is always an inevitable trade-off between principal axes rotation and the level of cross correlations. In 
	~\autoref{fig:TMS_eff} we illustrate this behaviour by tracking the value of the cross correlation elements $\braket{X_1Y_2}$ and $\braket{X_2Y_1}$.

	The change in the level of single-mode squeezing as synthetic flux is varied is embodied by the ratio of the variances of the squeezed and antisqueezed quadratures, obtained from the eigenvalues of 
	\begin{subequations}
		\begin{align}
			\sigma(\Phi)=& \frac{\gamma(2\bar{n}+1)}{\left(\gamma^{4}-4\gamma^{2}\eta^{2}+4J^{2}\left(\gamma^{2}-2\eta^{2}\right)+8\eta^{2}J^{2}\cos(\Phi)\right)}\left(\begin{array}{cccc}
				\frac{w_{+}}{4} & 0 & \eta J^{2}\sin(\Phi) & \gamma\eta J\cos\left(\frac{\Phi}{2}\right)\\
				0 & \frac{w_{+}}{4} & \gamma\eta J\cos\left(\frac{\Phi}{2}\right) & -\eta J^{2}\sin(\Phi)\\
				\eta J^{2}\sin(\Phi) & \gamma\eta J\cos\left(\frac{\Phi}{2}\right) & \frac{w_{-}}{4} & 0\\
				\gamma\eta J\cos\left(\frac{\Phi}{2}\right) & -\eta\sin(\Phi) & 0 & \frac{w_{-}}{4}
			\end{array}\right),\\
			w_\pm(\Phi)=&\gamma^{2}(\gamma-2\eta)+4J^{2}(\gamma-\eta)+4\eta J^{2}\cos(\Phi).
		\end{align}
	\end{subequations}
	
	We obtain the thermal-occupation-independent result,
	\begin{align}~\label{eq:var_ratio}
		\frac{\Delta R_\mathrm{sq.}^2}{\Delta R_\mathrm{a.}^2}(\Phi)= \frac{\gamma^{3}-2\eta\sqrt{\left(\gamma^{2}+4J^{2}\right)\left(\gamma^{2}-2J^{2}\cos(\Phi)+2J^{2}\right)}+4\gamma J^{2}}{\gamma^{3}+2\eta\sqrt{\left(\gamma^{2}+4J^{2}\right)\left(\gamma^{2}-2J^{2}\cos(\Phi)+2J^{2}\right)}+4\gamma J^{2}},
	\end{align}
	displayed in main text \autoref{fig:2}f in comparison with the experimental data.
	
	The variance ratios~\autoref{eq:var_ratio} are maximal (closest to $1$) at $\Phi=0$ and minimal (i.e. largest difference in variance) at $\Phi=\pi$. In the limit $J\gg\eta$, the reference value for this ratio reads $\frac{\Delta R_\mathrm{sq.}^2}{\Delta R_\mathrm{a.}^2}(0)=1-\frac{2 \eta }{J}$ and can be made arbitrarily close to 1 by increasing the ratio $J/\eta$, while the value at $\Phi=\pi$ is $J$-independent: $\frac{\Delta R_\mathrm{sq.}^2}{\Delta R_\mathrm{a.}^2}(\pi)=(\gamma -2 \eta)/(\gamma +2 \eta )$. 
\end{widetext}

\begin{figure*}
	\subfloat[Real parts of the eigenvalues]{%
		\includegraphics[width=0.8\linewidth]{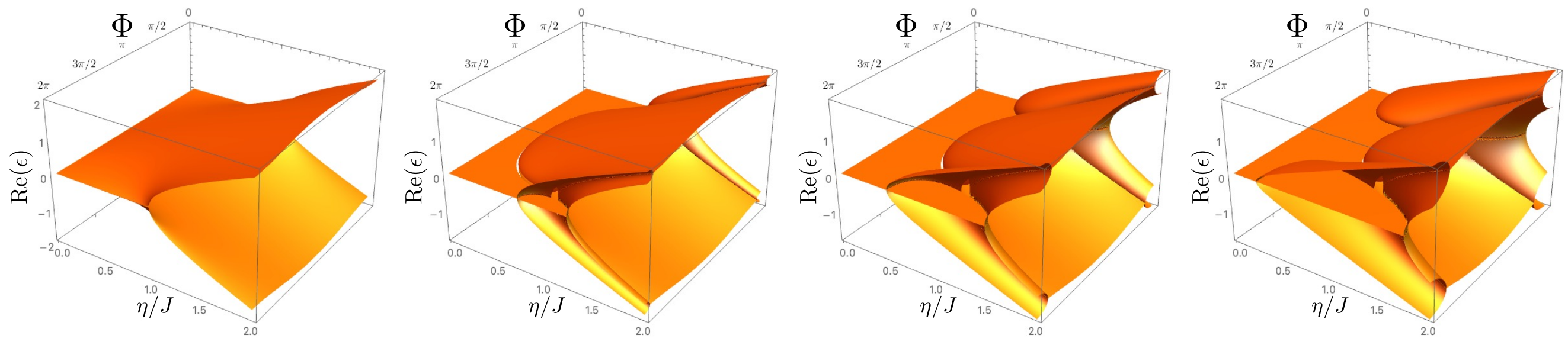}%
	}\\
	\subfloat[Imaginary parts of the eigenvalues]{%
		\includegraphics[width=0.8\linewidth]{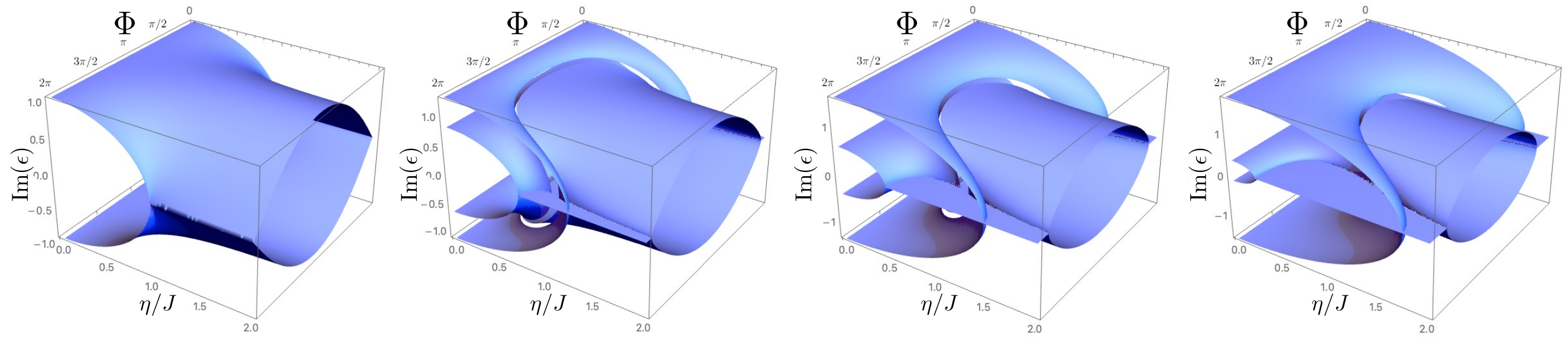}%
	}\\
	\subfloat[Phase diagram and cuts of real surfaces]{%
		\includegraphics[width=0.8\linewidth]{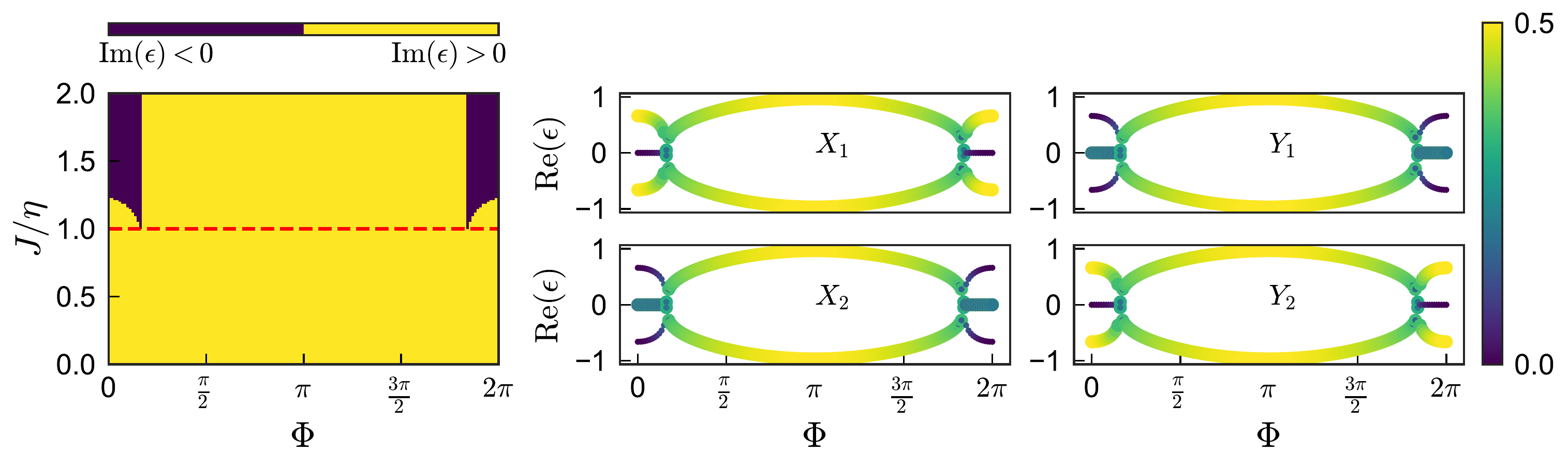}%
	}\\
	\subfloat[Phase diagram and cuts of imaginary surfaces]{%
		\includegraphics[width=0.8\linewidth]{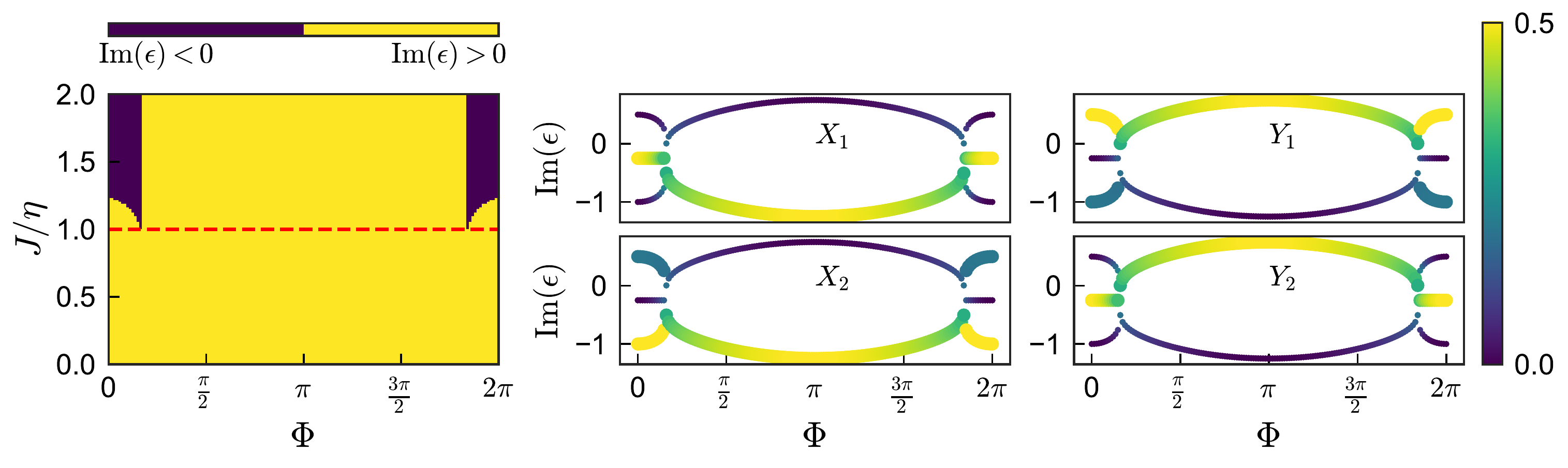}%
	}
	\caption{ \textbf{Asymmetric losses in the SD.} \textbf{a} Real and \textbf{b} imaginary complex surfaces in $\Phi-\eta/J$ space for $\eta=1$ kHz, as a function of increasing loss asymmetry (steps of $\Delta\gamma=0.5$ kHz from left to right, starting at the symmetric case). Imaginary parts show deviations with respect to the average loss rate $\bar{\gamma}$, here corresponding to $\Im(\epsilon)=0$. \textbf{c} (left) Linear stability phase diagram for the imaginary part of eigenenergies for a value of loss asymmetry $\Delta\gamma/\eta=1$. \textbf{c},\textbf{d} (right) Cuts of the real (imaginary) complex surfaces along the red-dashed trajectory in the phase diagram ($J/\eta=1$) show degeneracies associated with a $2^{\mathrm{nd}}$ order EC~\autoref{eq:eigv_asymmetric_loss}. The weights of the local quadratures $X_i,Y_i$ are shown in the colorscale. We employ different marker sizes to distinguish degenerate eigenfrequencies.} \label{fig:SD_delta_gamma}
\end{figure*}

\subsection{$\Phi$-tunable complex spectra of the SD: $a_i,a_i^{\dagger}$ basis}\label{sec:SD_as_AB}

Diagonalization of the (BdG) dynamical matrix for the SD can also be carried out in a particle-hole basis, where the relationship with the AB effect is more transparent. Here we assume a gauge where $\varphi_{12}=\Phi/2$ and $\theta_i=0$, and assume zero loss $\gamma=0$. Hence
\begin{align}
	\mathcal{A}=\left(
	\begin{array}{cc}
		0 & J e^{-i \Phi/2} \\
		J e^{i \Phi/2 } & 0 \\
	\end{array}
	\right),&& \mathcal{B}=\eta\mathbb{1}.
\end{align}  
The eigenvalues read $\{\epsilon_1,\epsilon_2,\epsilon_3,\epsilon_4\}=\{\epsilon,-\epsilon,-\epsilon^{*},\epsilon^{*}\}$ with
\begin{subequations}
	\begin{align}\label{eq:eigen_SD}
		\epsilon=&-\sqrt{J^{2}-\eta^{2}+2i\eta J\sin(\frac{\Phi}{2})}=	-i\sqrt{q_\Phi q_{2\pi-\Phi} },\\
		q_\Phi =&\left(\eta-Je^{i \Phi/2 }\right).
	\end{align}
\end{subequations}
Noting $\epsilon_1=\epsilon_2^*$ and $\epsilon_4=-\epsilon_1=\epsilon_2^*$ in~\autoref{eq:eigen_SD}, a complete basis with a $\Sigma_z$-norm~\cite{Flynn2020_} is formed by the rows of
\begin{subequations}\label{eq:eigenv_flux}
	\begin{align}
		T_\Phi=\left(\begin{array}{cccc}
			1 & -\frac{\sqrt{q_{2\pi-\Phi}^{*}}}{\sqrt{-q_{\Phi}^{*}}} & \frac{\sqrt{q_{2\pi-\Phi}^{*}}}{\sqrt{-q_{\Phi}^{*}}} & 1\\
			1 & \frac{\sqrt{q_{2\pi-\Phi}^{*}}}{\sqrt{-\text{\ensuremath{q_{\Phi}}}^{*}}} & -\frac{\sqrt{q_{2\pi-\Phi}^{*}}}{\sqrt{-q_{\Phi}^{*}}} & 1\\
			-1 & -\frac{\sqrt{q_{\Phi}}}{\sqrt{-q_{2\pi-\Phi}}} & -\frac{\sqrt{q_{\Phi}}}{\sqrt{-q_{2\pi-\Phi}}} & 1\\
			-1 & \frac{\sqrt{q_{\Phi}}}{\sqrt{-q_{2\pi-\Phi}}} & \frac{\text{\ensuremath{\sqrt{q_{\Phi}}}}}{\sqrt{-q_{2\pi-\Phi}}} & 1
		\end{array}\right),
	\end{align}
\end{subequations}
after $\Sigma_z$-normalisation.

The mathematical form of these eigenvectors embodies the interference between possible paths going from the resonator 1 to 2, while the nontrivial relative phases between $a_i,a_i^{\dagger}$ components demonstrate the breaking of $\mathcal{T}$ as a consequence of a synthetic flux in particle-hole space. Eigenvectors show a modulation in their resonator weights caused by the flux through the quantities $q_\Phi$ and $q_{2\pi-\Phi}$. These play the role of amplitudes of clockwise and counterclockwise processes. Accordingly, gain/attenuation for these modes, given by the imaginary parts of $\epsilon_n$ is tuned. 

To illustrate the effect of synthetic fluxes in the eigenstates, we consider first the case with $\Phi=0$. Here the eigenvalues, namely $\epsilon_{1,3}=\mp\sqrt{J^2-\eta ^2}$ and $\epsilon_{2,4}=\epsilon_{1,3}$, are real (i.e. no gain) and non-degenerate for $J>\eta$ and become unstable above the EP at $J=\eta$, where the real parts collapse to zero and the imaginary parts split ($J<\eta$). The change of behaviour in the eigenstates matches the breaking of  $\mathcal{P}_{X_iY_j}\mathcal{T}$ symmetry, detailed in the main text and Methods. Within the $\mathcal{P}_{X_iY_j}\mathcal{T}$ symmetric phase, 
\begin{align}\label{eq:eig_SD_simple}
	T_{\Phi=0}^{\eta< J}=\frac{1}{\sqrt{2}}\left(
	\begin{array}{cccc}
		-\cosh (r) & 1 & 0 & \sinh (r) \\
		1 & -\cosh (r) & \sinh (r) & 0 \\
		-\cosh (r) & -1 & 0 & \sinh (r) \\
		-1 & -\cosh (r) & \sinh (r) & 0 \\
	\end{array}
	\right),
\end{align}
with an effective two-mode squeezing parameter $r = \mathrm{arctanh}(\eta/J)$~\footnote{For $\eta\rightarrow0$  the eigenmodes for the coupled dimer with no parametric drive are recovered, i.e. $\psi_{1}-\psi_{3}=a_{1}-a_{2}$, $   \psi_{2}-\psi_{4}=a_{1}+a_{2}$.}. From~\autoref{eq:eig_SD_simple}, we observe that the eigenmodes only contain trivial phase differences between $a_i$ and $a_i^{\dagger}$ ($n\pi,n\in\mathbb{Z}$) if $J>\eta$ and correspond to hybrid quadratures. Crucially, in the opposite case above threshold $\eta> J$, relative phases become $\pm\pi/2$ and localisation into particle-hole combinations corresponding to the local quadratures $X_i,Y_i$ occurs: 
\begin{equation}
	T_{\Phi=0}^{\eta> J}=\left(\begin{array}{cccc}
		J/\eta & -i\sqrt{\eta^{2}-J^{2}}/\eta & 0 & 1\\
		-i\sqrt{\eta^{2}-J^{2}}/\eta & J/\eta & 1 & 0\\
		J/\eta & i\sqrt{\eta^{2}-J^{2}}/\eta & 0 & 1\\
		i\sqrt{\eta^{2}-J^{2}}/\eta & J/\eta & 1 & 0
	\end{array}\right).
\end{equation}

The coalescence of the eigenspectra of $\hbdg$ (EPs) can be assessed from the condition number $\mathrm{cond}(V^{-1})$ for the inverse eigenvector matrix $T$, which acquires larger values when $\hbdg$ is close to non-diagonalisability~\cite{moiseyev2011non_}. 

Our treatment relies on the found connections with non-Hermitian $\mathcal{PT}$-symmetric systems. Within this framework, nontrivial fluxes $\Phi\neq\{0,\pi\}$ are directly linked with an \textit{explicit} breaking of the symmetry and the removal of EPs. Note, however, that this effect can alternatively be regarded as a dynamical phase transition. Indeed, 
dynamical phase transitions in bosonic systems can occur in the absence of EPs, in events known as Krein collisions~\cite{Flynn2020_}. In these cases, degenerate real eigenvalues values split into non-real ones without loss of diagonalisability ---precisely as in the case of the eigenspectrum departing from trivial fluxes. Within this generalised notion of dynamical phase transitions, phase boundaries can be detected by suitably defining a phase-rigidity, which tracks the overlap of bi-orthogonal partners.

\subsection{Dynamical phases in a non-ideal SD: exceptional contours}\label{sec:ECS}

In our experiments on the SD, damping rates and parametric amplitudes typically present asymmetries, with $|\gamma_1 - \gamma_2|/\gamma_1 \approx \pm 0.3-\pm0.5$ and $|\eta_1-\eta_2|/\eta_1\approx \pm 0.1$. In this section we show how such asymmetries affect the occurrence and location of EPs.

The EPs still correspond to a dynamical phase transition, where now modified parity-time symmetries of the dynamical matrix break spontaneously, even though $\mathcal{M}$ is no longer invariant under $\mathcal{P}_{X_iY_j}\mathcal{T}$. These modified symmetries, remarkably, are now present for arbitrary fluxes. This fact, in particular, allows a rich pattern of intersections between complex surfaces for nontrivial $\Phi$, that correspond to boundaries of regions with broken symmetries. Namely, spontaneous-symmetry breaking in such generalised scenarios describe \textit{second order exceptional contours} (EC) at nontrivial fluxes $\Phi=\Phi_\mathrm{EP}$. As shown below, ECs tune due to the combined action of the AB effect over squeezing, beamsplitter and dissipative links.

With asymmetric loss rates ($\Delta\gamma = (\gamma_2-\gamma_1)/2\neq0$), the dynamical matrix for the SD becomes  quasi-$\mathcal{P}_{\bar{X}_i\bar{Y}_j}\mathcal{T}$-symmetric~\cite{Ornigotti2014_,li2020_} in the dynamically-offset quadrature basis $X_i\mapsto \bar{X}_i=X_ie^{-\bar{\gamma}t}$, $Y_i\mapsto \bar{Y}_i=Y_ie^{-\bar{\gamma}t}$ with average loss $\bar{\gamma}=(\gamma_1+\gamma_2)/2$. Note that in this case the open-system dynamical matrix (denoted, in the quadrature basis, as $\mathcal{M}_\text{SD}^{XY}$ and defined as $\mathcal{M}_\text{SD}^{XY}=\mathcal{H}_\text{SD}^{\text{XY}}-i\Gamma/2$) is no longer related with the BdG dynamical matrix by a rigid shift the imaginary parts. The determination of dynamical phases needs to be then formulated in terms of $\mathcal{M}_\text{SD}^{XY}$ instead of $\mathcal{H}_\text{SD}^{XY}$. In a gauge $\theta_i=\pi/2$ employed in all the SD calculations and the offset quadrature basis, the open-system dynamical matrix $\mathcal{M}_\text{SD}^{XY}$ reads
\begin{equation}\label{eq:dyn_delta_gamma}
	\hspace{-2mm}\bar{\mathcal{M}}_\text{SD}^{XY}=\left(\begin{array}{cccc}
		\frac{\Delta\gamma}{2}-\eta & J_{\parallel} & 0 & J_{\perp}\\
		-J_{\parallel} & -(\frac{\Delta\gamma}{2}+\eta) & J_{\perp} & 0\\
		0 & -J_{\perp} & \eta+\frac{\Delta\gamma}{2} & J_{\parallel}\\
		-J_{\perp} & 0 & -J_{\parallel} & \eta-\frac{\Delta\gamma}{2}
	\end{array}\right).\hspace{-2mm}
\end{equation}
$\mathcal{P}_{\bar{X}_i\bar{Y}_j}\mathcal{T}$ symmetry exists for nonzero fluxes. In particular, at $\Phi=\pi$,~\autoref{eq:dyn_delta_gamma} contains two blocks for the uncoupled dimers $X_1X_2$ and $Y_1Y_2$
\begin{align}
	\bar{\mathcal{M}}_{X_{1}X_{2}}=&\left(\begin{array}{cc}
		\frac{\Delta\gamma}{2}-\eta & J\\
		-J & -\left(\eta+\frac{\Delta\gamma}{2}\right)
	\end{array}\right),\nonumber\\
	\bar{\mathcal{M}}_{Y_{1}Y_{2}}=&\left(\begin{array}{cc}
		\eta+\frac{\Delta\gamma}{2} & J\\
		-J & \eta-\frac{\Delta\gamma}{2}
	\end{array}\right).
\end{align}
For arbitrary flux, the parameter dependency of the eigenfrequencies present a square root behaviour
\begin{subequations}
	\begin{align}
		\epsilon_{1}=&-\frac{\bar{\gamma}}{2}+\frac{1}{2}\sqrt{p_{\Phi}},\hspace{-2mm}&&\epsilon_{2}=-\frac{\bar{\gamma}}{2}-\frac{1}{2}\sqrt{p_{\Phi}},\nonumber\\
		\epsilon_{3}=&-\frac{\bar{\gamma}}{2}+\frac{1}{2}\sqrt{p_{\Phi}},&&\epsilon_{4}=-\frac{\bar{\gamma}}{2}-\frac{1}{2}\sqrt{p_{\Phi}},
	\end{align}
	with a factor
	\begin{equation}
		p_{\Phi}=\Delta\gamma^{2}+4\left(\eta^{2}-J^{2}\right)-i8\eta J\sqrt{\sin^{2}\frac{\Phi}{2}-\left(\frac{\Delta\gamma}{2J}\right)^{2}}.
	\end{equation}
\end{subequations}
Eigenfrequencies and eigenvectors of $\bar{\mathcal{M}}^{XY}_\text{SD}$, displayed in~\autoref{fig:SD_delta_gamma} and~\autoref{fig:SD_delta_gamma}c,d,  illustrate these ECs along $p_\Phi=0$  spanned along the flux dimension. These degeneracies coincide with the coalescence of eigenvectors, as an independent test of diagonalizability of the dynamical matrix is carried out. Two families of branch cuts exist: If the innermost square root vanishes ($|\sin(\frac{\Phi}{2})|=\Delta\gamma/(2J)$), then $\epsilon_1=\epsilon_3$ and $\epsilon_{2}=\epsilon_3$. These curves, independent of the parameter $\eta$, correspond to the AB tuning of the linewidth that would arise in a beam-splitter-coupled loop with asymmetric loss rates. If instead ($|\sin(\frac{\Phi}{2})|<\Delta\gamma/(2J)$) the condition for a 4-fold degeneracy (a pair of $2^{\text{nd}}$ order EPs) is
\begin{subequations}
	\begin{equation}\label{eq:eigv_asymmetric_loss}
		\sin\frac{\Phi}{2}=\sqrt{\left(\frac{\Delta\gamma}{2J}\right)^{2}-\left(\frac{4\left(J^{2}-\eta^{2}\right)\pm\Delta\gamma^{2}}{8\eta J}\right)^{2}}.
	\end{equation}
	Here the non-Hermitian AB effect modulates gain, connecting the $2^{\mathrm{nd}}$ order EPs for the $X_1-Y_2$ and $X_2-Y_1$ dimers ($J=\eta\pm\Delta\gamma/2$).
	The condition~\autoref{eq:eigv_asymmetric_loss} is only physical if $|\sin\left(\frac{\Phi}{2}\right)|\leq1$, i.e. if
	\begin{align}
		\begin{cases}
			\eta-\frac{\Delta\gamma}{2}\leq J\leq\frac{\Delta\gamma}{2}+\eta, & 0<\frac{\Delta\gamma}{2}<\eta\\
			\eta-\frac{\Delta\gamma}{2}<J\leq\frac{\Delta\gamma}{2}+\eta, & \frac{\Delta\gamma}{2}=\eta\\
			\frac{\Delta\gamma}{2}-\eta\leq J<\sqrt{\left(\frac{\Delta\gamma}{2}\right)^{2}-\eta^{2}}, & \frac{\Delta\gamma}{2}>\eta\\
			\sqrt{\left(\frac{\Delta\gamma}{2}\right)^{2}-\eta^{2}}<J\leq\frac{\Delta\gamma}{2}+\eta, & \frac{\Delta\gamma}{2}>\eta
		\end{cases}.
	\end{align}
\end{subequations}

\begin{figure*}
	\subfloat[Real parts of the eigenvalues]{%
		\includegraphics[width=0.8\linewidth]{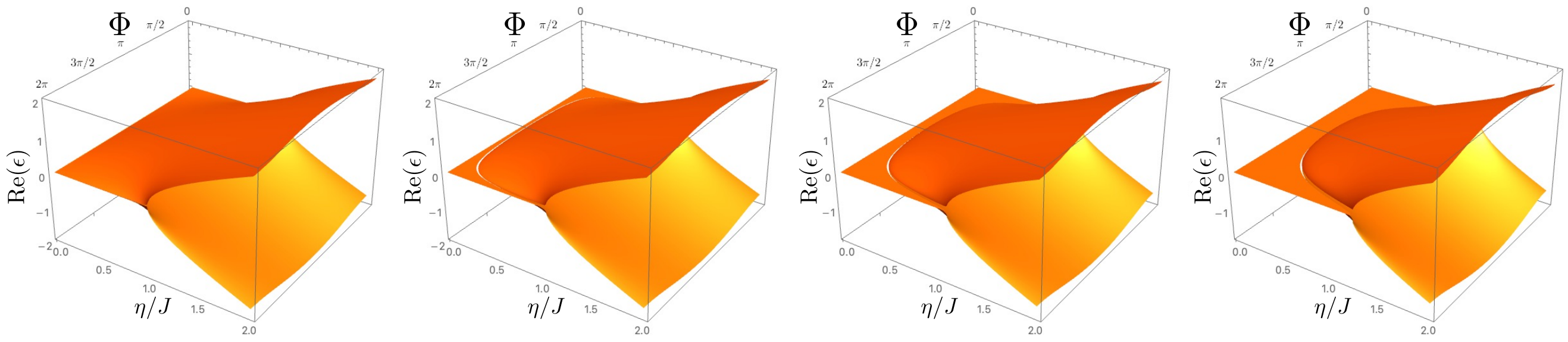}%
	}\\
	\subfloat[Imaginary parts of the eigenvalues]{%
		\includegraphics[width=0.8\linewidth]{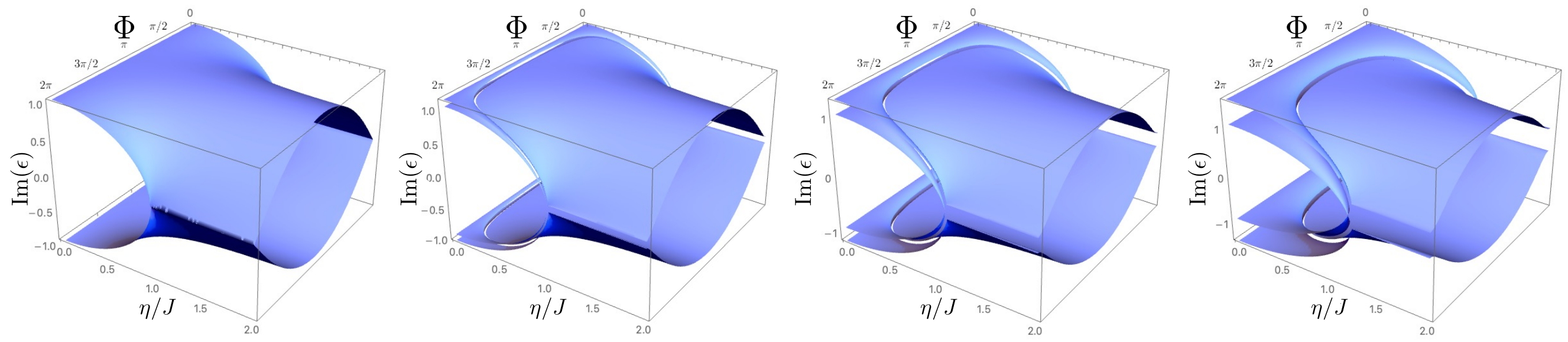}%
	}\\
	\subfloat[Phase diagram and cuts of real surfaces]{%
		\includegraphics[width=0.8\linewidth]{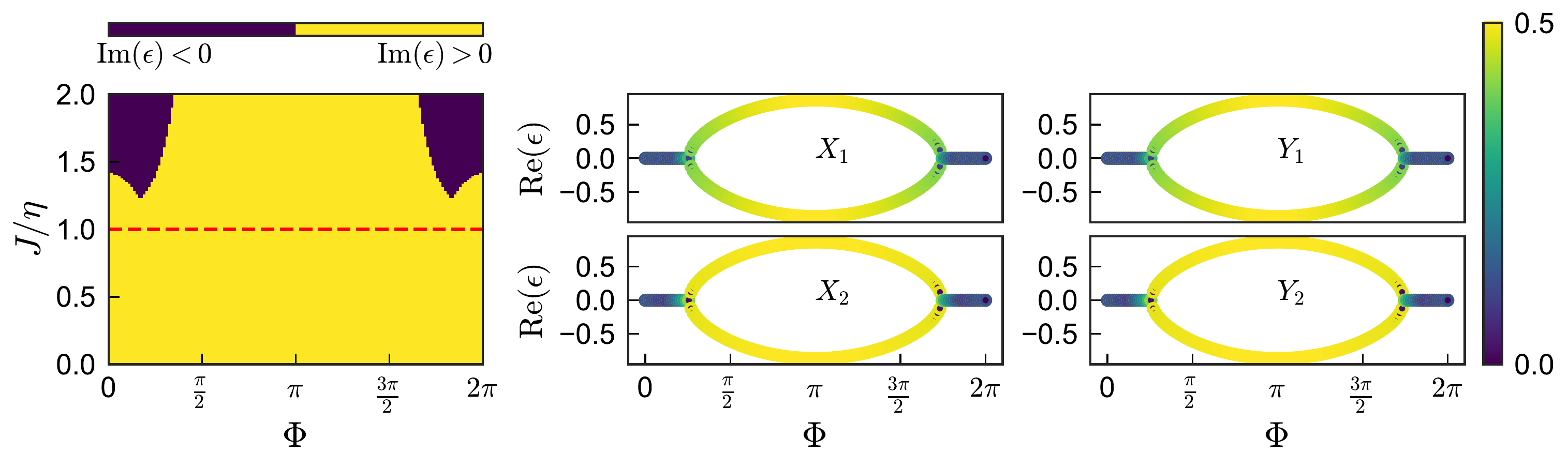}%
	}\\
	\subfloat[Phase diagram and cuts of imaginary surfaces]{%
		\includegraphics[width=0.8\linewidth]{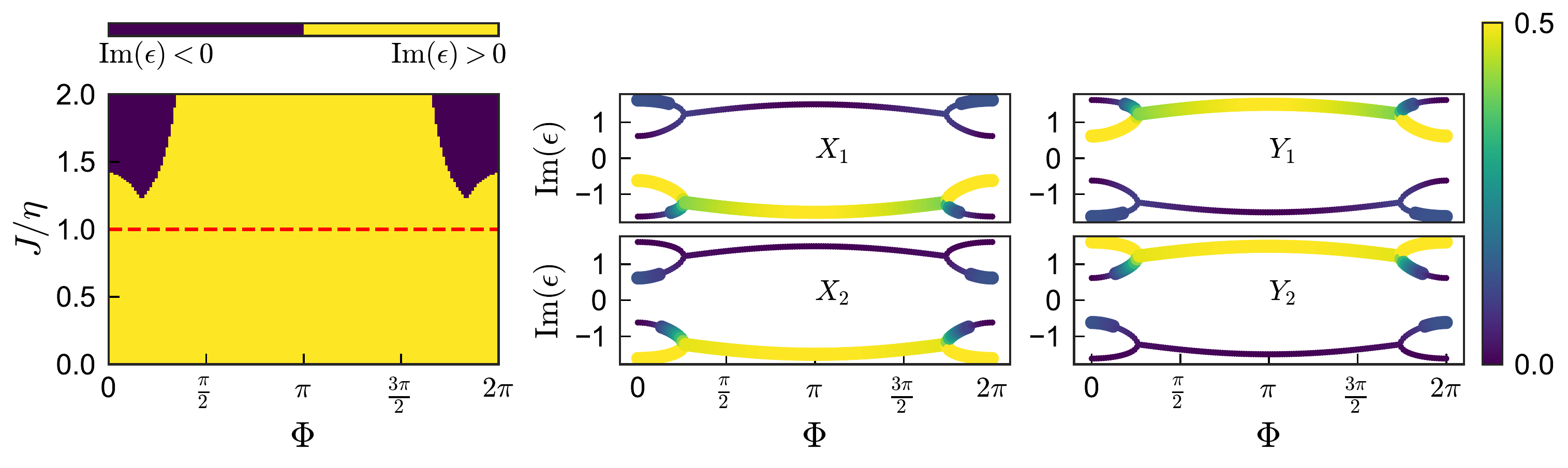}%
	}
	\caption{ \textbf{Asymmetric parametric driving amplitudes in the SD.} \textbf{a} Real and \textbf{b} imaginary complex surfaces in $\Phi-\eta/J$ space space for $\eta=1$ kHz, as a function of increasing loss asymmetry (values of $\Delta\eta=\{0,0.1,0.25,0.5\}$ kHz from left to right, starting at the symmetric case). Imaginary parts show deviations with respect to the equal loss rate $\gamma$. \textbf{c} (left) Linear stability phase diagram for the imaginary part of eigenenergies, for a value of parametric drive asymmetry $\Delta\eta/\eta=1$ and damping rate $\gamma_i=2$ kHz. A $2^{\mathrm{nd}}$ order EC is still appreciated in the eigenvalues, according to the roots of~\autoref{eq:eigv_asymmetric_eta}. \textbf{c,d} (right) Cuts of the real complex surfaces along the red-dashed trajectory in the phase diagram ($J/\eta=1$). The weights of the local quadratures $X_i,Y_i$ are shown in the colorscale. Different degenerate curves are represented with various marker sizes.} \label{fig:SD_delta_eta}
\end{figure*}

The eigenmodes present a similar topology to the case of asymmetric damping when parametric drives become asymmetric, due to the possibility of recovering effective $\mathcal{PT}$ symmetry by similar imaginary displacements of the modes. In this case, defining  $\bar{\eta}=(\eta_{1}+\eta_{2})/2$ and $\Delta\eta=(\eta_{2}-\eta_{1})/2$ we arrive to the eigenfrequencies
\begin{subequations}\label{eq:eigv_asymmetric_eta}
	\begin{align}
		\epsilon_{1}=-\frac{1}{2}\sqrt{y_{\Phi}^{-}}, &&	
		\epsilon_{2}=-\epsilon_{1}, &&
		\epsilon_{3}=-\frac{1}{2}\sqrt{y_{\Phi}^{+}},&&	
		\epsilon_{4}=-\epsilon_{3},
	\end{align}
	where now 
	\begin{align}
		y_{\Phi}^{\pm}=&4J^{2}-\Delta\eta^{2}-\bar{\eta}^{2}\pm\nonumber\\
		&2\sqrt{\Delta\bar{\eta}^{2}\bar{\eta}^{2}+2J^{2}\left(\bar{\eta}^{2}-\Delta\bar{\eta}^{2}\right)\cos(\Phi)-2J^{2}\left(\Delta\bar{\eta}^{2}+\bar{\eta}^{2}\right)}.
	\end{align}
\end{subequations}
The families of EP fall similarly along the zeros of the function $y_{\Phi}$, and appear as coalescences of eigenvalues and eigenvectors in~\autoref{fig:SD_delta_eta} and~\autoref{fig:SD_delta_eta}c,d. 

\begin{figure*}
	\subfloat[Phase diagram and cuts of real surfaces]{%
		\includegraphics[width=0.8\linewidth]{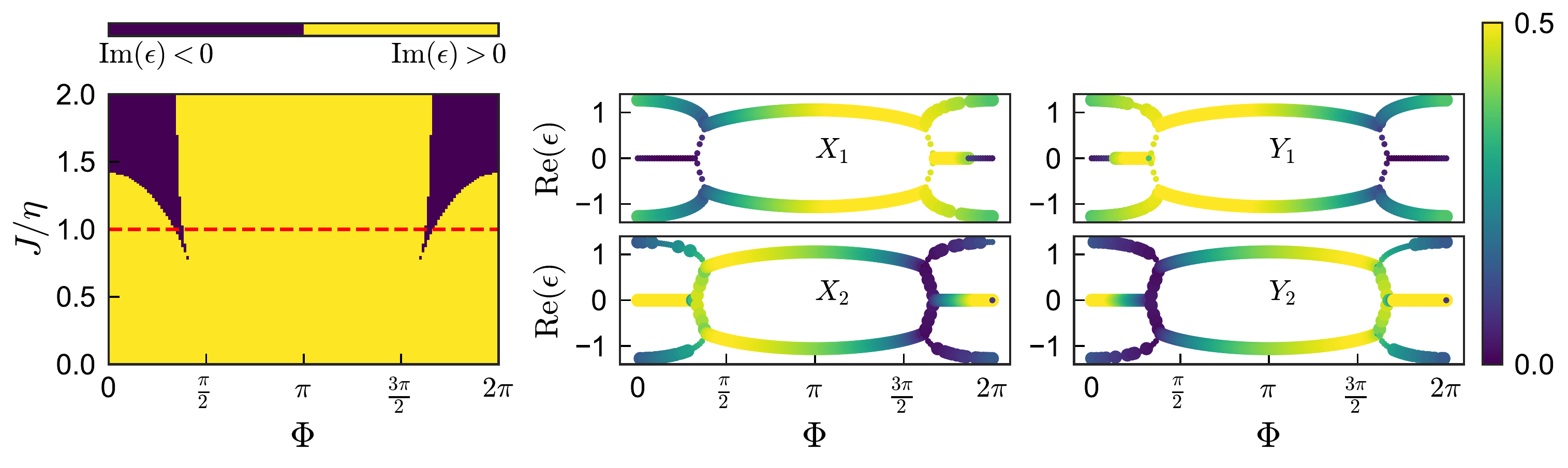}%
	}\\
	\subfloat[Phase diagram and cuts of real surfaces]{%
		\includegraphics[width=0.8\linewidth]{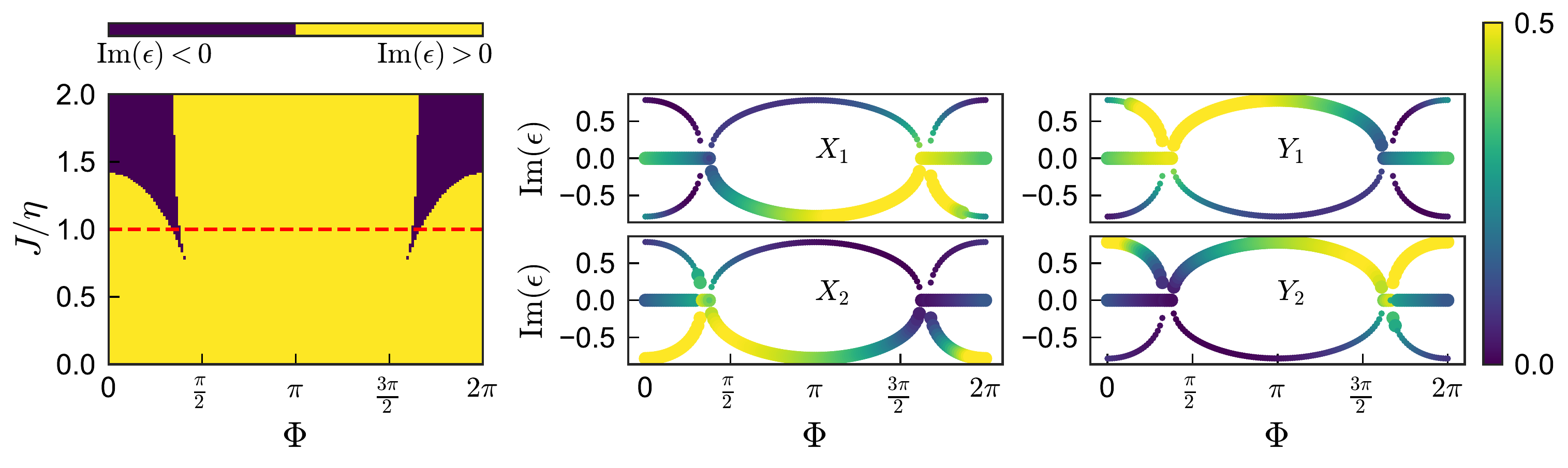}%
	}\\
	
	\caption{ \textbf{Effect of detuning from control fields in SD:} (left) Linear stability phase diagram for the imaginary part of eigenenergies, showing the stability to instability boundary in $J/\eta-\Phi$ space for a detuning $\delta=1$ kHz. (right) Cuts of the real complex surfaces along the red-dashed trajectory in the phase diagram ($\eta=1$ kHz) show degeneracies associated with a $2^{\mathrm{nd}}$ order EC (see~\autoref{eq:EC_delta}). The weights of the local quadratures $X_i,Y_i$ are shown in the colorscale. To represent degenerate curves curves, various marker sizes have been employed.} \label{fig:SD_detuning}
\end{figure*}

Finally, the last source of non-ideal behaviour we consider are finite detunings $\delta_i$ of the modulation frequencies from $\omega_i$. These induce the prefactors $e^{\pm i \delta_i}$ in the effective rotating-frame Hamiltonian (see Methods). These are removable by time-dependent gauge transformations $a_i\mapsto U_{\delta_i} a_i U_{\delta_i}^\dagger= a_ie^{\mp i\delta_i t}-\delta_i a_i^{\dagger}a_i$. These make explicit synthetic \textit{electric fields}~\cite{Lee2020_}, that produce
Stark shifts $-\delta_i a_i^{\dagger}a_i$. Such detuning on the control fields from either resonator produces a similar behaviour to other asymmetries, with exceptional contours spawned in parameter spaces involving $\Phi$. Treating for completeness the case $\delta_1\neq\delta_2$,
\begin{subequations}
	\begin{align}\label{eq:EC_delta}
		\epsilon_{1}=&-\frac{\sqrt{s-\sqrt{s'_{\Phi}}}}{\sqrt{2}},&&	
		\epsilon_{2}=-\epsilon_{1}, \\	
		\epsilon_{3}=&-\frac{\sqrt{s+\sqrt{s'_{\Phi}}}}{\sqrt{2}}, &&	
		\epsilon_{4}=-\epsilon_{3},
	\end{align}
\end{subequations}

where $s=\delta_{1}^{2}+\delta_{2}^{2}+2J^{2}-2\eta^{2}$ and
$s'_{\Phi}=\left(\delta_{1}+\delta_{2}\right){}^{2}\left(\left(\delta_{1}-\delta_{2}\right){}^{2}+4J^{2}\right)+8\eta^{2}J^{2}(\cos(\Phi)-1)$. The resulting complex surfaces with similar square-root topology are displayed in~\autoref{fig:SD_detuning}.

\subsection{Loop eigenmodes in the SCT network}\label{sec:loop_eig}
Here we proceed with the analytical diagonalisation of SCT. We discuss the arising spectral singularities --exceptional regions-- linked to the breaking of a suitably defined $\mathcal{P}_{gl}\mathcal{T}$ symmetry (see main text and Methods). Calculations are notably simplified after noting SCT features two disjoint loops (fluxes overall flux $\Phi=\varphi_{12}+\theta_{23}-\theta_{13}$ and $-\Phi$), rendering the matrix 	$\mathcal{H}_\mathrm{SCT}$  block diagonal. We choose a gauge where $\Phi=\varphi_{12}$ and $\theta_{23}=\theta_{13}=0$ and block-diagonalise $\mathcal{H}_\mathrm{SCT}$ by swapping $a_3\leftrightarrow a_3^{\dagger}$. This is implemented by the permutation matrix $G$:  $G\mathcal{H}_\mathrm{SCT}G=\mathrm{diag}(\mathcal{L},-\mathcal{L}^*)$, where 
\begin{align}\label{eq:dyn_disjoint}
	G=\left(
	\begin{array}{cccccc}
		1 & 0 & 0 & 0 & 0 & 0 \\
		0 & 1 & 0 & 0 & 0 & 0 \\
		0 & 0 & 0 & 0 & 0 & 1 \\
		0 & 0 & 0 & 1 & 0 & 0 \\
		0 & 0 & 0 & 0 & 1 & 0 \\
		0 & 0 & 1 & 0 & 0 & 0 \\
	\end{array}
	\right),&&\mathcal{L}=\left(
	\begin{array}{ccc}
		0 & J e^{-i \Phi } & \eta  \\
		J e^{i \Phi } & 0 & \eta  \\
		-\eta  & -\eta  & 0 \\
	\end{array}
	\right).
\end{align}
Given such sub-lattice symmetry in particle-hole space, the non-Hermitian dynamics of the SCT can be integrated by diagonalising $\mathcal{M}_{\mathcal{L}}=\mathcal{L}-i\frac{\Gamma_\mathcal{L}}{2}$ ($\Gamma_\mathcal{L}=\mathrm{diag}(\gamma_1,\gamma_2,\gamma_3)$) only, with eigenvalues $\epsilon_{n}$ and eigenvectors $\ket{\phi_n}$ related to the eigenvectors of the full open-system dynamical matrix $\mathcal{M}=\hbdg-i\Gamma/2$ by $\ket{\psi_n}=(\ket{\phi_n},0_3)^T$. The remaining half of $\mathcal{H}_\mathrm{SCT}$'s eigenmodes follows from $\mathcal{C}\ket{\psi_n}=(0_3,\ket{\tilde{\phi}_n})^T$ (eigenvalues $-\epsilon_{n}^*$), and will be orthogonal to $\ket{\psi_n}$. Similarly, eigenmodes only contain operators from within each of the loops.
\begin{align}
	\ensuremath{\hat{\psi}_{n}^{\dagger}=&\vec{\alpha}^{\dagger}\Sigma_{z}\ket{\psi_{n}}}=\vec{\alpha}_\mathcal{L}^{\dagger}\Sigma_{z}\ket{\phi_{n}},\nonumber\\
	\hat{\psi}_{n*}=&\bra{\psi_{n*}}\Sigma_{z}\vec{\alpha}=\bra{\phi_{n*}}\Sigma_{z}\vec{\alpha}_\mathcal{L}.
\end{align}	

\begin{figure*}
	\includegraphics[width=1\linewidth]{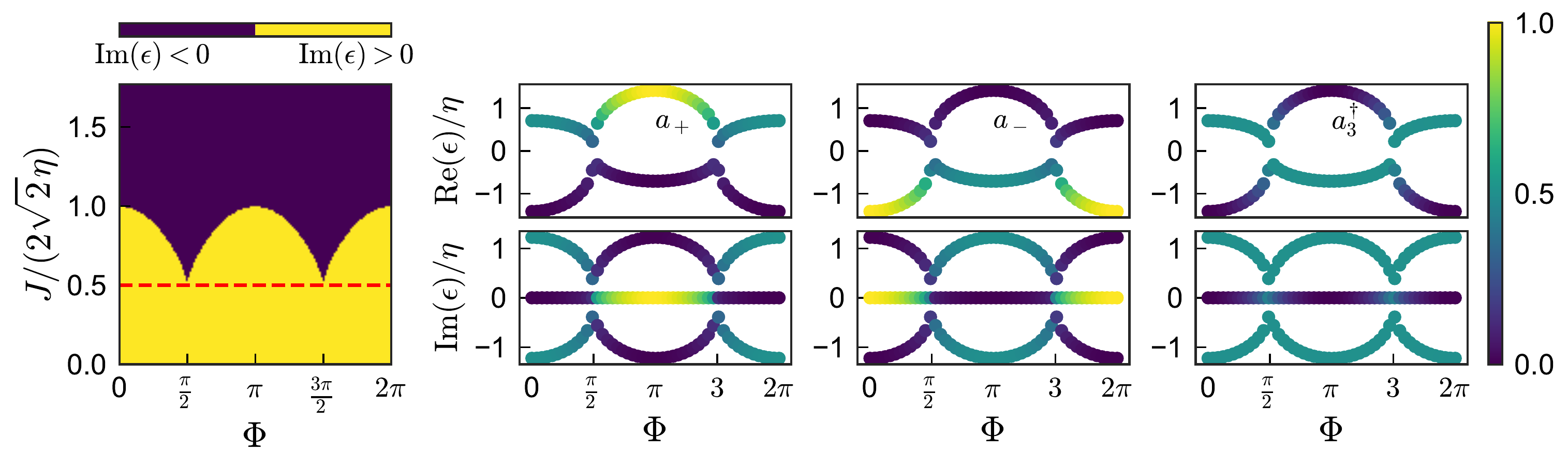}
	\includegraphics[width=1\linewidth]{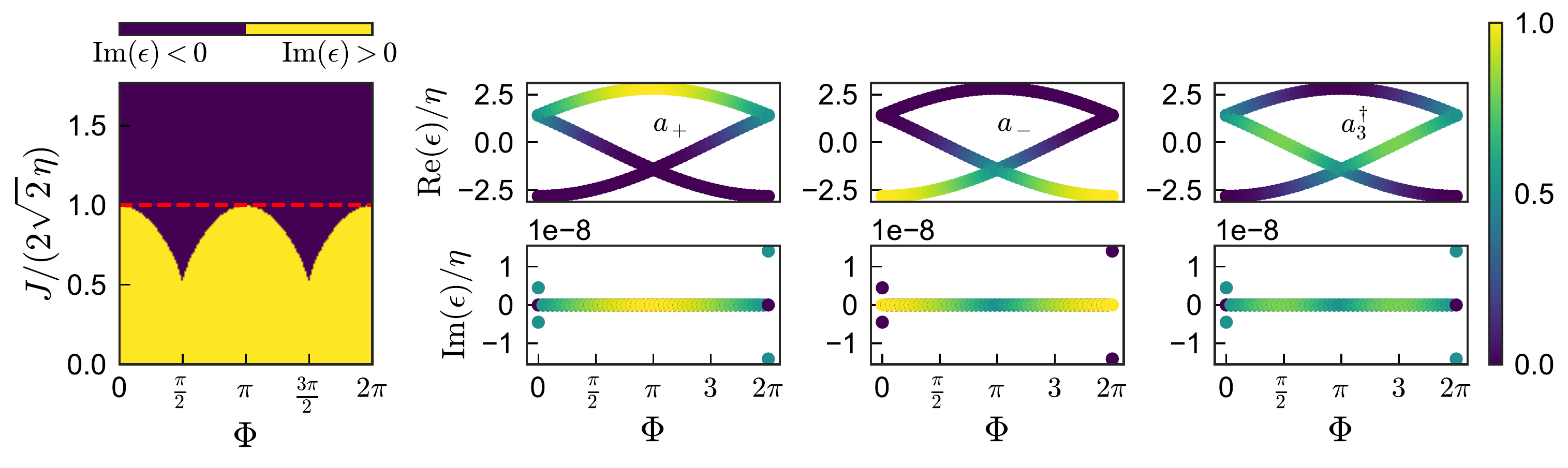}
	\caption{\textbf{ Eigenstates of the loop $a_1,a_2,a_3^{\dagger}$ in SCT:} (left) Linear stability phase diagram for the imaginary part of eigenenergies for $\gamma_i=0$, showing the stability to instability boundary in $\xi-\Phi$ space, where $\xi$ is the ratio ratio $\xi=J/(2\sqrt{2}\eta)$. Such boundary occurs in this case exactly along a $2^{\mathrm{nd}}$ order EC. (right, top) Cuts of the real complex surfaces along $\xi=1/2$, shown as a dashed trajectory in the phase diagram. The weights of the $\eta=0$ eigenstates, namely $a_+,a_-$ and $a_3^{\dagger}$ are shown in the colorscale. Similar data along the cut $\xi=1$ (dashed line in left, bottom plot), are shown in the right, bottom panels. Real and imaginary parts are re-scaled by $\eta$. \label{fig:SCT_phi_sweep}}
\end{figure*}	

For $\gamma=0$, the eigenvalues of the dynamical matrix for a given loop $\mathcal{L}$ (see~\autoref{eq:dyn_disjoint}) read, defined $\xi=J/\left(2\sqrt{2}\ensuremath{\eta}\right)$	 
\begin{subequations}\label{eq:evals_SCT}
	\begin{align}
		\epsilon_{1}=&-\frac{2\sqrt[3]{2}\left(\xi^{2}-1\right)}{\sqrt[3]{w_\Phi}}-\frac{\sqrt[3]{w_\Phi}}{3\sqrt[3]{2}},\\			\epsilon_{2}=&\frac{\sqrt[3]{2}\left(1+i\sqrt{3}\right)\left(\xi^{2}-1\right)}{\sqrt[3]{w_\Phi}}+\frac{\left(1-i\sqrt{3}\right)\sqrt[3]{w_\Phi}}{6\sqrt[3]{2}},\\
		\epsilon_{3}=&\epsilon_{2}^*,\\
		w_\Phi=&\sqrt{5832\xi^{2}\cos^{2}(\Phi)-864\left(\xi^{2}-1\right)^{3}}+54\sqrt{2}\xi\cos(\Phi).
	\end{align}
\end{subequations}
The corresponding eigenvectors can be similarly expressed as radical functions of $w_\Phi$, indicating a complex pole structure akin to the eigenvalues, in particular with a branch cut in the negative real axis for $\Phi=0$, departing from the roots of $w_\Phi=0$. Eigenvalues/eigenvectors of $\mathcal{L}$ display in this case a $2^{\mathrm{nd}}$ order EP at $\xi=1$. Namely $\epsilon_{2,3}=\frac{1}{2} \left(J\mp\sqrt{J^2-8 \eta ^2}\right)$, while simply $\epsilon_1=-J$, and 
\begin{equation}
	T_\Phi=\left(
	\begin{array}{ccc}
		-1 & 1 & 0 \\
		\frac{\sqrt{J^2-8 \eta ^2}-J}{4 \eta } & \frac{\sqrt{J^2-8 \eta ^2}-J}{4 \eta } & 1 \\
		-\frac{\sqrt{J^2-8 \eta ^2}+J}{4 \eta } & -\frac{\sqrt{J^2-8 \eta ^2}+J}{4 \eta } & 1 \\
	\end{array}
	\right).
\end{equation}

Finally, it is interesting to note that, for $\xi=1$, the eigenfrequencies of each of the disjoint loops of the SCT and the BST are identical in magnitude (see lower panels in~\autoref{fig:SCT_phi_sweep}). This suggests a duality between the two systems in this limit~\cite{flynn2020restoring_}.

Alternatively to the analysis of eigenvalues, the character of spectral singularities can be simply found from the discriminant of the characteristic polynomial of $\mathcal{L}$, namely
\begin{equation}
	D(P_{\mathcal{L}}(\epsilon))=4 \left(\left(J^2-2 \eta ^2\right)^3-27 \eta ^4 J^2 \cos ^2(\Phi )\right),
\end{equation}
which is \textit{i)} zero if and only if at least two roots degenerate, \textit{ii)} positive if the roots are three distinct real numbers, and \textit{iii)} negative if there is one real root and two complex conjugate roots. These conditions define a stability diagram displayed in \autoref{fig:SCT_phi_sweep} (left), where we depict the sign of the imaginary complex surfaces in terms of $\Phi$ and $J/(2\sqrt{2}\eta)$. 

Roots are imaginary for all $\Phi$  if  $J<\sqrt{2}\eta$ or, if $J>\sqrt{2}\eta$, within the region $|\cos(\Phi)|>\frac{\left(J^2-2 \eta ^2\right)^{3/2}}{3 \sqrt{3} \eta ^2 J}$. Interestingly, within the strip $\sqrt{2}\eta<J<2\sqrt{2}\eta$, the SCT performs multiple dynamical phase transitions into oscillatory and exponentially-evolving phases as  $\Phi$ is tuned.

\begin{widetext}	
	
	\subsection{Quadrature-independent excitation transport in the SCT}\label{sec:quad_dep}
	The presence of disjoint loops in the SCT implies quadrature-independent transport of excitations (see Methods). Here we show how the phononic population dynamics (dubbed $n_i(t)=(X_i^2+Y_i^2)/2$ and $\gamma_i=0$) is nonreciprocal for $\Phi\neq\{0,\pi\}$. For it, we first show explicit results for the time evolution of a vector  $\vec{R}(0)^{\mathrm{i.c.1}}\equiv(X_1(0),0,0,Y_1(0),0,0)^T$ or $\vec{R}(0)^{\mathrm{i.c.2}}\equiv(0,X_2(0),0,0,Y_2(0),0)^T$ for the rotating frame Hamiltonian (Methods) and $\Phi=-\pi/2$,
	\begin{subequations}
		\begin{align}
			n_1(t)=&n_1(0)\frac{\left(\eta ^2+\left(\eta ^2-J^2\right) \cosh \left(t \sqrt{2 \eta ^2-J^2}\right)\right)^2}{\left(J^2-2 \eta ^2\right)^2},\\
			n_2(t)=&n_1(0)\frac{\left(\eta ^2+J \sqrt{2 \eta ^2-J^2} \sinh \left(t \sqrt{2 \eta ^2-J^2}\right)+\eta ^2 \left(-\cosh \left(t \sqrt{2 \eta ^2-J^2}\right)\right)\right)^2}{\left(J^2-2 \eta ^2\right)^2},\\
			n_3(t)=&n_1(0)\frac{\eta ^2  \left(\sqrt{2 \eta ^2-J^2} \sinh \left(t \sqrt{2 \eta ^2-J^2}\right)+J \left(-\cosh \left(t \sqrt{2 \eta ^2-J^2}\right)\right)+J\right)^2}{\left(J^2-2 \eta ^2\right)^2},
		\end{align}
	\end{subequations}
	Keeping the same initial condition but reversing the flux ($\Phi=\pi/2$), one observes reduced transport to resonator 2:
	\begin{subequations}
		\begin{align}
			n_1(t)=&n_1(0)\frac{\left(\eta ^2+\left(\eta ^2-J^2\right) \cosh \left(t \sqrt{2 \eta ^2-J^2}\right)\right)^2}{\left(J^2-2 \eta ^2\right)^2},\\
			n_2(t)=&n_1(0)\frac{\left(-\eta ^2+J \sqrt{2 \eta ^2-J^2} \sinh \left(t \sqrt{2 \eta ^2-J^2}\right)+\eta ^2 \cosh \left(t \sqrt{2 \eta ^2-J^2}\right)\right)^2}{\left(J^2-2 \eta ^2\right)^2},\\
			n_3(t)=&n_1(0)\frac{\eta ^2  \left(\sqrt{2 \eta ^2-J^2} \sinh \left(t \sqrt{2 \eta ^2-J^2}\right)+J \cosh \left(t \sqrt{2 \eta ^2-J^2}\right)-J\right)^2}{\left(J^2-2 \eta ^2\right)^2}.
		\end{align}
	\end{subequations}
	
	The condition for nonreciprocity is summarised by the condition
	\begin{equation}
		\frac{n_2(t,\Phi)}{n_1(0)}^{\mathrm{i.c.1}}\neq \frac{n_1(t,\Phi)}{n_2(0)}^{\mathrm{i.c.2}},
	\end{equation}
	which states that the energy reaching resonator 2 when the system is initialised in 1 is different from the energy reaching resonator 1 when the system is initialised at 2 with the same input energy.  The roles of resonators 1 and 2 are exactly exchanged if the flux is reversed. 
	
	The above expressions are valid in either the $\mathcal{P}_{gl}\mathcal{T}$ symmetric or broken region (in the above expressions, $J^2>2\eta^2$ and $J^2<2\eta^2$ respectively). In particular, they faithfully describe the nonreciprocal unstable dynamics reported in \autoref{fig:4}e from the main text. In the experimentally unstable regime, large amplitudes inevitably lead to nanomechanical self-oscillations that are seeded thermally. In particular, in the attenuated region, their coherent amplitude averages to zero. A quantitative analysis of this region, which would require exploring the stochastic differential equations for the resonators that incorporate the optomechanical nonlinearity to the next order, goes beyond the scope of the current work.

	\subsection{Flux-asymmetries in thermomechanical spectra of the SCT}\label{sec:flux_a}
	Here we demonstrate how the thermomechanical spectrum is asymmetric under flipping the sign of synthetic flux $\Phi\mapsto\Phi$, even if resonators present the same thermal occupation or in the limit of zero temperature. From now on we assume $\bar{n}_i=\bar{n}$ to simplify our analysis.
	
	Within the stable regime $\Im(\epsilon)<\gamma$, where the steady-state within the linear theory exists, the spectrum is readily obtained from~\autoref{eq:noise_sp}, applied to a single block~\autoref{eq:dyn_disjoint}. For arbitrary frequency and flux, the noise spectrum $\mathcal{S}_{ii}(\omega,\Phi)$ can be given in a closed form in terms of lengthy rational, trigonometric expressions.  For the current analysis, it is sufficient to consider the resonant case (in the rotating frame $\omega_i= \omega = 0$)

	\begin{subequations}
		\begin{align}
			\mathcal{S}_{11}(0,\Phi)=& \frac{4 \gamma  \left(-4 \eta ^2 \left(\gamma ^2 (\bar{n}+2)-4 J^2 \bar{n}\right)+\gamma ^2 (\bar{n}+1) \left(\gamma ^2+4 J^2\right)+16 \gamma  \eta ^2 J (2 \bar{n}+1) \sin (\Phi )+32 \eta ^4 (\bar{n}+1)\right)}{\left(\gamma ^3-8 \gamma  \eta ^2\right)^2+16 \gamma ^2 J^4+8 J^2 \left(\gamma ^2-4 \eta ^2\right)^2+128 \eta ^4 J^2 \cos (2 \Phi )},\\
			\mathcal{S}_{22}(0,\Phi)=&\frac{4 \gamma  \left(-4 \eta ^2 \left(\gamma ^2 (\bar{n}+2)-4 J^2 \bar{n}\right)+\gamma ^2 (\bar{n}+1) \left(\gamma ^2+4 J^2\right)-16 \gamma  \eta ^2 J (2 \bar{n}+1) \sin (\Phi )+32 \eta ^4 (\bar{n}+1)\right)}{\left(\gamma ^3-8 \gamma  \eta ^2\right)^2+16 \gamma ^2 J^4+8 J^2 \left(\gamma ^2-4 \eta ^2\right)^2+128 \eta ^4 J^2 \cos (2 \Phi )},\\
			\mathcal{S}_{33}(0,\Phi)=&\frac{4 \gamma  \left(\gamma ^2+4 J^2\right) \left(4 J^2 \bar{n}+\gamma ^2 \bar{n}+8 \eta ^2 (\bar{n}+1)\right)}{\left(\gamma ^3-8 \gamma  \eta ^2\right)^2+16 \gamma ^2 J^4+8 J^2 \left(\gamma ^2-4 \eta ^2\right)^2+128 \eta ^4 J^2 \cos (2 \Phi )}.
		\end{align}
	\end{subequations}
	
	Note that, while $\mathcal{S}_{33}(0,\Phi)=\mathcal{S}_{33}(0,-\Phi)$ (in fact, for all $\omega$), the noise spectra for resonators 1 and 2 display asymmetries in flux,
	\begin{subequations}
		\begin{align}
			\frac{\mathcal{S}_{11}(0,\Phi)}{\mathcal{S}_{11}(0,-\Phi)} =& \frac{-4 \eta ^2 \left(\gamma ^2 (\bar{n}+2)-4 J^2 \bar{n}\right)+\gamma ^2 (\bar{n}+1) \left(\gamma ^2+4 J^2\right)+16 \gamma  \eta ^2 J (2 \bar{n}+1) \sin (\Phi )+32 \eta ^4 (\bar{n}+1)}{-4 \eta ^2 \left(\gamma ^2 (\bar{n}+2)-4 J^2 \bar{n}\right)+\gamma ^2 (\bar{n}+1) \left(\gamma ^2+4 J^2\right)-16 \gamma  \eta ^2 J (2 \bar{n}+1) \sin (\Phi )+32 \eta ^4 (\bar{n}+1)},\\
			\frac{\mathcal{S}_{22}(0,\Phi)}{\mathcal{S}_{22}(0,-\Phi)} =& \frac{-4 \eta ^2 \left(\gamma ^2 (\bar{n}+2)-4 J^2 \bar{n}\right)+\gamma ^2 (\bar{n}+1) \left(\gamma ^2+4 J^2\right)-16 \gamma  \eta ^2 J (2 \bar{n}+1) \sin (\Phi )+32 \eta ^4 (\bar{n}+1)}{-4 \eta ^2 \left(\gamma ^2 (\bar{n}+2)-4 J^2 \bar{n}\right)+\gamma ^2 (\bar{n}+1) \left(\gamma ^2+4 J^2\right)+16 \gamma  \eta ^2 J (2 \bar{n}+1) \sin (\Phi )+32 \eta ^4 (\bar{n}+1)}.
		\end{align}
	\end{subequations}
	These asymmetries emerge as a combination of squeezing interactions ($\eta>0$) and chirality ($\Phi\neq\{0,\pi\}$), which is maximal at $\Phi=\pm \pi/2$ (note that by following a spectral decomposition of the susceptibility matrix $\chi_m$ in sec.IC., the noise spectra at a given frequency $\omega$ can be seen to be the consequence of excitations of the resonators along multiple, interfering paths). Remarkably, asymmetries persist at zero temperature ($\bar{n}\ll 1$), where only contributions from two-mode squeezed vacuum fluctuations exist:
	\begin{align}
		\frac{\mathcal{S}_{11}(0,\Phi)}{\mathcal{S}_{11}(0,-\Phi)} = \frac{1}{\frac{1}{2}-\frac{8 \gamma  \eta ^2 J \sin (\Phi )}{\gamma ^4-8 \gamma ^2 \eta ^2+32 \eta ^4+4 \gamma ^2 J^2}}-1,&&
		\frac{\mathcal{S}_{22}(0,\Phi)}{\mathcal{S}_{22}(0,-\Phi)} = \frac{1}{\frac{1}{2}+\frac{8 \gamma  \eta ^2 J \sin (\Phi )}{\gamma ^4-8 \gamma ^2 \eta ^2+32 \eta ^4+4 \gamma ^2 J^2}}-1.
	\end{align}
	
	In addition, asymmetries are optimal upon the matching condition $J=(\sqrt{\gamma ^4-8 \gamma ^2 \eta ^2+32 \eta ^4})/(2 \gamma)$.  
\end{widetext}

\end{document}